\documentclass[notitlepage, nofootinbib, amsmatE,amssymb, aps, prd, superscriptaddress]{revtex4-1}

\usepackage{graphicx,grffile}
\usepackage{dcolumn}
\usepackage{relsize}
\usepackage{verbatim}
\usepackage{bm,physics}
\usepackage{amsthm}
\usepackage{ulem}
\usepackage{lipsum}
\usepackage{appendix}

\usepackage{graphicx,xcolor,float}
\usepackage[export]{adjustbox}
\usepackage[caption=false]{subfig}
\usepackage{tikz}
\usepackage[text= black , arrow = gray ]{callouts}
\usepackage{hyperref}

\usepackage{braket}
\usetikzlibrary{matrix}
\begin{document}
 
\title{Phase Transitions of Correlations in Black Hole Geometries}

\author{Sristy Agrawal}
\email{sristy.agrawal@colorado.edu}
\affiliation{Department of Physics, University of Colorado, Boulder CO 80309, USA}
\affiliation{JILA, University of Colorado/NIST, Boulder, CO, 80309, USA}

\author{Oliver DeWolfe}
\email{oliver.dewolfe@colorado.edu}
\affiliation{Department of Physics and Center for Theory of Quantum Matter, University of Colorado, Boulder CO 80309, USA}

\author{Joshua Levin}
\email{joshua.t.levin@colorado.edu}
\affiliation{Department of Physics, University of Colorado, Boulder CO 80309, USA}
\affiliation{JILA, University of Colorado/NIST, Boulder, CO, 80309, USA}

\author{Graeme Smith}
\email{graeme.smith@colorado.edu}
\affiliation{Department of Physics and Center for Theory of Quantum Matter, University of Colorado, Boulder CO 80309, USA}
\affiliation{JILA, University of Colorado/NIST, Boulder, CO, 80309, USA}

\date{\today}
             
\begin{abstract}
We study the holographic realization of optimized correlation measures -- measures of quantum correlation that generalize elementary entropic formulas -- in two-dimensional thermal states dual to spacetimes with a black hole horizon. We consider the symmetric bipartite optimized correlation measures: the entanglement of purification, Q-correlation, R-correlation, and squashed entanglement, as well as the mutual information, a non-optimized correlation measure, and identify the bulk surface configurations realizing their geometric duals over the parameter space of boundary region sizes and the black hole radius. This parameter space is divided into phases associated with given topologies for these bulk surface configurations, and first-order phase transitions occur as a new topology of bulk surfaces becomes preferred. The distinct phases can be associated with different degrees of correlation between the boundary regions and the thermal environment. The Q-correlation has the richest behavior, with a structure of nested optimizations leading to two topologically distinct bulk surface configurations being equally valid as geometric duals at generic points in the phase diagram.
\end{abstract}

\maketitle

\section{Introduction}

Understanding the nature of correlations that exist in quantum states is of fundamental importance to quantum information theory, and quantum physics more generally.  Some simple measures of correlation, such as the mutual information, consist of a linear combination of entropies.  Other more complex measures of correlation, such as the entanglement of purification ($E_P$) \cite{EP02} and the squashed entanglement ($E_{\rm sq}$) \cite{Tucci,Christandl}, result from considering all possible purifications of a quantum state, and optimizing a linear combination of entropies involving the state and its extensions. Such optimized  information measures are able to parse finer details of the correlations in a quantum state.   A particularly interesting class of such measures also has the property of montonicity under local processing -- they can never increase when degrees of freedom are removed -- and these have been called {\it optimized correlation measures} (OCMs) \cite{LS20}.

Meanwhile, in the context of the AdS/CFT correspondence \cite{maldacena1999} it has become clear that the geometry of certain quantum gravitational systems is encoded in the entanglement structure of the dual quantum field theory. In particular, the entanglement entropy of a spacetime region is realized as an extremal surface in the gravity dual. Essential properties of information measures such as sudadditivity emerge geometrically from the extremization process, and quantities like the mutual information can undergo sudden transitions as the size of the spacetime region is varied.

It is natural to inquire how the more general class of optimized correlation measures is realized in terms of geometry, and what this can teach us about quantum gravity. Several steps in this direction have been taken in a case-by-case basis. In \cite{TU18,Swing18}, it was proposed that the gravity dual of $E_P$ is a spacetime surface called the entanglement wedge cross-section (EWCS).\footnote{In addition to $E_P$, several other quantities have been proposed as the boundary dual of the EWCS; in \cite{dutta2019canonical} it was argued, using ideas from \cite{engelhardt2019coarse,engelhardt2018decoding}, that the EWCS is dual to the {\it reflected entropy}.  In \cite{kudler2019entanglement}, the {\it log negativity} \cite{vidal2002computable} was proposed as the boundary dual of the EWCS, while \cite{tamaoka2019entanglement} argued that the {\it odd entanglement entropy} is the correct dual.  Indeed, investigations of the EWCS and its proposed boundary duals have been thoroughly studied in a large body of work \cite{bao2018holographic,hirai2018towards, espindola2018entanglement,bao2019conditional, bao2019entanglement,cheng2020optimized,ghodrati2019connection, chu2020generalizations,velni2019some,velni2020evolution,agon2019geometric,caputa2019holographic, kudler2019holographic,jokela2019notes,kudler2020quantum, kusuki2019dynamics,kusuki2019derivation,jeong2019reflected, kusuki2020entanglement,umemoto2019quantum,akers2020,bao2019multipartite,Marolf2019CFTSA,mollabashi2020field,du2020inequalities,ghodrati2021correlations}.  Many of these proposed dualities coincide with $E_P$ in the limit of classical geometry, and their identification need not be mutually exclusive.} The $E_P$ proposal and the EWCS have been generalized to multipartite systems in \cite{umemoto2018entanglement}, which also argued that in holographic theories, $E_{\rm sq}$ coincides with (and shares its holographic dual with) half the mutual information, though in generic quantum states this need not be true.  

To shed more light on these issues, we have undertaken a program to both construct and classify optimized correlation measures, and to formulate a systematic procedure for realizing them in a holographic context. In \cite{LS20}, bipartite OCMs were systematically enumerated as the rays of a convex polyhedral cone, and two new bipartite OCMs were identified, the $Q$-correlation and the $R$-correlation.  To realize these measures holographically, it was necessary to develop a prescription for purifying states and evaluating their entropies in a geometric context. This was begun in \cite{LSD20}, where the $Q$-correlation and $R$-correlation were realized holographically in the AdS$_3$ vacuum using a prescription inspired by the surface-state correspondence (SSC) of Miyaji and Takayanagi \cite{SurfaceState}; the geometric presentations of $E_P$ and $E_{\rm sq}$ proposed in \cite{TU18,Swing18,umemoto2018entanglement} were also reproduced by this method. The $R$-correlation coincided with the $E_P$ as the EWCS, while the $Q$-correlation was realized as a novel sum and difference of certain surfaces. The prescription was elaborated in \cite{LDS20}, where tripartite OCMs were systematically classified, and their holographic duals (some of which had multiple presentations that could be quite intricate) were again determined in the AdS$_3$ vacuum.

In this paper, we extend this work beyond the vacuum to thermal states of a (1+1)-dimensional holographic CFT, corresponding to (2+1)-dimensional asymptotically AdS bulk geometries containing a black hole horizon, the BTZ spacetimes \cite{BTZ}. We find that the parameter space corresponding to the radius of the black hole and the sizes and locations of the boundary regions divides itself into {\it phases} where the bulk surface configurations which realize the OCMs take particular forms.  The phase transitions are associated with transitions in the bulk surface configurations which realize the OCMs, and at phase boundaries the correlation measures are continuous but with discontinuous derivatives, indicating the transitions may be thought of as first order. In some cases, the transitions are also associated with various inequalities between the OCMs becoming equalities, or vice versa. It was already known that the EWCS, and hence the holographic realization of $E_P$, can transition from continuous to discontinuous as the radius of the black hole grows larger, or as the two boundary regions change sizes or locations \cite{TU18}. The $Q$-correlation is found to have a considerably richer structure, being holographically described by one of four different {\it pairs} of equivalent configurations of bulk surfaces, depending on the size and angular separation of the boundary regions and the horizon radius. The fact that this correlation measure is generically represented by two different-looking configurations of surfaces, that nonetheless calculate the same value, can be understood in terms of the ``minimization within a minimization" structure where each entropy inside the formula for the correlation measure is calculated holographically by an extremization process, before the whole formula is extremized over all purifications; an entropy appearing in the formula with a minus sign will find itself optimized where two bulk surface configurations are numerically equal.

We first study thermal states, where the boundary space is divided into two complementary regions in the presence of a black hole. We observe phases of total correlation of one region with the thermal environment, with the other region completely uncorrelated, as well as phases of mixed correlation of both regions with the environment, and phases where one or both the $Q$-correlation and $E_P$ are discontinuous. We then turn to the more intricate space of reduced thermal states, where the two regions do not span the boundary and hence additional degrees of freedom are traced out. Here new phases appear, a phase of partial correlation where one region is uncorrelated with the environment but the other is not totally correlated, as well as trivial phases where the horizon does not affect the bulk surface configuration or the entanglement wedge is disconnected and the correlation measures vanish. In the reduced thermal states the $Q$-correlation also takes its more intricate form of being represented by two geometrically different but numerically identical bulk surface configurations in each phase.

The structure of this paper is as follows.  In Sec.~\ref{Sec:OCMAdSCFT} we review bipartite optimized correlation measures, their holographic presentations, and how the geometric realization of entanglement entropy is affected by the presence of a horizon. In Sec.~\ref{OCMs thermal} we present our results on the geometric evaluation of bipartite OCMs for thermal states, including the one-dimensional phase diagram for fixed horizon size, and corresponding bulk surface configurations.  In Sec.~\ref{OCMs reduced thermal}, we address the more general case of reduced thermal states, presenting several two-dimensional slices of the phase diagram and the corresponding bulk surface configurations, including matching pairs for the $Q$-correlation. Finally, in Sec.~\ref{Discussion} we wrap up with some concluding remarks.

\section{Optimized Correlation Measures in the AdS/CFT Correspondence}
\label{Sec:OCMAdSCFT}

\subsection{Optimized Correlation Measures}\label{OCMs}
Among the simplest information measures for a quantum system are those made from linear combinations of entropies for subsystems; for example for a bipartite state $\rho_{AB}$ with associated reduced density matrices $\rho_A$ and $\rho_B$ we can consider the mutual information $I(A:B) \equiv S(\rho_A) + S(\rho_B) - S(\rho_{AB})$ or the conditional entropy $S(A|B) \equiv S(\rho_{AB}) - S(\rho_B)$. (In what follows we will generally write $S(A), S(AB)$ for $S(\rho_A), S(\rho_{AB})$ and so on, with the dependence on the state understood.) However, other measures of information in quantum states besides these linear entropic formulas can be useful.

Optimized information measures are functions of a multipartite quantum state calculated by considering all possible extensions or purifications of the state, and extremizing some linear entropic formula (depending in general on both the degrees of freedom of the state and of the extension) over all possible extensions.
For a bipartite state $\rho_{AB}$, we may consider all possible pure states $\ket\psi_{ABx}$, where $x$ denotes additional degrees of freedom added to the Hilbert space such that ${\rm Tr}_x \ket\psi \bra\psi = \rho_{AB}$. Moreover we can consider all possible partitions of the extension $x$ into ancilla degrees of freedom associated to $A$, which we call $a$, and ancilla degrees of freedom associated to $B$, which we call $b$. Then let $f(\ket\psi_{ABab})$, called the objective function, be a linear entropic formula in $A$, $B$, $a$, $b$. The optimized information measure $E(\rho_{AB})$ is then an infimum of the objective function over all such purifications:
\begin{align}
E(\rho_{AB}) = \inf_{\ket\psi_{ABab}}f(\ket\psi_{ABab})\,.\label{OCM}
\end{align}
Different objective functions produce different optimized information measures. One may always choose to use the purity of $\ket\psi$ to remove explicit dependence of the objecetive function on one of the ancilla, for example $b$, and then one may think of the optimization as being over extensions $a$, but we will use the language of purifications. If the objective function is chosen independent of the extensions $a$ and $b$, the optimization is trivial and the expression reduces to an ordinary linear entropic formula. A generalization to multipartite systems is straightforward. 

Several bipartite optimized information measures have been defined in the literature with various useful properties. One is the entanglement of purification, denoted $E_P(\rho_{AB})$ or in a compact notation $E_P(A:B)$, and defined as \cite{EP02}
\begin{align}
    E_P(A:B) \equiv \inf_{\ket\psi} S(Aa) \,,\label{EP}
\end{align}
which minimizes the entanglement between subsystems $Aa$ and $Bb$. Its regularized version,
$E_P^\infty(A:B) \equiv \lim_{n\to\infty}\frac{1}{n}E_P(\rho_{AB}^{\bigotimes n})$, 
has the operational interpretation as the number of EPR pairs per copy of the state $\rho_{AB}$ required to form $n$ copies of $\rho_{AB}$ using local operations and vanishing communication, in the limit of large $n$ \cite{EP02}. In most cases, this quantity is difficult to calculate due to the likely nonadditivity of $E_P$ \cite{NonAdd}.  But since $E_P$ is expected to be additive for holographic CFT states \cite{TU18}, one has in that case $E_P^\infty = E_P$, and $E_P$ takes on this operational interpretation itself. 

Another bipartite optimized information measure is the squashed entanglement $E_{\rm sq}(A:B)$ \cite{Tucci,Christandl},

\begin{align}
    E_{\rm sq}(A:B) \equiv \inf_{\ket\psi} {1 \over 2} \left(S(Aa) + S(Ba) - S(a) - S(b)\right)  \,,
\end{align}
which minimizes $I(A:B|a)$, the  mutual information between $A$ and $B$ conditioned on the extension $a$, and has the property that it vanishes on separable states, and thus does not measure classical correlations, but quantum entanglement only.

Both the entanglement of purification and the squashed entanglement also share the property of monotonicity under local processing, that is
\begin{eqnarray}
\label{Monotonicity}
	E_\alpha(\rho_{(A_1A_2)B}) \geq E_\alpha(\rho_{A_1B}) \,,
\end{eqnarray}
and similarly for $B$. An optimized information measure that is also monotonic was called an {\it optimized correlation measure} (OCM) in \cite{LS20}, where bipartite OCMs were classified. The list of such measures symmetric between the two parties consists of the mutual information $I(A:B)$ (a case with trivial optimization), $E_P$, $E_{\rm sq}$, and two additional measures, the $Q$-correlation,
\begin{align}
\label{eqn:EQ}
    E_Q(A:B) \equiv \inf_{\ket\psi} {1 \over 2} \left( S(A) + S(B) + S(Aa) - S(Ba)  \right)\,,
\end{align}
and the $R$-correlation,
\begin{align}
    E_R(A:B) \equiv \inf_{\ket\psi}  {1 \over 2} \left( 2 S(Aa) +S(AB) - S(a) - S(b)  \right)\,.
\end{align}
All these correlation measures are non-negative, a consequence of monotonicity together with being bounded below, and this can be made manifest by writing the objective functions as conic combinations of mutual informations and conditional mutual informations \cite{LDS20}.

These five bipartite OCMs can be shown to satisfy a hierarchy of inequalities,
\begin{eqnarray}
	E_{\rm sq}(A:B) \leq {1 \over 2} I(A:B) \leq E_R(A:B), E_Q(A:B) \leq E_P(A:B) \leq \min(S(A), S(B)) \,,
\end{eqnarray}
the proof of which is reviewed in Appendix A. The $Q$-correlation and $R$-correlation also share with $E_P$ the inequality
\begin{align}
E_\alpha(A:BC) \geq {1 \over 2} (I(A:B) + I(A:C)) \,,
\end{align}
for $\alpha = P, Q, R$. The operational interpretations of the $Q$-correlation and $R$-correlation are still open questions, although the $Q$-correlation was found to be related to a bipartite entanglement measure called the symmetric side-channel assisted distillable entanglement $I^{ss}$ by \cite{LSD20}
\begin{align}
    S(A) = E_Q(A:B) + I^{ss}(E\rangle A)\,,
\end{align}
for pure states $\rho_{ABE}$.
$I^{ss}(A\rangle B)$ is defined as the number of EPR pairs distillable from $\rho_{AB}$ using local operations and classical communication (LOCC) and the assistance of a symmetric channel from $A$ to $B$ \cite{SSW06}.  Thus, $E_Q(A:B)$ can be understood as the part of $A$'s entropy which cannot be distilled as entanglement with $E$ given a symmetric channel from $E$ to $A$.

Given two OCMs, optimizing a conic combination of their objective functions will always produce an OCM, since conic combinations preserve the monotonicity property. OCMs whose objective functions cannot be written as a conic combination of others may be called ``extremal". The five quantities $I(A:B)$, $E_P$, $E_{\rm  sq}$, $E_Q$ and $E_R$ are the complete list of symmetric OCMs on two parties that are extremal, and thus all other such measures can be built as combinations of their objective functions.

\subsection{Ryu-Takayanagi Formula and Entanglement Entropy}\label{RT}

The AdS/CFT correspondence states that a quantum gravitational theory living in an asymptotically anti-de Sitter spacetime is precisely dual
to a lower-dimensional, non-gravitational quantum field theory living on the spacetime boundary. Each state of the boundary field theory (in the most symmetric case, a conformal field theory (CFT)) corresponds to a geometry in the gravity dual.
Since AdS/CFT is a duality, all information contained in the boundary state is also present in the dual gravity theory.  In particular, the entanglement structure of the boundary field theory is encoded in the bulk geometry.

\begin{figure}[H]

\centering
\begin{tikzpicture}
\node[inner sep=0pt] at (5,0)
    {\includegraphics[scale = 0.32]{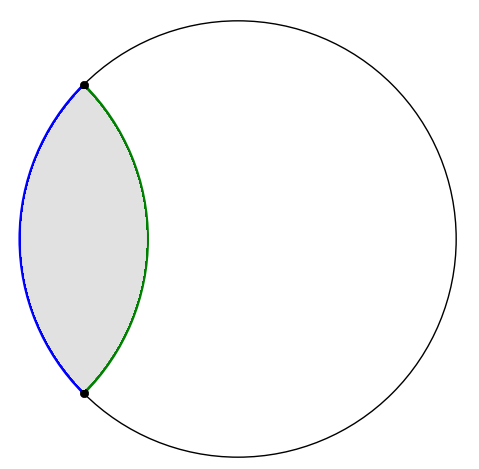}};
\note{2.7,0}{\large{$A$}}    
\note{ 3.6,0}{\large{$\Omega_A$}}
\note{ 4.6,1.2}{\large{$\Gamma_A$}}
   \draw[black, thick, ->] (4.5,1)->(4.15, 0.68);
\end{tikzpicture}
\caption{Boundary region $A$ is shown in blue, and its RT-surface $\Gamma_A$ is shown in green.  The shaded region labelled $\Omega_A$ is the entanglement wedge of $A$. }
\label{fig:AdS_A}
\end{figure}

The simplest expression of this occurs for a time-independent geometry, in the limit where the bulk gravity theory is classical, corresponding to the limit of a large number $N$ of degrees of freedom for the boundary theory. Then, the entanglement entropy of a spatial region $A$ of the boundary, in a given state, can be calculated as a minimal area surface in a constant-time slice of the corresponding bulk geometry, according to the Ryu-Takayanagi (RT) formula \cite{RT},
\begin{align}
\label{RTEqn}
    S(A) = \min_{C_A}\frac{{\rm Area}(C_A)}{4 G_N} \,,
\end{align}
where the minimum is taken over all bulk surfaces $C_A$ which share the same boundary as $A$, and are homologous to $A$, meaning the union of $A$ and $C_A$ forms the boundary of a submanifold of the spatial slice. $G_N$ is Newton's gravitational constant in the bulk theory, and the expression (\ref{RTEqn}) should be understood as the leading order term in an expansion in $1/G_N$.\footnote{This expression can be generalized away from constant-time geometries to a covariant expression \cite{Hubeny:2007xt}, and quantum corrections include entanglement in the bulk, in general shifting the RT surface to a quantum extremal surface \cite{Faulkner:2013ana, engelhardt2015quantum}.  Also see \cite{nakata2021new} for more recent work on time-dependent generalizations of holographic entanglement entropy. In this work, we will stick to the classical, time-independent case.} 
The surface $\Gamma_A$ achieving the minimum is called the RT surface of $A$, as shown in Fig.~\ref{fig:AdS_A}. In this paper we will focus on the AdS$_3$/CFT$_2$ case; then a constant time-slice of the bulk is a two-dimensional hyperbolic geometry, and the RT surfaces become geodesic curves.

The spatial region $\Omega_A$ bounded by $A$ and $\Gamma_A$ is the entanglement wedge of $A$\footnote{Technically this is a spatial slice of the entanglement wedge, which forms the spacetime domain of dependence of this region, but we will abuse terminology slightly to refer to the spatial slice as the wedge.}. The entanglement wedge represents the bulk physics that can be reconstructed by the boundary region $A$ alone, and everything in the bulk beyond the RT surface $\Gamma_A$ requires additional degrees of freedom other than those contained in $A$ to construct; by computing $S(A)$, the area of $\Gamma_A$ counts the entanglement of $A$ with these additional degrees of freedom.

\begin{figure}[H]

\centering
\begin{tikzpicture}
\node[inner sep=0pt] at (5,0)
    {\includegraphics[scale = 0.32]{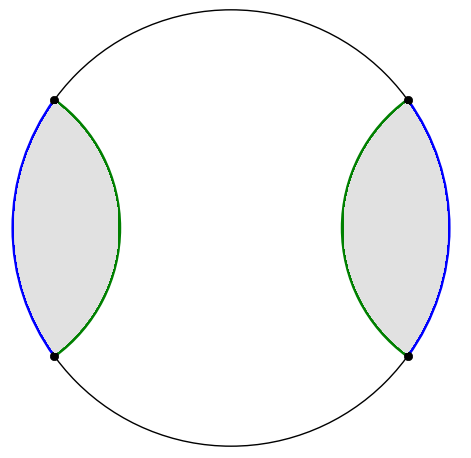}};
\node[inner sep=0pt] at (11,0)
    {\includegraphics[scale = 0.32]{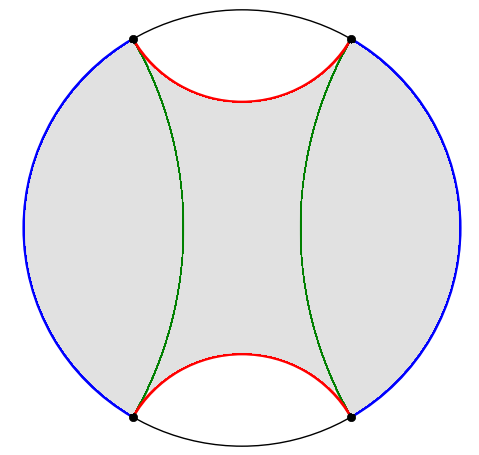}};
\note{2.7,0}{\large{$A$}}    
\note{ 3.6,0}{\large{$\Omega_A$}}
\note{7.1,0}{\large{$B$}}    
\note{ 6.3 ,0}{\large{$\Omega_B$}}

\note{8.8,0}{\large{$A$}}  
\note{13.2,0}{\large{$B$}}  
\note{ 11,0.5}{\large{$\Omega_{AB}$}}
\note{ 11,-2.5}{\large{(b)}}
\note{ 5,-2.5}{\large{(a)}}
\note{ 4.6,1.2}{\large{$\Gamma_A$}}
\note{ 5,-1.2}{\large{$\Gamma_B$}}
\note{ 11,1.25}{\small{$\Gamma_{AB}$}}
\note{ 11,-1.3}{\small{$\Gamma_{AB}$}}
   \draw[black, thick, ->] (10.58,0.3)->(10.05, -0.2);
      \draw[black, thick, ->] (11.48,0.3)->(11.95, -0.2);
    \draw[black, thick, ->] (11,0.2)->(11, -0.4);
    
\draw[black, thick, ->] (4.5,1)->(4.15, 0.68);    \draw[black, thick, ->] (5.2,-1)->(5.7,- 0.58);
\end{tikzpicture}
\caption{(a) Regions $A$ and $B$ comprising the bipartite region $AB$ are separated enough that $\Gamma_{AB} = \Gamma_A \cup \Gamma_B$, and hence the entanglement wedge $\Omega_{AB} = \Omega_A \cup \Omega_B$ is disconnected and the mutual information $I(A:B)$ vanishes. (b) Regions $A$ and $B$ are close enough that the joint entanglement wedge $\Omega_{AB}$ is larger than $ \Omega_A \cup \Omega_B$, and hence connected. The RT surface $\Gamma_{AB}$, shown in red, is distinct from $ \Gamma_A \cup \Gamma_B$, and the non-zero mutual information $I(A:B)$ is given by the sum of the green curves minus the sum of the red curves.}
\label{fig:AdS_AB}
\end{figure}
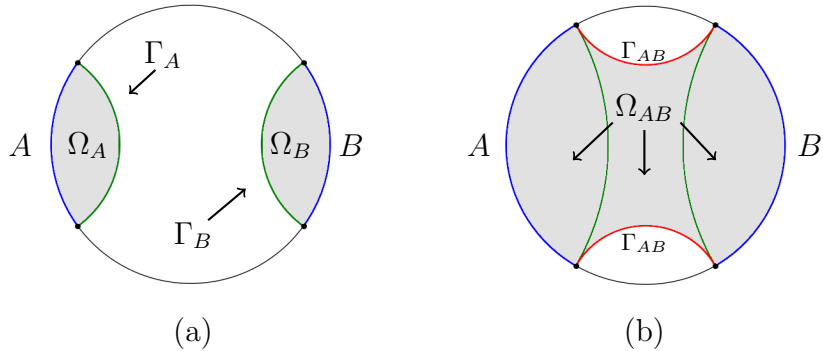

We can  use the RT formula to compute the entropy of a bipartite boundary region $AB$, where $AB$ is the union of two regions $A$ and $B$,  as shown in Fig.~\ref{fig:AdS_AB}.  If the two regions $A$ and $B$ are sufficiently far apart on the boundary, as in Fig.~\ref{fig:AdS_AB}a, the RT-surface $\Gamma_{AB}$ is just the union of the RT-surfaces  $\Gamma_A$ and $\Gamma_B$ (green in the figure), and likewise the entanglement wedge $\Omega_{AB}$ is the union of the entanglement wedges $\Omega_A$ and $\Omega_B$ (gray in the figure) and is therefore disconnected.  In this case, we see that $S(AB) = S(A) + S(B)$, and therefore the mutual information $I(A:B) \equiv S(A) + S(B) - S(AB)$ vanishes, so $\rho_{AB} = \rho_A\otimes\rho_B$ and the states on $A$ and $B$ are totally uncorrelated (to leading order in $1/G_N$).  But if the regions $A$ and $B$ are sufficiently close together, as in Fig.~\ref{fig:AdS_AB}b, then $\Gamma_{AB}$ is not the union of $\Gamma_A$ and $\Gamma_B$, but instead takes the form of the curves shown in red.  In this case the entanglement wedge  $\Omega_{AB}$ is connected, and contains more than just $\Omega_A \cup \Omega_B$; the regions $A$ and $B$ together can reconstruct more bulk physics than the two on their own. In terms of entropies we then have $S(AB) < S(A) + S(B)$, so the mutual information (represented in the figure as the green curves, minus the red curves) is nonzero, $I(A:B)>0$.

In an asymptotically AdS metric, the length of any curve extending to the boundary is divergent. Correspondingly, entanglement entropies in the field theory have short-distance divergences. To make entropies finite, we must introduce a regularization scheme in which we remove the divergent parts of curves that extend to the boundary.  This is usually done with a radial cutoff, wherein the arc length integrals used to calculate lengths of curves will all be cut off at some maximum radius $r_{cut}$,  corresponding to a maximum energy scale in the boundary field theory.  This makes all entanglement entropies finite but cutoff-dependent. However, the mutual information for disconnected regions is independent of the cutoff; this is visible geometrically in Fig.~\ref{fig:AdS_AB}b, since every curve extending to the boundary is balanced by another curve extending to the same boundary point with the opposite sign. Were $A$ and $B$ to share a boundary point, however, cutoff-dependence would reappear in $I(A:B)$.

Quantum informational identities such as the positivity of the mutual information (weak subadditivity) and positivity of the conditional mutual information (strong subadditivity) follow from geometrical considerations using the RT formula as well. In addition, the RT formula leads to further constraints on entropies that need not hold for a generic quantum state.  In particular, the tripartite mutual information $I_3(A:B:C) \equiv S(A) + S(B) + S(C) - S(AB) - S(AC) - S(BC) + S(ABC)$ has no definite sign in general, but for holographic states can be shown to be non-positive $I_3(A:B:C)\leq 0$,
called the monogamy of mutual information (MMI) \cite{Hayden:2011ag}. This indicates that the {\it holographic entropy cone} (in the limit corresponding to classical gravity) is more constrained than for a general quantum system \cite{HEC15}.

\subsection{Surface-State Correspondence and Optimized Correlation Measures}\label{SSC}

In general, any linear entropic formula may be calculated in the AdS/CFT correspondence using the Ryu-Takayanagi formula. To evaluate optimized correlation measures however, we must obtain a geometric presentation of purifying our states, and we must have a rule for assigning entropies to the extensions. Our prescription for this is inspired by the surface-state correspondence.

The surface-state correspondence (SSC) \cite{SurfaceState} is an extension of the AdS/CFT correspondence in which a state in a quantum Hilbert space is associated not only to the AdS boundary, but to surfaces in the interior of the geometry as well.  A pure state is dual to a spacelike closed convex surface of spacetime codimension 2, while a mixed state is dual to a strict subset of a closed convex surface with a boundary, i.e.~a pure state with a subsystem traced out (erased). The entropy of a mixed state is given by the area of the minimal surface which closes the surface; note that for states whose dual surfaces lie in the AdS boundary, this reduces exactly to the RT formula.  Heuristically, the closed and open surfaces corresponding to pure and mixed states are like smaller versions of the AdS boundary or a component of it, nested inside the original boundary.

In light of the SSC, the RT formula can be understood in a slightly different but equivalent way.  Given a boundary region $A$, the SSC tells us that the RT surface of $A$ can be viewed as a purifying system for $A$. Since complementary systems of a pure state have the same entropy, we can find the entropy of $A$ by computing the entropy of its purifying system.  Since the RT surface always lies along a minimal area surface, the entropy of its SSC dual is just its area, which yields the RT formula.  This alternate way of conceptualizing the computation of entropies will be useful in what follows.

We can use the SSC entropy formula as a prescription for the evaluation of OCMs as follows. We begin with boundary regions $A$, $B$ associated to a density matrix $\rho_{AB}$. There is then a class of geometric purifications consisting of closing up the boundary region $AB$ with all possible convex surfaces sharing the boundary of $AB$.  One such purification goes along the AdS boundary; others extend into the bulk. The extremal such purification forms a closed surface by connecting $AB$ precisely along the RT surface $\Gamma_{AB}$. The optimum purification must then be divided into ancilla $a$ and $b$, so the curves representing the purifying system must be partitioned, and the objective function of the correlation measure minimized over all such partitions, with the entropies calculated using the SSC prescription.

In principle one can imagine other purifications that cannot be realized geometrically. We will make the assumptions that the optimum purification for our correlation measures is geometric \cite{cheng2020optimized}, and moreover that it is the purification of extremal length; note to go beyond the extremal surface violates the SSC prescription for convexity. Thus in practice, the geometric realization of the optimized correlation measures involves partitioning the RT surface $\Gamma_{AB}$ into portions $a$ and $b$, calculating the objective function $f^\alpha(A,B,a,b)$ for each such partition, and minimizing.

\begin{figure}[H]
\centering
\begin{tikzpicture}
\node[inner sep=0pt] at (5,0)
    {\includegraphics[scale = 0.32]{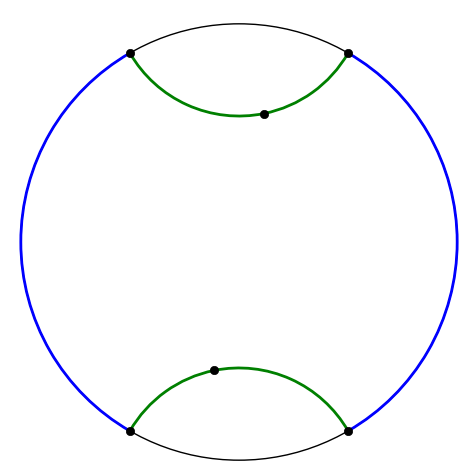}};
\node[inner sep=0pt] at (11,0)
    {\includegraphics[scale = 0.32]{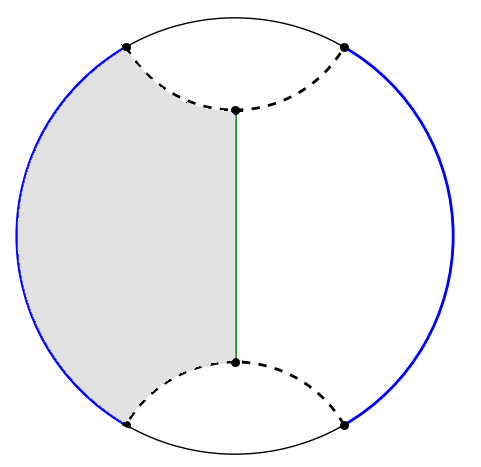}};
\note{2.9,0}{\large{$A$}}
\note{7.1,0}{\large{$B$}}
\note{4.5,0.9}{\small{$a$}}
\note{5.7,1.0}{\small{$b$}}
\note{4.3,-1}{\small{$a$}}
\note{5.5,-.9}{\small{$b$}}
\note{8.7,0}{\large{$A$}}  
\note{13.2,0}{\large{$B$}}  
\note{10,0}{\large{$\Omega_{Aa}$}}
\note{10.4,1}{\small{$a$}}
\note{10.4,-1}{\small{$a$}}
\note{11.6,1}{\small{$b$}}
\note{11.6,-1}{\small{$b$}}
\note{ 11,-2.5}{\large{(b)}}
\note{ 5,-2.5}{\large{(a)}}
\end{tikzpicture}
\caption{(a) The RT-surface for $AB$ can be thought of as a system which purifies $\rho_{AB}$, and can be partitioned into two pieces $a$ and $b$.  Entropies of systems that do not lie fully in the boundary can then be computed using the SSC. (b) Partitioning the purifying system and computing $S(Aa)$ according to the SSC, we find it is given by the length of the curve connecting to the two partitioning points.  Minimizing over partitions, we find that $E_P(A:B)$ is the entanglement wedge cross section.}
\label{fig:AdS_SSC}
\end{figure}

As an example, let us compute $E_P(A:B)$ for two regions $A$ and $B$ on the boundary which are large/close enough that the entanglement wedge $\Omega_{AB}$ is connected, in the field theory vacuum state dual to pure three-dimensional anti-de Sitter space.  We first purify the state $\rho_{AB}$,  partition the purifying system into $a$ and $b$, and minimize  the objective function $S(Aa)$ over all such partitions. Taking the optimal purification to be the RT surface $\Gamma_{AB}$, we show one possible partition into $a$ and $b$ in  Fig.~\ref{fig:AdS_SSC}a; we call the points dividing the partition {\it optimization points}. Then, the SSC prescription says that $S(Aa)$ is given by the length of the geodesic connecting the optimization points, since this is the minimal length closure of the open curve defined by $Aa$. The minimizing configuration is shown in Fig.  \ref{fig:AdS_SSC}b; for $A$ and $B$ equal size and diametrically opposite, the optimization points end up at the centers of the arcs of $\Gamma_{AB}$. This curve, the minimal-length path dividing $\Omega_{AB}$ into a part containing all of $A$ and a part containing all of $B$, is called the entanglement wedge cross-section (EWCS). We will refer to the geometric realization of our information measures as a {\it bulk surface configuration} (BSC). Thus our prescription states that the BSC dual to $E_P(A:B)$ is precisely the EWCS, as originally proposed in \cite{TU18,Swing18}.

\begin{figure}
\begin{tikzpicture}
\node[inner sep=0pt] at (5,0)
    {\includegraphics[scale = 0.32]{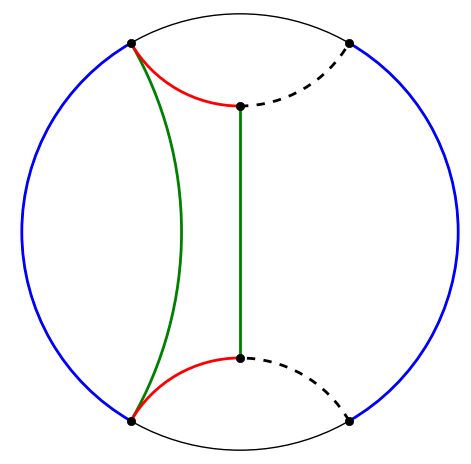}};
\node[inner sep=0pt] at (11,0)
    {\includegraphics[scale = 0.31]{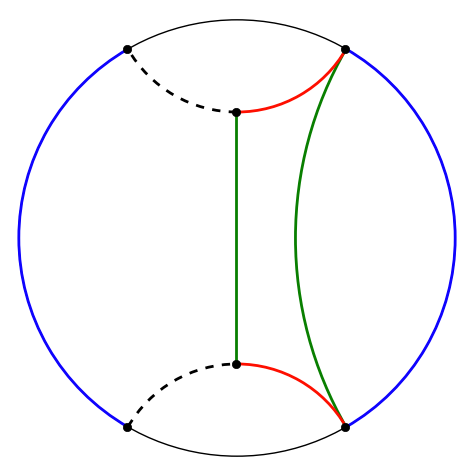}};
\note{2.9,0}{\large{$A$}}
\note{7.1,0}{\large{$B$}}
\note{4.6,1.3}{\small{$a$}}
\note{5.4,1.35}{\small{$b$}}
\note{4.6,-1.3}{\small{$a$}}
\note{5.5,-1.3}{\small{$b$}}

\note{10.6,1.3}{\small{$a$}}
\note{11.4,1.35}{\small{$b$}}
\note{10.6,-1.3}{\small{$a$}}
\note{11.5,-1.3}{\small{$b$}}
\note{8.9,0}{\large{$A$}}  
\note{13.1,0}{\large{$B$}}  
\note{ 11,-2.5}{\large{(b)}}
\note{ 5,-2.5}{\large{(a)}}
\end{tikzpicture}
\caption{Two equivalent representations for $E_Q(A:B)$, derived using the prescription of the SSC.  The fact that the two representations are numerically equal fixes the location of the partitioning points.}\label{AdS EQ}
\end{figure}

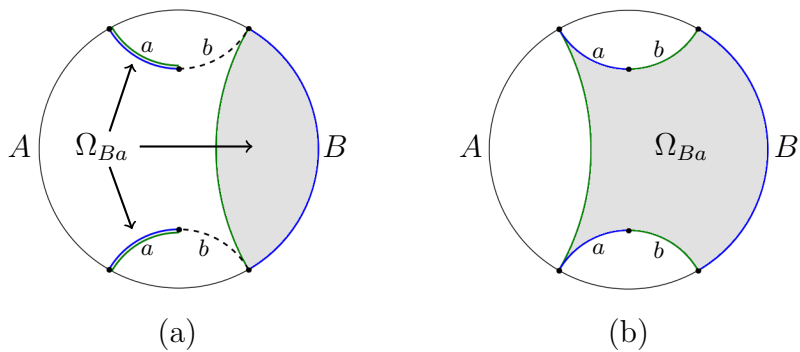
\begin{figure}[H]
\centering
\begin{tikzpicture}
\node[inner sep=0pt] at (5,0)
    {\includegraphics[scale = 0.32]{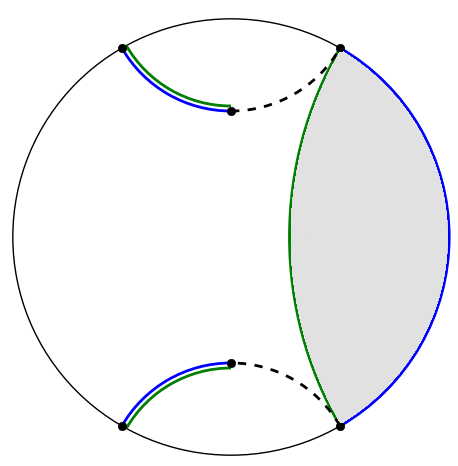}};
\node[inner sep=0pt] at (11,0)
    {\includegraphics[scale = 0.32]{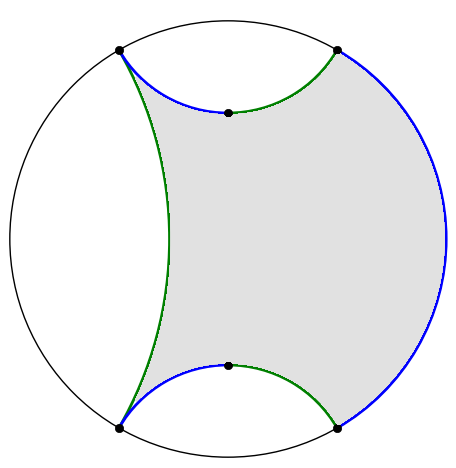}};
\note{2.9,0}{\large{$A$}}
\note{7.1,0}{\large{$B$}}

\note{ 4,0}{\large{$\Omega_{Ba}$}}
\note{ 11.7,0}{\large{$\Omega_{Ba}$}}
\note{8.9,0}{\large{$A$}}  

\draw[black, thick, ->] (4.1,0.2)->(4.4, 1.1);
\draw[black, thick, ->] (4.1,-0.27)->(4.4, -1.1);
\draw[black, thick, ->] (4.5,0)->(6,0);

\note{4.6,1.32}{\small{$a$}}  
\note{4.6,-1.37}{\small{$a$}}

\note{5.4,1.33}{\small{$b$}}  
\note{5.35,-1.35}{\small{$b$}}

\note{10.6,1.25}{\small{$a$}}  
\note{10.6,-1.36}{\small{$a$}}

\note{11.4,1.3}{\small{$b$}}  
\note{11.4,-1.37}{\small{$b$}}

\note{13.1,0}{\large{$B$}}  
\note{ 11,-2.5}{\large{(b)}}
\note{ 5,-2.5}{\large{(a)}}
\end{tikzpicture}
\caption{Two candidate presentations of the entropy $S(Ba)$. In (a), $B$ is purified by its RT surface $\Gamma_B$, while $a$ is purified by curves degenerate with itself. In (b), curves following $\Gamma_A$ and $b$ purify $Ba$. Note that the same two presentations with roles exchanged appear for $S(Ab)$, equal to $S(Ba)$ since $ABab$ is pure.}
\label{Fig:SBa}
\end{figure}

Let us now consider the slightly more complicated case of $E_Q(A:B)$, again in the vacuum state. The objective function of $E_Q$ (\ref{eqn:EQ}) contains four terms, although $S(A)$ and $S(B)$ are independent of the optimization and are calculated simply by the RT formula. We can use the SSC prescription to evaluate the remaining terms, $S(Aa)$ and $S(Ba)$, for a given purification of $\rho_{AB}$ and a partition of the purifying system into $a$ and $b$, and then minimize over purifications. We find two equivalent BSCs, shown in Fig. \ref{AdS EQ}, each involving adding the EWCS to the RT surface for one of the regions (green curves) and subtracting the portion of $\Gamma_{AB}$ that connects these surfaces (red curves). Both BSCs are numerically equal, which can be viewed as a consequence of symmetry. However, a way to think about it that will generalize to other cases is that the two presentations for $E_Q$ result from there being two numerically equal presentations of $S(Ba)$, as shown in Fig.~\ref{Fig:SBa}. If the optimization points are moved, one candidate for $S(Ba)$ becomes numerically smaller, and is hence preferred in the RT optimization that produces $S(Ba)$; however as $S(Ba)$ is subtracted from the objective function, this would make $E_Q$ bigger, and hence is not preferred in the optimization leading to $E_Q$. Thus this ``minimization within a minimization" guides the optimization points to a location where two equivalent BSCs exist. We will see this again in cases with black hole horizons, though in general the shapes of the two equivalent BSCs will not be related by symmetry, and may depend on the sizes and locations of the regions $A$ and $B$.

Tensor network models of AdS/CFT (see e.g., Refs. \cite{Swing12, Happy15, hayden2016holographic,nezami2020multipartite}) were the original motivation for the SSC, and can provide additional insight into the meaning of the SSC. The process of deforming a boundary region $A$ into its bulk RT surface $\Gamma_A$ can be thought of as a process of entanglement distillation, wherein the boundary degrees of freedom that live in $A$ are sorted into those entangled with degrees of freedom in the complement of $A$ (tensors adjacent to the RT surface) and those that are not entangled with the complement of $A$, and are left behind as the surface advances into the bulk. For a pure state the RT surface for $AB$ is also the RT surface for the complement of $AB$ on the boundary, and thus we can view selecting $\Gamma_{AB}$ as the optimal purification as a process of distilling the minimum set of degrees of freedom needed to provide the entanglement with the degrees of freedom in the entanglement wedge (and hence still represent a purification).

In the holographic context, several of the bipartite OCMs satisfy new equalities and inequalities. An important implication of the SSC is that a state which is dual to a geodesic in AdS must have no correlation between any of its subsystems; dividing the geodesic into disjoint pieces $a$ and $b$ and applying the entropy rule, one sees that $I(a:b) = 0$.  Since the objective functions of $E_P$ and $E_R$ differ by exactly $\frac{1}{2}I(a:b)$, we will have
\begin{equation}
  E_R = E_P \,,  
\end{equation}
 whenever the optimal purification for both quantities is given by the RT-surface of $AB$, which we take to be the case.

Furthermore, the MMI inequality, $I_3(A:B:C)\leq 0,$ implies (as reviewed in Appendix A)
\begin{align}
    E_Q\leq E_R \,,
\end{align} in any holographic state. MMI is also equivalent to the statement $I(A:B|C)\geq I(A:B)$, 
and as pointed out by  \cite{umemoto2018entanglement}, since the objective function of $E_{\rm sq}$ is $\frac{1}{2} I(A:B|a)$, the minimizing partition will be the one for which $a=0$, which leaves us with 
 \begin{align}
    E_{\rm sq} = \frac{1}{2}I(A:B)\,,
\end{align} for holographic states.

Thus using MMI and the properties of the optimization procedure coming from the SSC, the five bipartite correlation measures now satisfy
\begin{eqnarray}
	E_{\rm sq} = {1 \over 2} I(A:B) \leq E_Q \leq  E_R = E_P \leq \min(S(A), S(B)) \,,
\end{eqnarray}
and only $E_P$ and $E_Q$, along with the non-optimized $I(A:B)$, are independent.

\subsection{Entropies in black hole spacetimes }\label{RT_BTZ}

We turn now to a discussion of OCMs in geometries with horizons. In particular, we will consider a black hole in three-dimensional anti-de Sitter spacetime, the Ba\~nados-Teitelboim-Zanelli (BTZ) geometry \cite{BTZ}. Uniquely to three dimensions, these black holes are quotients of empty anti-de Sitter space, and thus locally maximally symmetric. It is convenient to calculate geodesics in the black hole geometries by undoing this quotient and calculating in pure AdS, before returning to BTZ; we review this map in Appendix B.

A non-rotating black hole geometry is characterized by its horizon radius $r_H$, related to the black hole mass $m$ by $r_H = \ell \sqrt{m}$, with $\ell$ the radius of anti-de Sitter space. Far from the black hole the geometry approaches anti-de Sitter space, including its boundary. The black hole has an associated Hawking temperature \cite{BTZ}
\begin{equation}
    T_H = {r_H \over 2\pi} \,,
\end{equation}
and the field theory state dual to the black hole is a thermal density matrix at the Hawking temperature.

\begin{figure}[H]
\centering
\begin{tikzpicture}
\node[inner sep=0pt] at (5,0)
    {\includegraphics[scale = 0.32]{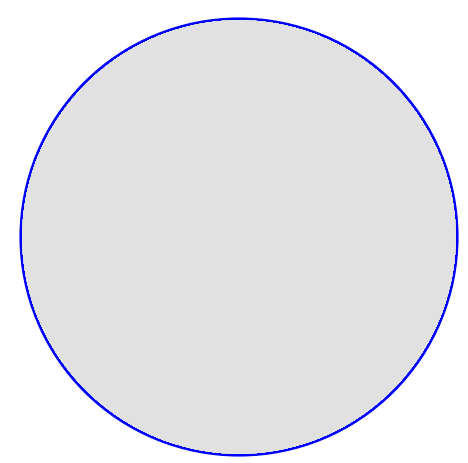}};
\node[inner sep=0pt] at (11,0)
    {\includegraphics[scale = 0.31]{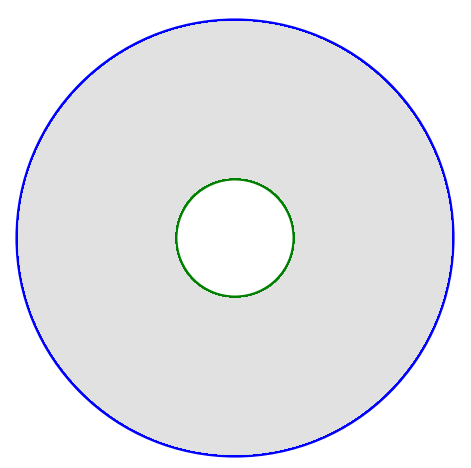}};
\note{2.9,0}{\large{$A$}}

\note{8.9,0}{\large{$A$}}  

\note{ 11,-2.5}{\large{(b)}}
\note{ 5,-2.5}{\large{(a)}}
\end{tikzpicture}
\caption{(a) In the absence of a horizon, the entanglement wedge for the whole boundary fills the space, and the RT surface minimizes to zero size; the boundary state is pure. (b) In a black hole geometry, the entanglement wedge of the whole boundary stops at the horizon, and the RT surface spans the horizon. The boundary state is thermal, with entropy the horizon area.}
\label{Fig:PureMixed}
\end{figure}

Let us now review how a black hole horizon affects the nature of the dual state. Without a horizon or other nontrivial spatial topology, if we try to calculate the entropy of the complete boundary using the RT formula, the candidate RT surfaces can shrink to a point while remaining homologous to the boundary. Thus the RT surface vanishes, and the entanglement wedge of the complete boundary fills the space; the boundary system thus has zero entropy, and is in a pure state (see Fig.~\ref{Fig:PureMixed}(a)). In the presence of a horizon, however, the candidate RT surface stops at the horizon, and the entropy of the boundary is given by the horizon area. The horizon stands for the degrees of freedom that must be added to purify the system. The system is in a mixed state, with thermal density matrix characterized by the black hole's Hawking temperature (see Fig.~\ref{Fig:PureMixed}(b)).

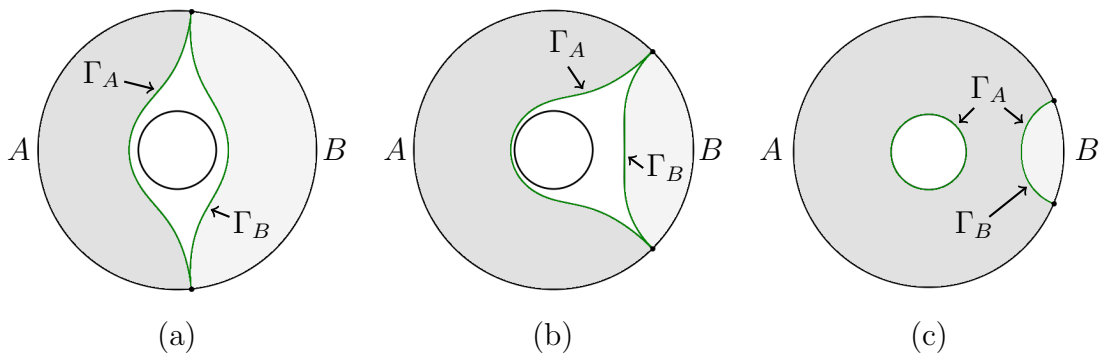
\begin{figure}[H]
\centering
\begin{tikzpicture}

\node[inner sep=0pt] at (0,0)
    {\includegraphics[scale = 0.32]{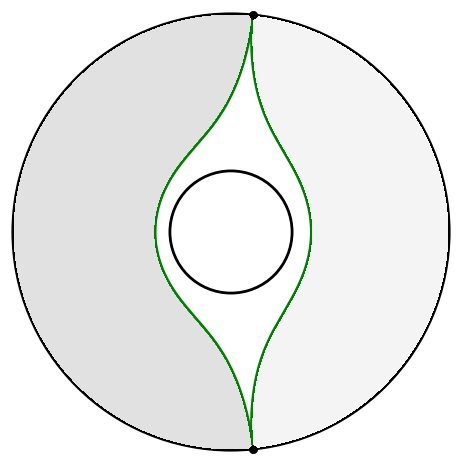}};
    
\node[inner sep=0pt] at (5,0)
    {\includegraphics[scale = 0.32]{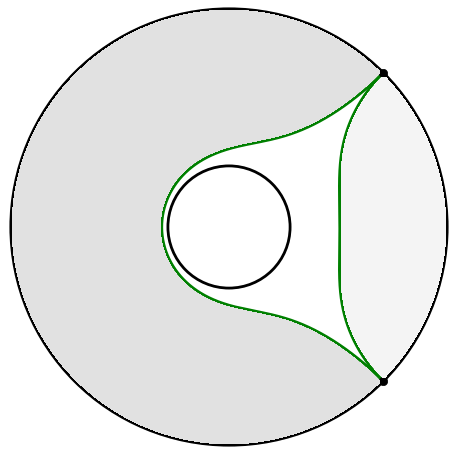}};
\node[inner sep=0pt] at (10,0)
    {\includegraphics[scale = 0.31]{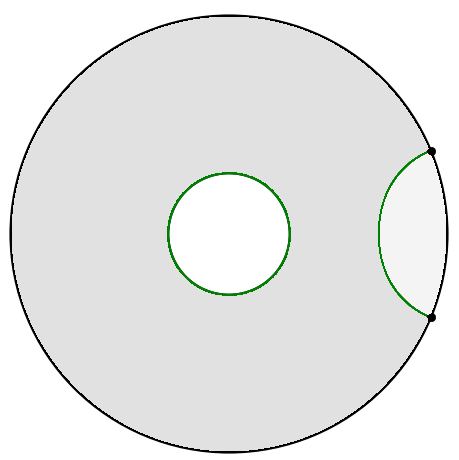}};

\note{-1,1}{\large{$\Gamma_A$}}
\note{1,-1}{\large{$\Gamma_B$}}
\draw[black, thick, ->] (-0.7,0.9)->(-0.3, 0.8);
\draw[black, thick, ->] (0.7,-.9)->(0.45, -0.8);

\note{6.5,-0.25}{\large{$\Gamma_B$}}
\draw[black, thick, ->] (6.2,-0.25)->(6, -0.1);

\note{10.6,-1}{\large{$\Gamma_B$}}
\draw[black, thick, ->] (10.8,-0.9)->(11.3,-0.5);

\note{-2.1,0}{\large{$A$}}
\note{2.1,0}{\large{$B$}}

\note{2.9,0}{\large{$A$}}
\note{7.1,0}{\large{$B$}}

\note{7.9,0}{\large{$A$}}  
\note{5.2,1.4}{\large{$\Gamma_A$}} 

   \draw[black, thick, ->] (5.2,1.1)->(5.4, 0.8);
   
\note{10.8,0.8}{\large{$\Gamma_A$}} 

\draw[black, thick, ->] (10.6,0.55)->(10.4, 0.35);
\draw[black, thick, ->] (11,0.55)->(11.25, 0.35);
   
\note{12.1,0}{\large{$B$}} 
\note{ 10,-2.5}{\large{(c)}}
\note{ 5,-2.5}{\large{(b)}}
\note{ 0,-2.5}{\large{(a)}}
\end{tikzpicture}
\caption{RT surfaces $\Gamma_A$ and $\Gamma_B$ for $A$ and $B$ together spanning the boundary, in the presence of a black hole horizon. In (a) and (b), the black hole horizon is not contained by either $\Omega_A$ or $\Omega_B$. In (b), the RT surface $\Gamma_A$ must go around the horizon to remain homologous to $A$. In (c), $A$ has become large enough that $\Gamma_A$ has ``jumped the horizon" and is now the union of $\Gamma_B$ and the horizon, and the horizon is surrounded by $\Omega_A$.}
    \label{BTZ A}
\end{figure}

Consider now two subregions $A$ and $B$ of the boundary, whose union $AB$ is the entire boundary. If $A$ is sufficiently small, the presence of the horizon will modify the precise area of the RT surface $\Gamma_A$ but not its topology, and the entanglement wedge $\Omega_A$ does not include the horizon. The case of $A$ and $B$ approximately the same size is shown in Fig.~\ref{BTZ A}a, where we can see each RT surface $\Gamma_A$ and $\Gamma_B$ avoiding the horizon and remaining homologous with $A$ and $B$, respectively. As $A$ grows, $\Gamma_A$ will partially encircle the horizon, and this will remain the preferred RT surface even though a smaller area surface connecting the boundary of $A$ exists, namely the surface $\Gamma_B$, because due to the horizon this smaller surface is not homologous to $A$; see Fig.~\ref{BTZ A}b. 

As $A$ continues to get bigger, eventually the area of the global minimum surface connecting the endpoints of $A$ ($\Gamma_B$) plus the horizon, becomes smaller than the surface connecting the endpoints of $A$ without enclosing the horizon. At this point the RT surface $\Gamma_A$ ``jumps the horizon",  becoming the union of the horizon and the global minimum surface (see Fig.~\ref{BTZ A}c). The entanglement wedge $\Omega_A$ is no longer simply connected, and the RT surface $\Gamma_A$ has two disconnected components, $\Gamma_A = \Gamma_B \cup E$ with $E$ the horizon; $\Gamma_B$ represents some traced-out degrees of freedom on the boundary, while $E$ represents  more unknown degrees of freedom due to the original thermal state, all necessary to purify the reduced state of $A$. In the limit of $A$ filling the boundary, we revert to Fig.~\ref{Fig:PureMixed}(b).

The size of $A$ for which $\Gamma_A$ jumps the horizon represents a transition in the information quantities. The area of the horizon is $S(AB) = S(E)$ with $E$ the environment purifying the thermal state. 
Before the horizon is jumped, we have $S(A) < S(E) + S(B) = S(AB) + S(B)$. However, 
 once $A$ is large enough that the horizon is jumped and $\Gamma_A = E+ \Gamma_B$, then $S(A) = S(AB) + S(B)$. This equality can be expressed in a number equivalent ways: that the mutual information of $A$ and $B$ saturates $I(A:B) = 2S(B)$, 
 that $A$ and $B$ or alternately $A$ and the environment $E$ saturate the Araki-Lieb inequality, or that $B$ is sufficiently small that it has no mutual information with the environment, 
\begin{align}
\label{Eqn:ArakiLieb}
  I(A:B) = 2 S(B) \quad \leftrightarrow \quad   S(AB) - S(A) = - S(B)\quad \leftrightarrow \quad  S(AE) - S(A) = - S(E)   \quad \leftrightarrow \quad   I(B:E) =0  \,. 
\end{align}
When $\Gamma_A$ has not jumped the horizon, we have  $I(A:B) < 2S(B)$, $S(AB) - S(A) > -S(B)$, $S(AE) - S(A) > -S(E)$, and $I(B:E) > 0$.

\section{Optimized correlation measures in thermal states}\label{OCMs thermal}

We are now ready to discuss the results of this paper, the geometric realization of bipartite OCMs in geometries with a black hole horizon. In this section we will begin by discussing the case of thermal states, where $B$ is the complement of $A$ on the boundary. In the next section we describe the reduced thermal states, where $A$ and $B$ do not together span the boundary.

The thermal state is characterized by a temperature, related to the single parameter of the black hole spacetime, the mass or equivalently the horizon radius $r_H$. There is only one further parameter characterizing how we partition the boundary into $A$ and $B$, which we may take to be  $\theta_{A,{\rm wid}}$, the angular width of region $A$, and then $\theta_{B,{\rm wid}}$,  the angular width   of $B$, is determined by $\theta_{A,{\rm wid}} + \theta_{B,{\rm wid}} = 2 \pi$.

As we noted, there are only three distinct OCMs in (classical) holographic cases, which we may take as $\frac{1}{2}I(A:B)$, $E_P(A:B)$ and $E_Q(A:B)$; then $E_{\rm sq} = \frac{1}{2} I(A:B)$ and $E_R(A:B) = E_P(A:B)$. They satisfy the inequalities
\begin{equation}
\label{Inequalities}
    {1 \over 2} I(A:B) \leq E_Q(A:B) \leq E_P(A:B) \leq {\rm min}(S(A), S(B))\,.
\end{equation}

\begin{figure}[H]

\centering
\begin{tikzpicture}
\node[inner sep=0pt] at (5,0)
    {\includegraphics[scale = 0.35, trim = 0 55 0 55, clip]{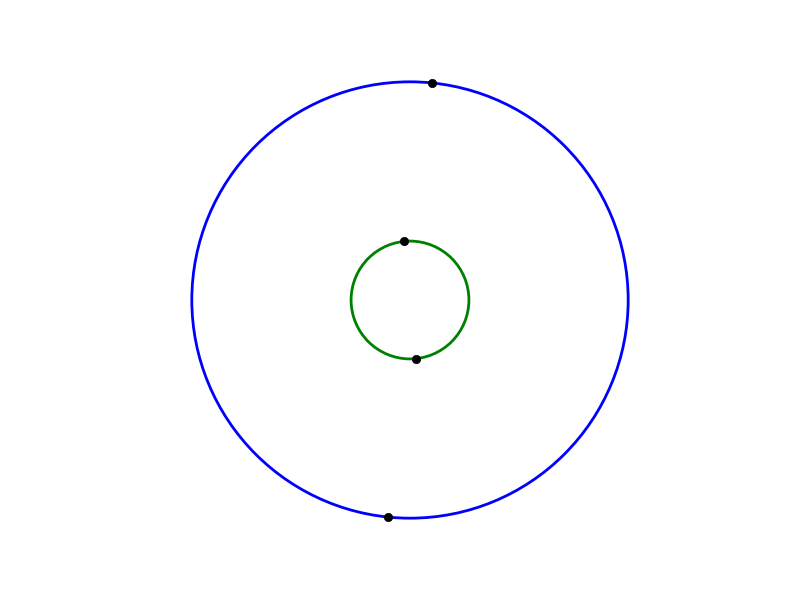}};
\note{2.7,0}{\large{$A$}}   
\note{7.4,0}{\large{$B$}}  
\note{4.35,0}{\large{$a$}}   
\note{5.8,0}{\large{$b$}}  
\end{tikzpicture}
\caption{Purification of a thermal state involves partitioning the RT surface, in this case just the horizon, into $a$ and $b$ portions. It is allowed for one of $a$ and $b$ to vanish and the other to comprise the complete horizon.}
\label{Fig:HorPartition}
\end{figure}

As described previously, holographically realizing correlation measures involves partitioning the RT surface $\Gamma_{AB}$ into portions assigned to $a$ and portions assigned to $b$, and minimizing the objective function over all such partitions. For these states, $\Gamma_{AB}$ is just the horizon. Thus optimization involves partitioning the horizon into $a$ and $b$ portions, as in Fig.~\ref{Fig:HorPartition}. The horizon may be shared, or it is possible for one of $a$ and $b$ to vanish, and the entire horizon to be associated to the other ancilla. 

\begin{figure}[H]
    \centering
    \includegraphics[scale = 0.48]{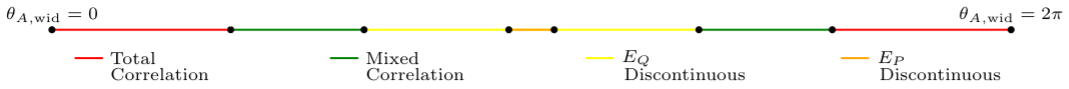}
    \caption{The four phases for the thermal states as function of $\theta_{A, wid}$. $E_P$ is disconnected only in the $E_P$-discontinuous phase (marked in orange), while $E_Q$ is disconnected in both the $E_Q$- and $E_P$-discontinuous phases (yellow and orange).}
    \label{Diagonal}
\end{figure}

\begin{figure}[H]\label{Diagonal phases}
    \centering
    \includegraphics[scale = 0.545]{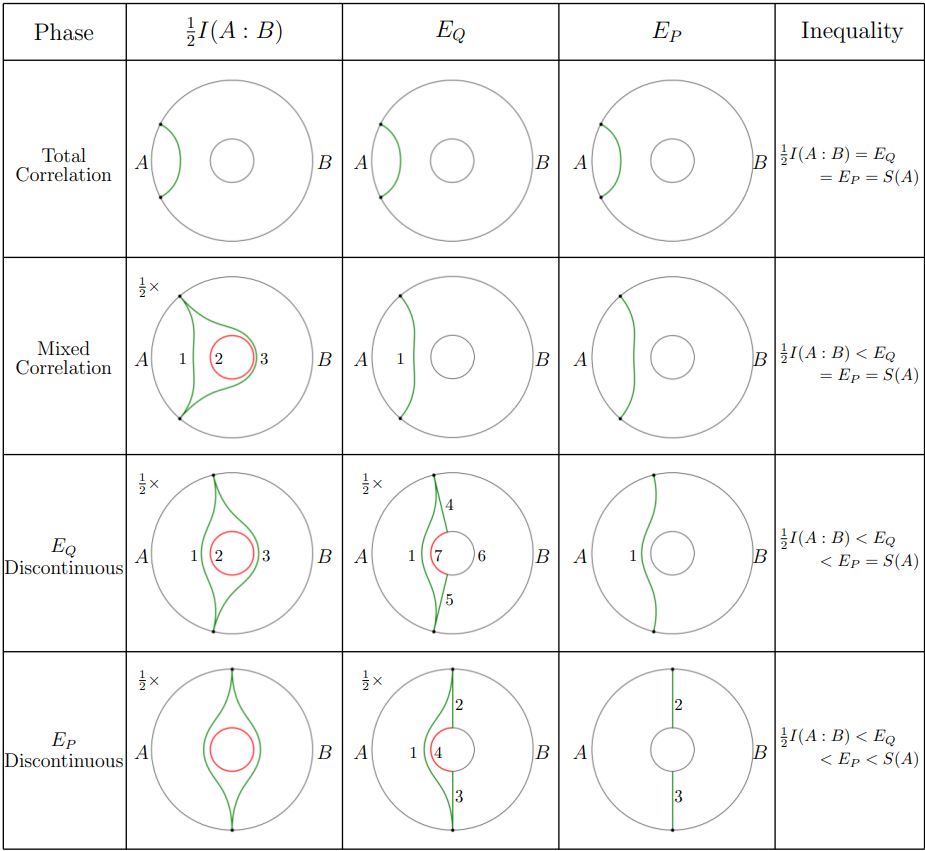}
    \caption{The four phases for thermal states, the BSCs of the correlation measures, and their (in)equalities. In the total correlation phase, $A$ is small enough that $\Omega_B$ encloses the horizon, and all the measures are degenerate with $S(A)$; the horizon belongs to $b$. In the mixed correlation phase, $A$ has grown enough that $\Omega_B$ does not enclose the horizon, and $\frac{1}{2} I(A:B)$ is now strictly smaller than the other measures, which still coincide with $S(A)$. The horizon is still entirely $b$. In the $E_Q$-discontinuous phase, $A$ is large enough that $E_Q$ can find a new minimum by giving $a$ part of the horizon, but $E_P$ is still minimized by giving all of the horizon to $b$ and still equals $S(A)$. Finally, the $E_P$-discontinuous phase, $E_P$ also minimizes by giving some of the horizon to $a$, and the EWCS is now discontinuous; all the information measures are different from each other and smaller than $S(A)$.}
    \label{Fig:ThermalResults}
\end{figure}

In what follows, for each choice of $r_H$ and of $\theta_{A,{\rm wid}}$ we numerically optimized over purifications by varying optimization points and finding the minimum value of the objective function. For a fixed $r_H$, as we vary $\theta_{A,{\rm wid}}$ we find up to four distinct phases, with different associated BSCs for the three independent measures. Each different phase can be associated to different combinations of the inequalities in (\ref{Inequalities}) being either equalities or strict inequalities. The phase diagram for $r_H$ sufficiently large that all four phases are present is shown in Fig.~\ref{Diagonal}, and the BSCs and (in)equalities are displayed in  Fig.~\ref{Fig:ThermalResults}. The phase diagram is symmetric between exchange of the sizes of $A$ and $B$; we will assume without loss of generality that $\theta_{A,{\rm wid}}\leq \theta_{B,{\rm wid}}$ so that $S(A)\leq S(B)$, and start with $A$ small and progress to the point where the regions are of equal size. 

\bigskip
\noindent
{\bf Total correlation phase:}

We begin with the smallest sizes of $A$. As described in the previous subsection, if region $A$ is so small that $B$'s RT surface has jumped the horizon, then $I(B:E) = 2S(E)$ and $I(A:E) = 0$. We call this the \textit{total correlation} phase, and it is shown in red in Fig.~\ref{Diagonal}. Here, we also have ${1 \over 2} I(A:B) = {\rm min}(S(A), S(B))$. This then collapses all the inequalities in (\ref{Inequalities}) to equalities, and all three independent correlation measures are just equal to the entropy of the smaller region, for us $S(A)$. Indeed, numerically evaluating $E_Q(A:B)$ and $E_P(A:B)$ shows that their BSCs minimize simply to the RT surface $\Gamma_A$. In both cases, $a=0$ and the entire horizon belongs to $b$. This is shown in the first row of Fig.~\ref{Fig:ThermalResults}.


\bigskip
\noindent
{\bf Mixed correlation phase:}

As the size of $A$ grows, we eventually reach the point where $\Gamma_B$ no longer jumps the horizon, and $E$ is no longer contained either in $\Omega_A$ or $\Omega_B$. This is represented by the green region in Fig.~\ref{Diagonal}. Here, we have $0<I(A:E)<2S(E)$ and $0<I(B:E)<2S(E)$, so neither boundary region is maximally nor minimally correlated with the environment, and hence we refer to this as the \textit{mixed correlation} phase. The presentation $\Gamma_A + \Gamma_B - E$ of $\frac{1}{2} I(A:B)$ now changes, since now $\Gamma_B \neq \Gamma_A + E$, and we now have $\frac{1}{2} I(A:B) < S(A)$, but for this range of $\theta_{A,{\rm wid}}$ both $E_Q(A:B)$ and $E_P(A:B)$ still have $\Gamma_A$ as their BSC, and thus $E_Q = E_P = S(A)$. The optimization again assigns all of the horizon to $b$. This is shown in the second row of Fig.~\ref{Fig:ThermalResults}. We can show that $\frac{1}{2}I(A:B) < E_Q$ strictly  in this phase by showing that (using the labelling of curves in Fig.~\ref{Fig:ThermalResults})  $|1| + |3| - |2| < 2|1|$, which is equivalent to $|3| < |1| + |2|$, or $\Gamma_B$ is shorter than $\Gamma_A \cup E$, which is the defining characteristic of not being in the total correlation phase.

\bigskip
\noindent
{\bf $E_Q$-discontinuous phase:}

Both $E_Q$ and $E_P$ involve $S(Aa)$ in their objective functions, which for $E_P$ is realized geometrically as the EWCS. In the total and mixed correlation phases, $S(Aa)$ for both measures manifests as a continuous curve, avoiding the horizon. As the size of $A$ increases, we reach a point where the optimization of $E_Q$ can find a lower minimum by choosing an $a$ that includes part of the horizon, so $S(Aa)$ can stretch to the horizon and break into two disconnected pieces. Values of $\theta_{A, {\rm wid}}$ for which this occurs but $E_P$ still manifests as the continuous curve $\Gamma_A$ constitute the $E_Q$-\textit{discontinuous} phase, the yellow region in Fig.~\ref{Diagonal}, with the BSCs shown in the third row of Fig.~\ref{Fig:ThermalResults}. We see in this phase qualitatively different optimizations can be preferred for the two different measures. As far as information inequalities, now have $\frac{1}{2}I(A:B) < E_Q < E_P = S(A)$. To show that $\frac{1}{2}I(A:B) < E_Q$ in this phase, we need to show that $|1| + |3| - |2| < |1| + |4| + |5| - |7|$ with the numbers given in Fig.~\ref{Fig:ThermalResults}, which is equivalent to $|3| < |4| + |5| + |6|$.  This is true because curve 3 is a geodesic connecting two boundary points, while the piece-wise curve 4-6-5 connects the same two points and is not a geodesic.  To show that $E_Q<E_P$ in this phase, we need to show that $|1| + |4| + |5| - |7| < 2|1|$, which is equivalent to $|4| + |5| - |7| < |1|$.  This follows from a defining property of the $E_Q$ discontinuous phase, namely that the $S(Aa)$ contribution to $E_Q$ is a discontinuous curve (if the inequality was false, then $f_Q$ would achieve its minimum by replacing $|4| + |5| - |7|$ with $|1|$, as in the mixed correlation phase). 

\bigskip
\noindent
{\bf $E_P$-discontinuous phase:}

As $A$ continues to grow, eventually we reach the $E_P$-\textit{discontinuous} phase, the orange region in Fig.~\ref{Diagonal}, where the presentation of $S(Aa)$ becomes discontinuous both for $E_Q$ and $E_P$; the horizon is split between $a$ and $b$ in both cases. The BSCs are shown in the fourth row of Fig.~\ref{Fig:ThermalResults}.  We now have strict inequalities $\frac{1}{2} I(A:B) < E_Q(A:B) < E_P(A:B) < S(A)$. $E_P$ is constant within this phase, since the lengths of geodesics connecting the boundary to the horizon are independent of the size of $A$. The proof that $\frac{1}{2}I(A:B) < E_Q$ is identical to $E_Q$-discontinuous phase.  To show that $E_Q<E_P$, we need $|1| + |2| + |3| - |4| < 2|2| + 2|3|$, or $|1| < |2| + |3| + |4|$.  This is true because curve 1 is a geodesic connecting two boundary points and the piecewise curve 2-3-4 connects the same two points and is not a geodesic.  Finally, to show that $E_P<S(A)$, we need that $|2| + |3|< |1|$.  But this follows from the defining property of the $E_P$-discontinuous phase that the bulk curve dual to $E_P$ is discontinuous (if the inequality was false, then $f_P$ would achieve its minimum using the continuous curve 1 instead of the discontinuous union of curves 2 and 3).

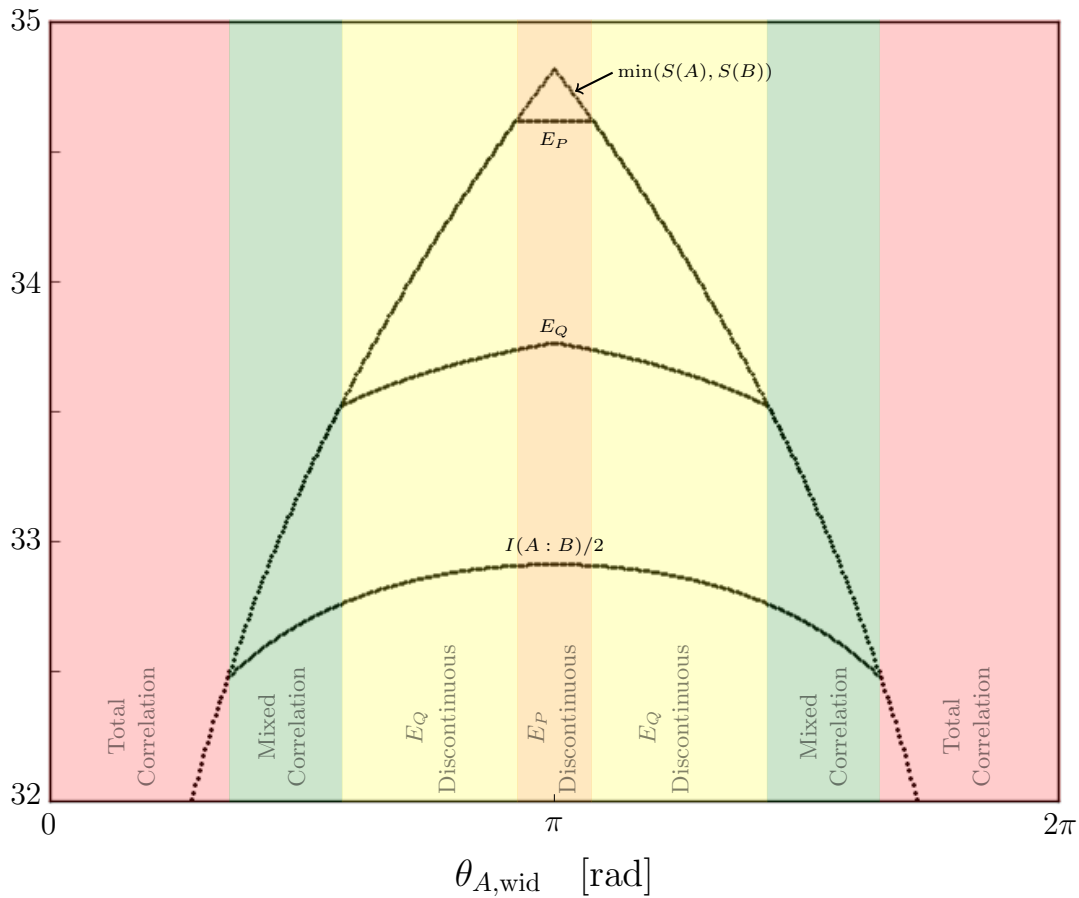
\begin{figure}[H]

\centering
\begin{tikzpicture}
\node[inner sep=0pt] at (0,0)
    {\includegraphics[scale = 0.85]{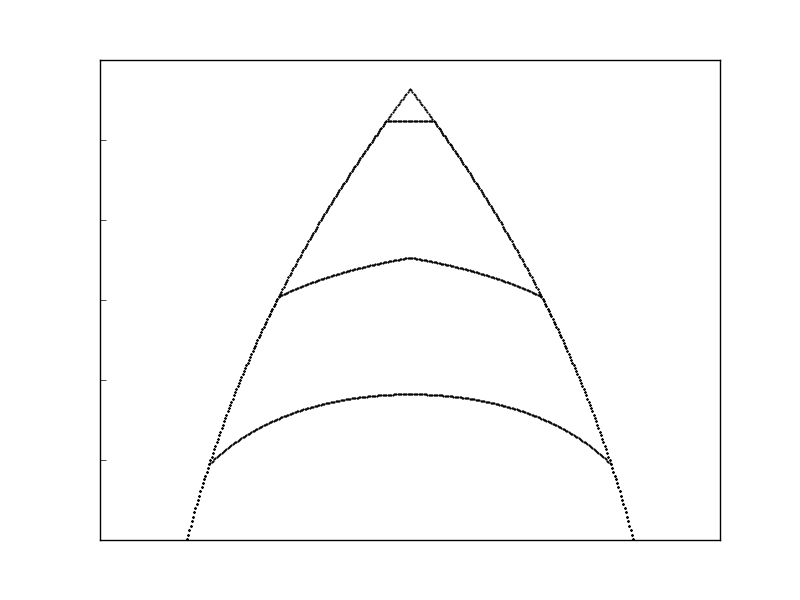}};\
    \definecolor{darkgreen}{rgb}{0,0.5,0}
\definecolor{yellow2}{rgb}{1,1,0}
\definecolor{orange2}{rgb}{1,0.643,0}
    \draw [ultra thin, draw=red, fill=red, opacity=0.2]
       (-6.5,-5.2) -- (-4.1,-5.2) -- (-4.1,5.2) -- (-6.5,5.2);
    \draw [ultra thin, draw=red, fill=red, opacity=0.2]
       (4.55,-5.2) -- (6.95,-5.2) -- (6.95,5.2) -- (4.55,5.2);
    
    \draw [ultra thin, draw=darkgreen, fill=darkgreen, opacity=0.2]
       (-4.1,-5.2) -- (-2.6,-5.2) -- (-2.6,5.2) -- (-4.1,5.2);
    \draw [ultra thin, draw=darkgreen, fill=darkgreen, opacity=0.2]
       (3.05,-5.2) -- (4.55,-5.2) -- (4.55,5.2) -- (3.05,5.2);
    \draw [ultra thin, draw=yellow2, fill=yellow2, opacity=0.2]
       (-2.6,-5.2) -- (-0.27,-5.2) -- (-0.27,5.2) -- (-2.6,5.2);
    \draw [ultra thin, draw=yellow2, fill=yellow2, opacity=0.2]
       (0.72,-5.2) -- (3.05,-5.2) -- (3.05,5.2) -- (0.72,5.2);
    \draw [ultra thin, draw=orange2, fill=orange2, opacity=0.25]
       (-0.27,-5.2) -- (0.72,-5.2) -- (0.72,5.2) -- (-0.27,5.2);
     \draw[black, ultra thin] (0.225, -5.2)--(0.225,-5.1);

    \note{ -6.5,-5.5}{\large{$0$}}
    \note{ 0.225,-5.5}{\large{$\pi$}}
    \note{ 6.95,-5.5}{\large{$2\pi$}}
    \note{ -6.8,-5.1}{\large{$32$}}
    \note{ -6.8,-1.7}{\large{$33$}}
    \note{ -6.8,1.7}{\large{$34$}}
    \note{ -6.8,5.2}{\large{$35$}}
    
        \note{ 0.225,3.6}{$E_P$}
        \note{ 0.225,1.1}{$E_Q$}
        \note{ 0.225,-1.8}{$I(A:B)/2$}
        
        \note{2.1,4.5}{$\min(S(A),S(B))$}
        \draw[black, thick, ->] (1,4.5)->(0.5, 4.25);
    
    \node[rotate=90, text opacity = 0.5] at (-5.6,-4.2) {Total};
    \node[rotate=90, text opacity = 0.5] at (-5.2,-4.2) {Correlation};
    
    \node[rotate=90, text opacity = 0.5] at (-3.6,-4.2) {Mixed};
    \node[rotate=90, text opacity = 0.5] at (-3.2,-4.2) {Correlation};
    
    \node[rotate=90, text opacity = 0.5] at (-1.6,-4.2) {$E_Q$};
    \node[rotate=90, text opacity = 0.5] at (-1.2,-4.1) {Discontinuous};
    
    \node[rotate=90, text opacity = 0.5] at (0,-4.2) {$E_P$};
    \node[rotate=90, text opacity = 0.5] at (0.4,-4.1) {Discontinuous};
    
    \node[rotate=90, text opacity = 0.5] at (1.5,-4.2) {$E_Q$};
    \node[rotate=90, text opacity = 0.5] at (1.9,-4.1) {Discontinuous};
    
    \node[rotate=90, text opacity = 0.5] at (3.6,-4.2) {Mixed};
    \node[rotate=90, text opacity = 0.5] at (4,-4.2) {Correlation};
    
    \node[rotate=90, text opacity = 0.5] at (5.5,-4.2) {Total};
    \node[rotate=90, text opacity = 0.5] at (5.9,-4.2) {Correlation};
    
\note{0.225,-6.2}{\Large{$\theta_{A,\rm wid}$ ~ [rad]}}
\note{0.225,6}{\LARGE{Correlation measures in thermal states}}
\end{tikzpicture}
\caption{Optimized correlation measures $E_P$ and $E_Q$, along with their lower and upper bounds $I(A:B)/2$ and $\min(S(A), S(B))$, respectively, plotted for thermal states with $r_H\approx 0.61$ as a function of $\theta_{A,\rm wid}$.}
\label{EQ_EP_thermal}
\end{figure}

\bigskip
The numerical values of $I(A:B)/2$, $E_Q(A:B)$, $E_P(A:B)$, and $\min(S(A),S(B))$ are plotted in Fig.~\ref{EQ_EP_thermal} for $r_H \approx 0.61$ (we have taken $4G_N = 1$).  The validity of Ineq.~(\ref{Inequalities}) is clearly visible, as well as the fact that each phase boundary corresponds to exactly one of these inequalities transitioning between a strict inequality and an equality, as described in the right column of Fig.~\ref{Fig:ThermalResults}.  Also note that the phase transitions are accompanied by discontinuities in the first derivative (with respect to region size) of one OCM, as the minimizing bulk surface configuration changes, so the transitions may be thought of as first order. We will see this behavior again when we look at reduced thermal states in the next section. We note that all the quantities plotted in Fig.~\ref{EQ_EP_thermal} are cutoff dependent; as we described in Sec.~\ref{RT}, when $A$ and $B$ share a boundary point the mutual information is sensitive to the cutoff, and this is true for $E_Q$ and $E_P$ as well, as can be seen in their BSCs in Fig.~\ref{Fig:ThermalResults}.

For $r_H$ (and thus the temperature) sufficiently large, all four phases are present. As $r_H$ decreases the discontinuous phases disappear, first the $E_P$-discontinuous phase disappearing at $r_H \approx 0.58$, and then the $E_Q$-discontinuous phase disappearing at $r_H \approx 0.37$. Both the total correlation phase and mixed correlation phase exist for all values of $r_H$.

\section{Optimized Correlation Measures in Reduced Thermal States}\label{OCMs reduced thermal}

The thermal states studied in the last section may be the simplest holographic states in which the three correlation measures $\frac{1}{2}I(A:B)$, $E_Q(A:B)$ and $E_P(A:B)$ measure different things; in the pure vacuum state they are all just $S(A) = S(B)$. As reviewed earlier, the measures are also distinct in the reduced state we get from tracing out some degrees of freedom in the pure state. We now turn to the more complicated case that combines both, tracing out some boundary degrees of freedom from the thermal states, which we refer to as ``reduced thermal states". This corresponds to $A$ and $B$ not spanning the boundary, in the presence of a black hole. As we shall see, the BSCs become more intricate.

\subsection{Reduced states with a horizon}

\begin{figure}[H]
\centering
\begin{tikzpicture}
\node[inner sep=0pt] at (3,0)
    {\includegraphics[scale = 0.32]{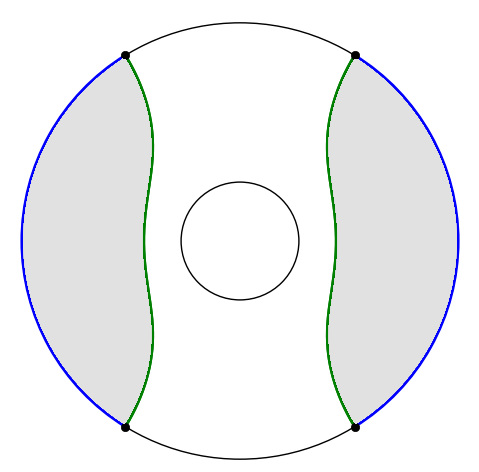}};
\node[inner sep=0pt] at (8.5,0)
    {\includegraphics[scale = 0.31]{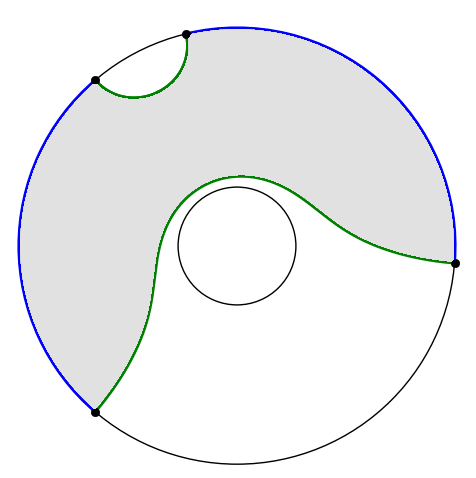}};
\node[inner sep=0pt] at (14,0)
    {\includegraphics[scale = 0.31]{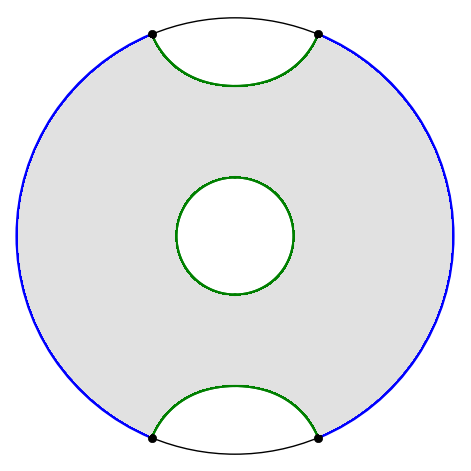}};
    
\note{0.8,0}{\large{$A$}}
\note{5.2,0}{\large{$B$}}

\note{6.3,0}{\large{$A$}}
\note{9.7,1.8}{\large{$B$}}

\note{11.8,0}{\large{$A$}}
\note{16.2,0}{\large{$B$}}

\note{3,1.2}{\large{$\Gamma_{AB}$}} 
\draw[black, thick, ->] (3.35,1)->(3.65, 0.8);
\draw[black, thick, ->] (2.65,1)->(2.35, 0.8);
   
\note{12.9,0.5}{\large{$\Gamma_{AB}$}} 
\draw[black, thick, ->] (13.1,0.7)->(13.5, 1.3);
\draw[black, thick, ->] (12.9,0.15)->(13.6, -1.2);
\draw[black, thick, ->] (13.15,0.25)->(13.5, 0.17);  

\note{8.8,1.2}{\large{$\Gamma_{AB}$}} 
\draw[black, thick, ->] (8.4,1.25)->(8.1, 1.3);
\draw[black, thick, ->] (8.8,0.9)->(8.7, 0.6);

\note{ 8.5,-2.5}{\large{(b)}}
\note{ 3,-2.5}{\large{(a)}}
\note{ 14,-2.5}{\large{(c)}}
\end{tikzpicture}

\caption{Types of entanglement wedges for $AB$ when $A$ and $B$ do not fill the boundary. In (a), $\Omega_{AB}$ is disconnected. In (b) $\Omega_{AB}$ is connected but does not enclose the horizon, while in $(c)$ it is connected and does enclose the horizon.}
\label{BTZ AB}
\end{figure}
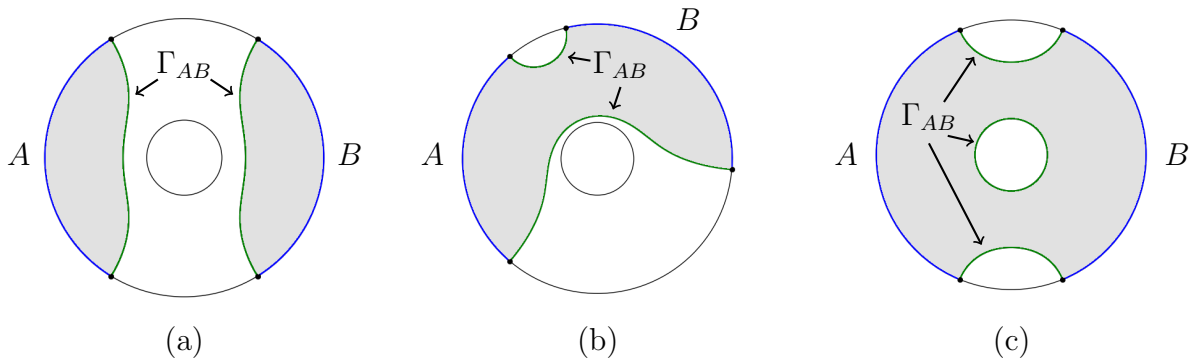

Like in the AdS case, if regions $A$ and $B$ are sufficiently far apart on the boundary on the BTZ spacetime, then the RT-surface $\Gamma_{AB}$ is simply the union of the RT surfaces $\Gamma_A$ and $\Gamma_B$ and the entanglement wedge $\Omega_{AB} = \Omega_A \cup \Omega_B$ is disconnected, as shown in Fig.~\ref{BTZ AB}a. In this case $I(A:B) =0$, and in fact all the other bipartite correlation measures vanish as well. From the geometrical perspective, one can see this because here it is always possible to take $a = \Gamma_A$ and $b = \Gamma_B$, which purifies $A$ and $B$ into two uncorrelated product states; this implies relations on the entropies like $S(a) = S(A)$, $S(Aa) = 0$ and $S(Ba) = S(B) + S(a)$, which cause all the objective functions to vanish. Since the objective functions are nonnegative, the correlation measures then vanish as well\footnote{This vanishing of the objective functions follows from monotonicity, which requires \cite{LDS20} that $A$ and $a$ must appear in $f^\alpha$ either as $Aa$ or as $S(AX) - S(aY)$ with $X$ and $Y$ not including $A$ or $a$. Given $Aa$ pure and existing in a product state with all other parties, all $Aa$-dependence cancels out, and the same happens recursively for all other parties, leaving zero.}.

When the regions $A$ and $B$ are sufficiently close together, the entanglement wedge becomes connected and $I(A:B)$ becomes nonzero, again like the AdS case.  However, in the presence of a horizon there are two ways this can happen. If $A$ and $B$ are close enough to one side of the boundary, $\Omega_{AB}$ may be connected but not include the black hole, as in Fig.~\ref{BTZ AB}b. In these instances the BSCs are qualitatively similar to the pure AdS cases, as reviewed in Sec.~\ref{Sec:OCMAdSCFT}, and are less novel.

\begin{figure}[H]
\centering
\begin{tikzpicture}
\node[inner sep=0pt] at (0,0)
    {\includegraphics[scale = 0.32]{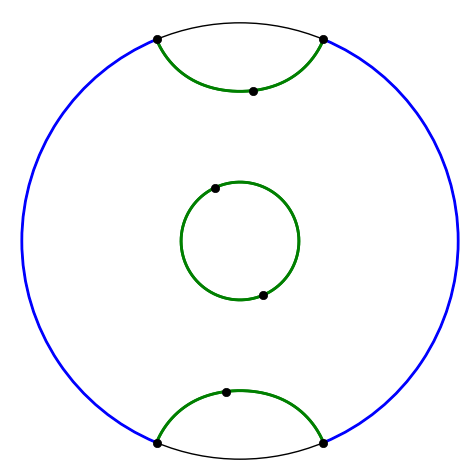}};
\note{-2,0}{\large{$A$}}
\note{2,0}{\large{$B$}}

\note{-.45,1.2}{\small{$a$}}
\note{-.45,-1.2}{\small{$a$}}
\note{-0.6,-0.2}{\small{$a$}}
\note{0.5,1.2}{\small{$b$}}
\note{0.45,-1.15}{\small{$b$}}
\note{0.6,0.2}{\small{$b$}}
\end{tikzpicture}
\caption{Boundary regions $A$ and $B$ are shown in blue, with $AB$'s RT surface (a purifying system for $AB$) shown in green.  The purifying system has been partitioned into pieces $a$ and $b$, in accordance with Eq. (\ref{OCM}).  For this example, we have chosen to partition the horizon into two pieces, one belonging to $a$ and the other to $b$.  Partitions where the horizon belongs entirely to $a$ or to $b$ are also included in the optimization of Eq. (\ref{OCM}).}
    \label{fig:ABab}
\end{figure}

More nontrivially, it can be the case that the horizon is surrounded by the entanglement wedge of $AB$, and is part of the RT  surface
$\Gamma_{AB}$ as in Fig.~\ref{BTZ AB}c. This gives the entanglement wedge a hole, and thus a nontrivial topology, which leads to novel structure in the OCMs. Enclosing the horizon means the complement of $AB$ on the boundary has no mutual information with the horizon $E$, or equivalently $AB$ and $E$ saturate the Araki-Lieb inequality, 
\begin{eqnarray}
\label{Eq:HorizonInWedge}
S(ABE) - S(AB) = - S(E)\,. 
\end{eqnarray}
We will largely focus on this case. Then since $A$ and $B$ do not span the boundary of BTZ, the entanglement wedge has boundary components both along the horizon, and associated to missing degrees of freedom in the BTZ boundary. Purification and optimization involves assigning all segments of these surfaces to ancilla $a$ or $b$, as shown in Fig.~\ref{fig:ABab}, in all possible ways.

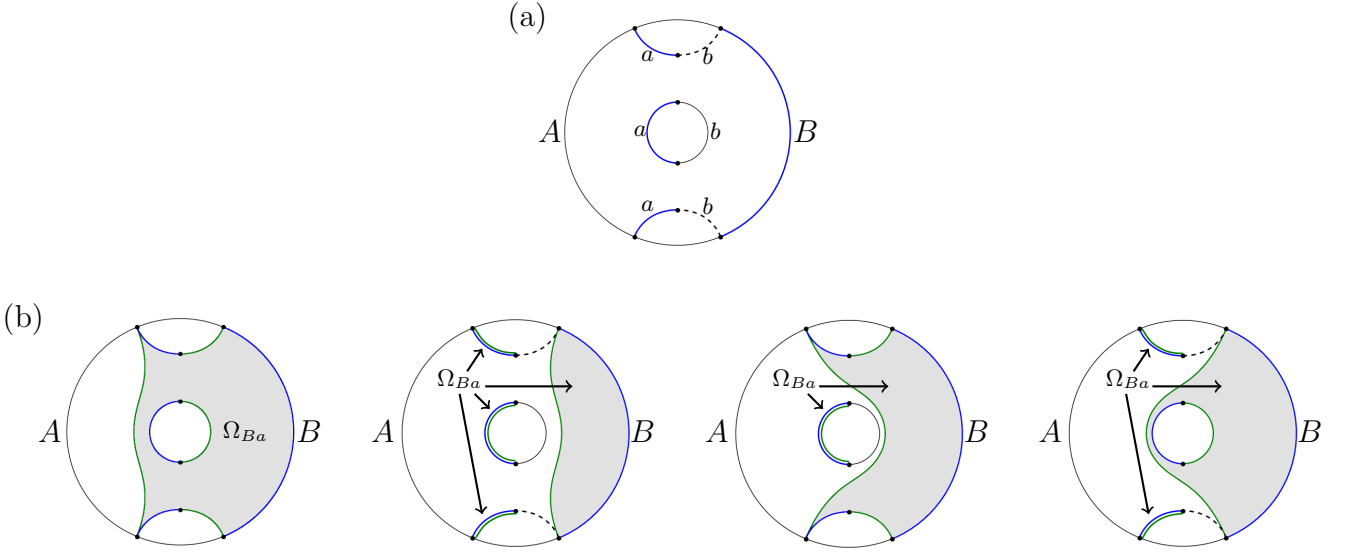
\begin{figure}[H]
\centering
\begin{tikzpicture}

\node[inner sep=0pt] at (8.2,2)
    {\includegraphics[scale = 0.27]{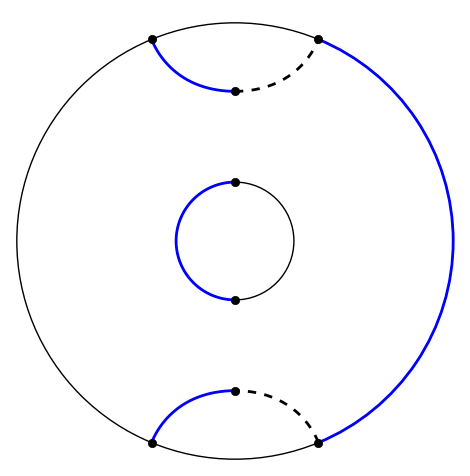}};
\node[inner sep=0pt] at (1.6,-2)
    {\includegraphics[scale = 0.26]{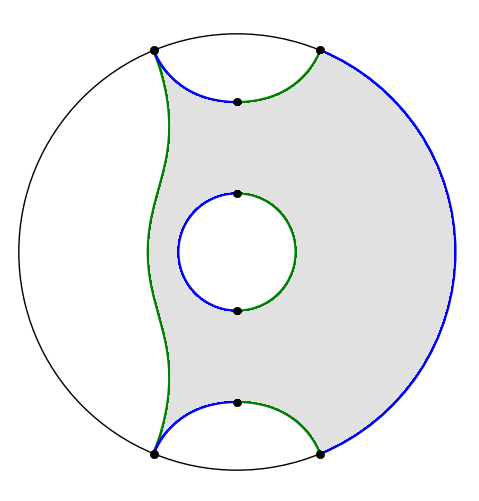}};
\node[inner sep=0pt] at (6.03,-2)
    {\includegraphics[scale = 0.26]{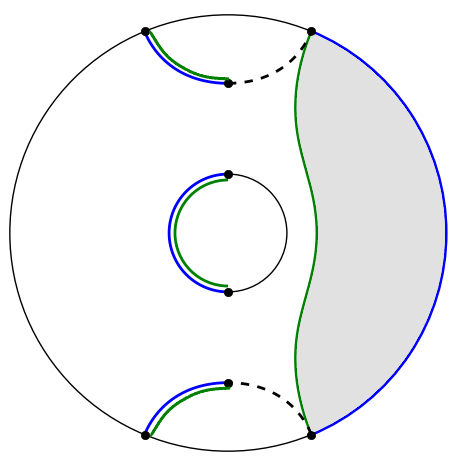}};
\node[inner sep=0pt] at (10.47,-2)
    {\includegraphics[scale = 0.26]{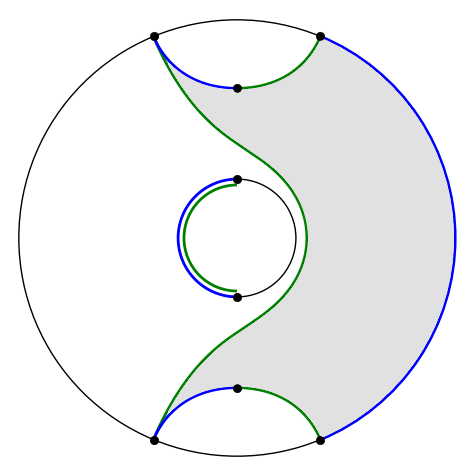}};
\node[inner sep=0pt] at (14.9,-2)
    {\includegraphics[scale = 0.26]{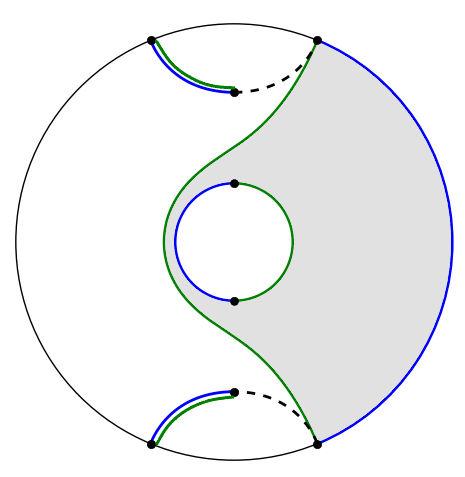}};
    
\note{6.5,2}{\large{$A$}}
\note{9.9,2}{\large{$B$}}  

\note{7.8,3}{\small{$a$}}
\note{7.7,2}{\small{$a$}}
\note{7.8,1}{\small{$a$}}
\note{8.6,3}{\small{$b$}}
\note{8.7,2}{\small{$b$}}
\note{8.6,1}{\small{$b$}}

\note{-.15,-2}{\large{$A$}}
\note{3.3,-2}{\large{$B$}}
\note{2.45,-2}{\small{$\Omega_{Ba}$}}

\note{4.28,-2}{\large{$A$}}
\note{7.73,-2}{\large{$B$}}
\note{5.3,-1.3}{\small{$\Omega_{Ba}$}}
\draw[black, thick, ->] (5.4,-1.2)->(5.6, -0.9);
\draw[black, thick, ->] (5.5,-1.5)->(5.7, -1.7);
\draw[black, thick, ->] (5.3,-1.5)->(5.6, -3.1);
\draw[black, thick, ->] (5.64,-1.4)->(6.8, -1.4);

\note{8.72,-2}{\large{$A$}}
\note{12.17,-2}{\large{$B$}}
\note{9.76,-1.3}{\small{$\Omega_{Ba}$}}
\draw[black, thick, ->] (9.93,-1.5)->(10.13, -1.7);
\draw[black, thick, ->] (10.08,-1.4)->(11, -1.4);

\note{13.15,-2}{\large{$A$}}
\note{16.6,-2}{\large{$B$}}
\note{14.19,-1.3}{\small{$\Omega_{Ba}$}}
\draw[black, thick, ->] (14.26,-1.2)->(14.46, -0.9);
\draw[black, thick, ->] (14.16,-1.5)->(14.46, -3.1);
\draw[black, thick, ->] (14.51,-1.4)->(15.43, -1.4);

\note{ 6.2,3.5}{\large{(a)}}
\note{ -0.5,-0.5}{\large{(b)}}

\end{tikzpicture}

\caption{(a) System $Ba$ is shown in blue (b) Four possible purifying systems for $Ba$, each one shown in green.  In each case, the sub manifold whose boundary consists of $Ba$ and its purifying system is labeled $\Omega$.  In the first purification, $\Omega$ is connected and 2D. But in the other three, $\Omega$ has multiple connected components, some of which are 1D.  Each connected component of $\Omega$ corresponds to a tensor factor in a pure product state.  The 1D components are degenerate cases of purification via SSC, wherein the surface representing the purifying system lies on top of the surface it purifies.  These are shown as doubled curves with one blue and the other green.}
    \label{fig:Ba}
\end{figure}

To illustrate the process, consider the calculation of $S(Ba)$, which appears in the objective function of $E_Q$ and which we reviewed in the case without a horizon in Sec. \ref{SSC}.  The SSC prescription instructs us to first find all possible surfaces that purify $Ba$, and then choose the one with smallest total area.  Fig.~\ref{fig:Ba} shows a system $Ba$ and the four purifications whose total areas are candidates for $S(Ba)$.  Each one, when combined with the surface representing system $Ba$, forms the boundary of a submanifold of the space (disconnected in some cases).  The first case forms a single connected region by closing $Ba$ along $\Gamma_A$ and $b$. Each of the others involves at least one disconnected part. Moreover, when a surface that is already minimal like $a$ and $b$ purifies itself, the purifying surface is degenerate with the surface being purified. Thus in the second configuration, each component of $a$ purifies itself, while in the other two, part of $a$ purifies itself while the rest is purified along with $B$. We saw this happen already in the case with no horizon shown in Fig.~\ref{Fig:SBa}; the existence of the horizon now presents more possibilities.

Consider now the parameter space of configurations with two boundary regions $A$ and $B$. As before, the black hole geometry has one parameter, the mass or equivalently  the radius $r_H$, dual to the temperature of the state. Choosing $A$ and $B$ on the boundary then involves four parameters, the locations of the centers $\theta_{A,{\rm cen}}$ and $\theta_{B,{\rm cen}}$, and the angular widths   $\theta_{A,{\rm wid}}$ and $\theta_{B,{\rm wid}}$. Without a black hole, one can use the boundary limit of the AdS isometries to fix three of these parameters, leaving only one; for example one can choose $A$ and $B$ diametrically opposite and having the same size, and this size is the one parameter. In black hole spacetimes however, the symmetries are reduced, as most AdS isometries would either scale the horizon size or translate the black hole away from the center of the geometry. The only symmetry remaining is total rotations of the space, which we can use to fix one parameter; we will fix the center of region A to be at $\theta_{A,{\rm cen}}= \pi$. We then have the three remaining parameters $\theta_{B,{\rm cen}}$, $\theta_{A,{\rm wid}}$ and $\theta_{B,{\rm wid}}$:
\begin{eqnarray}\label{params}
0 \leq \theta_{B,{\rm cen}}\leq \pi \,, \quad\quad
0 \leq \theta_{A,{\rm wid}} \leq 2\pi - 2\theta_{B,{\rm cen}}  \,, \quad \quad 
0\leq \theta_{B,{\rm wid}} \leq 2 \pi -2\theta_{B,{\rm cen}} - \theta_{A,{\rm wid}} \,.
\end{eqnarray}
(Strictly speaking we are double-counting configurations by not always declaring for example $\theta_{A, {\rm wid}} \leq \theta_{B, {\rm wid}}$, but we do this to make our phase diagrams intuitively clearer.)

We now numerically determine the BSCs for the mutual information (which also gives the squashed entanglement), the entanglement of purification (which also gives the $R$-correlation) and the $Q$-correlation for various values of the parameters $r_H$, $\theta_{B,{\rm cen}}$, $\theta_{A,{\rm wid}}$ and $\theta_{B,{\rm wid}}$. We will chart out the regions, or phases, on parameter space where each measure has a BSC with a fixed topology. There are then phase transitions, where one BSC jumps into another for one or more measures. As we will see, the BSCs of $I(A:B)$ and $E_P$ behave similarly to how they do in the pure thermal case, while $E_Q$ shows considerably more structure.

In what follows we will focus on cases where $A$ and $B$ are diametrically opposite, so $\theta_{B, {\rm cen}} = 0$. For these cases, when the entanglement wedge of $AB$ is connected, it always includes the black hole horizon, as in Fig.~\ref{BTZ AB}c. Configurations with $\theta_{B, {\rm cen}}$ at other values can lead to connected entanglement wedges not including the horizon as in Fig.~\ref{BTZ AB}b, but the BSCs of the correlation measures then have the same form as in the cases with no horizon reviewed in section~\ref{Sec:OCMAdSCFT}. All the novel structure will be visible for $A$ and $B$ diametrically opposite.

\subsection{Phases for reduced thermal states}

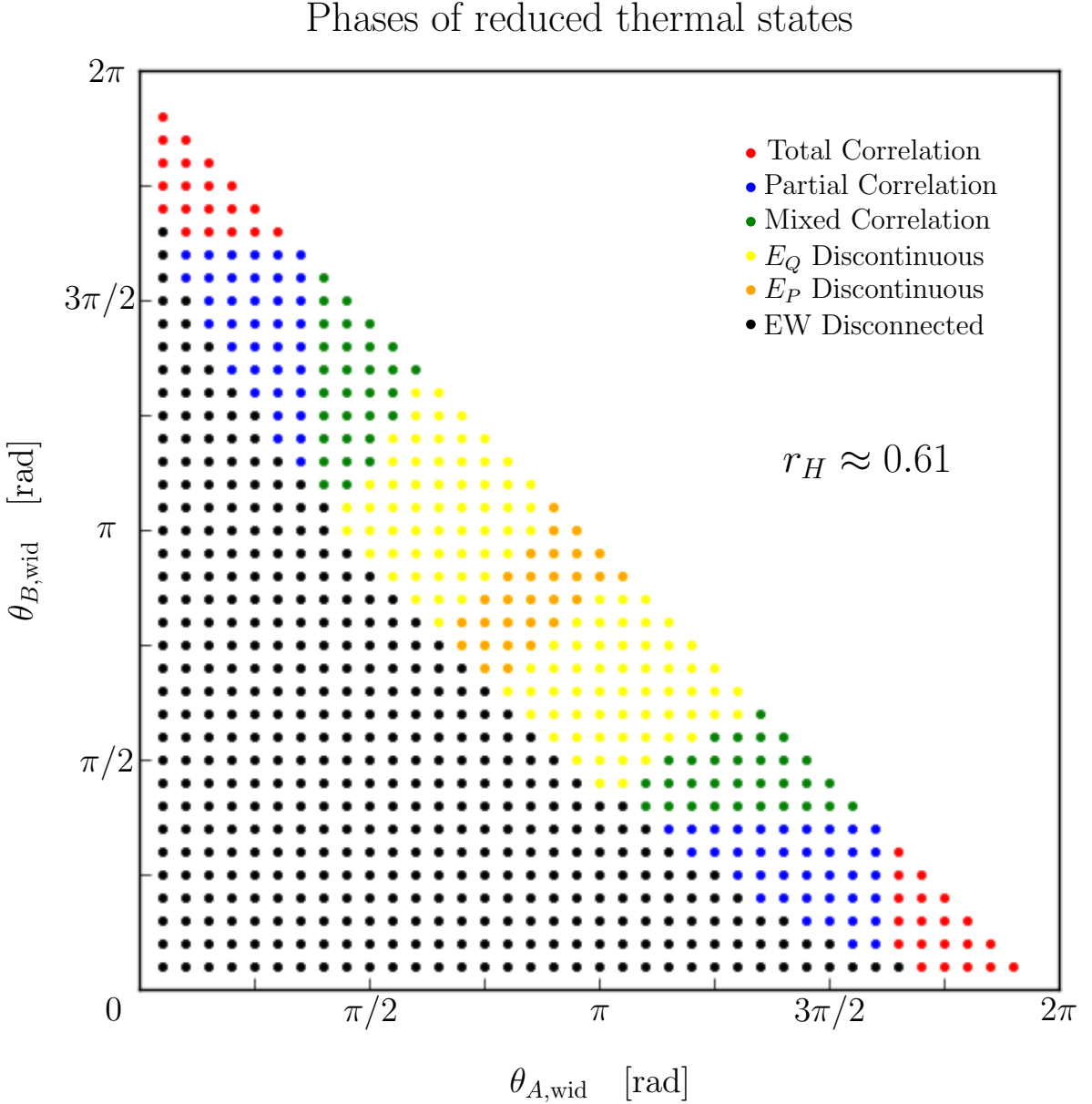
\begin{figure}[H]
\centering
\begin{tikzpicture}

\node[inner sep=0pt] at (0,0)
    {\includegraphics[scale = 1.1, trim = 100 0 100 0, clip]{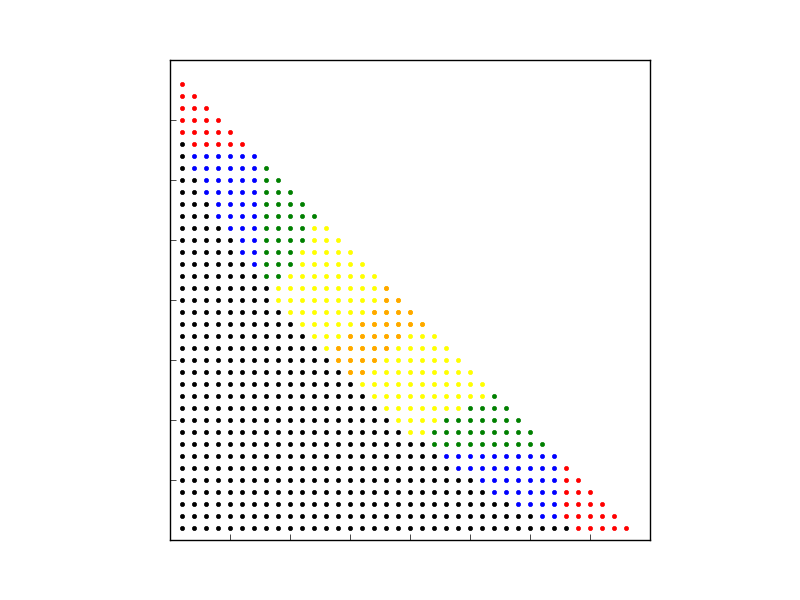}};
\definecolor{darkgreen}{rgb}{0,0.5,0}
\definecolor{yellow2}{rgb}{1,1,0}
\definecolor{orange2}{rgb}{1,0.643,0}
\fill[red] (2.5,5.5) circle (2.0pt);
\fill[blue] (2.5,5.0) circle (2.0pt);
\fill[darkgreen] (2.5,4.5) circle (2.0pt);
\fill[yellow2] (2.5,4.0) circle (2.0pt);
\fill[orange2] (2.5,3.5) circle (2.0pt);
\fill[black] (2.5,3.0) circle (2.0pt);
\note{4.3,5.55}{\large{Total Correlation}}
\note{4.4,5.05}{\large{Partial Correlation}}
\note{4.35,4.55}{\large{Mixed Correlation}}
\note{4.3,3.95}{\large{$E_Q$ Discontinuous}}
\note{4.3,3.5}{\large{$E_P$ Discontinuous}}
\note{4.3,3.0}{\large{EW Disconnected}}
\note{4.2,1}{\LARGE{$r_H \approx 0.61$}}

\note{-6.8,-7}{\Large{$0$}}
\note{7,-7}{\Large{$2\pi$}}
\note{-6.9,6.7}{\Large{$2\pi$}}
\note{0.3,-7.05}{\Large{$\pi$}}
\note{-6.9,0}{\Large{$\pi$}}
\note{-6.9,-3.4}{\Large{$\pi/2$}}
\note{-7,3.35}{\Large{$3\pi/2$}}
\note{-3.05,-7.05}{\Large{$\pi/2$}}
\note{3.65,-7.05}{\Large{$3\pi/2$}}

\note{0.3,-8.1}{\Large{$\theta_{A,\rm wid}$ ~ [rad]}}
\node[rotate=90] at (-8.1,0) {\Large{$\theta_{B,\rm wid}$ ~ [rad]}};
\note{0,7.5}{\LARGE{Phases of reduced thermal states}}

\end{tikzpicture}
\caption{Phase diagram for $r_H \approx 0.61$ and $\theta_{B,cen} = 0$.  At these values of $r_H$ and $\theta_{B,cen}$, all five phases are present.  The two triple points are visible on the diagonal $\theta_{A,wid} + \theta_{B,wid} = 2\pi$.} 
    \label{fig:EQ Phase r0.28}
\end{figure}

The phases we saw for thermal states still exist for the reduced thermal states with diametrically opposite regions: total correlation, mixed correlation, $E_Q$-discontinuous and $E_P$-discontinuous. A new phase also appears between the total and mixed correlation phases, which we call partial correlation, when one region is small enough that its complement has jumped the horizon, but the other region is not large enough to have done so. For sufficiently small regions we also find the case of a disconnected entanglement wedge. The phase diagram for black hole radius $r_H \approx 0.61$ is shown in Fig.~\ref{fig:EQ Phase r0.28}.

The behavior of $I(A:B)$ and $E_P$ show similar transitions between bulk surface configurations in reduced thermal states as in thermal states. In particular, the mutual information has a transition between its BSC not surrounding the horizon in the total correlation phase, and surrounding it in all other nontrivial phases. $E_P$, meanwhile, again has a BSC (the enanglement wedge cross section) that is continuous for most of the parameter space, while it becomes discontinuous for the $E_P$-discontinuous phase. 

The new structure in the case of reduced thermal states comes from the $Q$-correlation. Throughout the parameter space, $E_Q$ has two equivalent BSCs which look different but calculate the same value. This was seen in case without a black hole, where the two equivalent presentations were simply parity reflections of each other (see Fig.~\ref{AdS EQ}). In this more general context however, the two equivalent presentations can look quite different. The two equivalent BSCs exist for the reason outlined in sec.~\ref{Sec:OCMAdSCFT}: $S(Ba)$ has several different potential presentations, as shown in Fig.~\ref{fig:Ba}, each of which may be preferred for a given candidate purification. However, because $S(Ba)$ is subtracted in the $E_Q$ objective function, the minimization that leads to $E_Q$ singles out the purification for which two of these
potential presentations of $S(Ba)$ are equal; away from this balance point the minimization leading to $S(Ba)$ would prefer the smaller of the two, which then would make $E_Q$ larger.  Thus the ``minimization within a minimization" with a subtracted entropy in the objective function produces  two equivalent BSCs for $E_Q$. Each phase in the phase diagram is associated with a particular pair of equal BSCs. In the thermal state limit of $A$ and $B$ filling the boundary, the two equivalent presentations degenerate to one of the single configurations studied in the last section.

\begin{figure}

\begin{tikzpicture}
\node[inner sep=0pt] at (8.25,15)
    {\includegraphics[scale = 0.5]{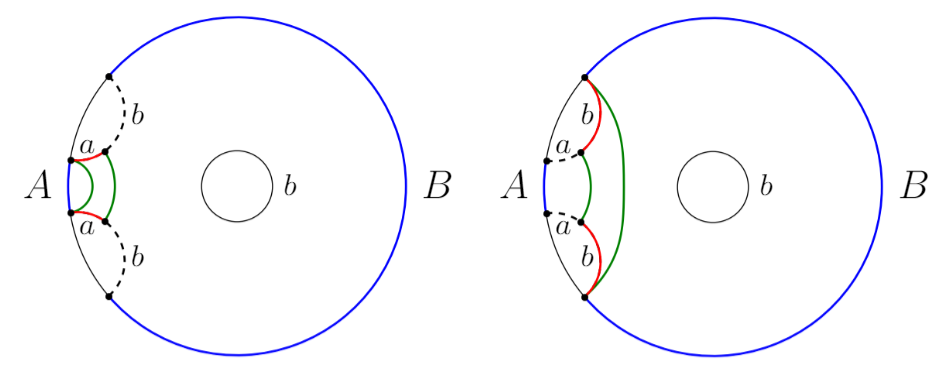}};
\node[inner sep=0pt] at (8.25,11)
    {\includegraphics[scale = 0.5]{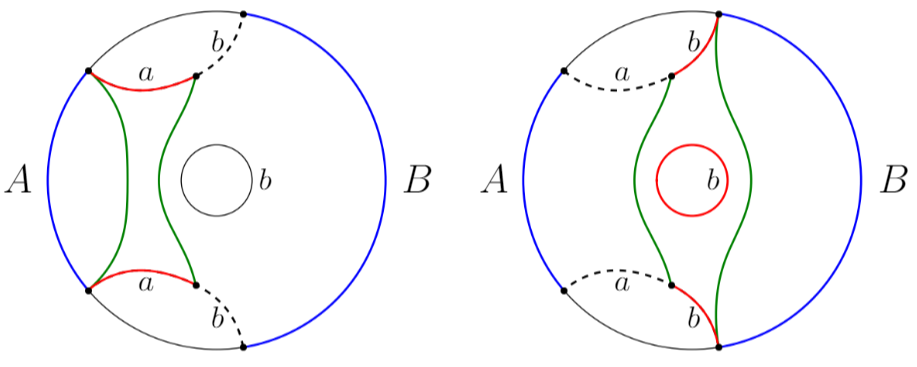}};
\node[inner sep=0pt] at (8.25,7)
    {\includegraphics[scale = 0.5]{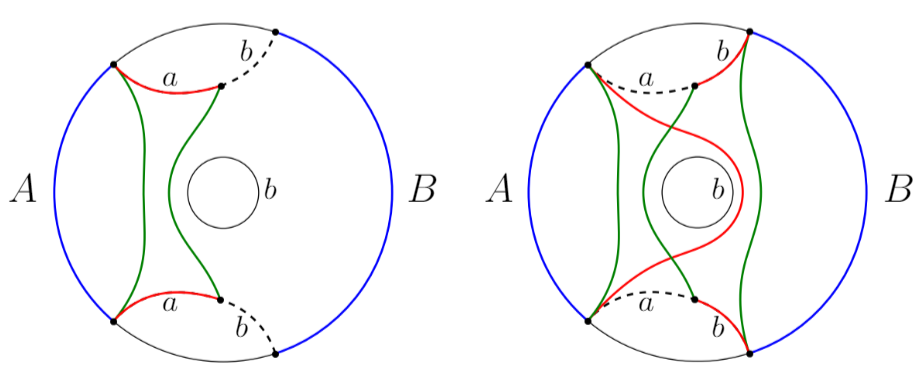}};
\node[inner sep=0pt] at (8.25,3)
    {\includegraphics[scale = 0.5]{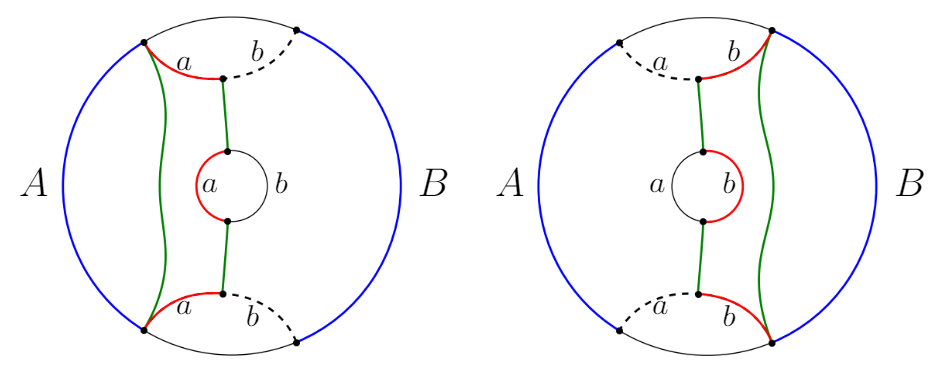}};

\note{1.5,15.3}{\large{Total}}
\note{1.5,14.9}{\large{Correlation}}

\note{1.5,11.3}{\large{Partial}}
\note{1.5,10.9}{\large{Correlation}}

\note{1.5,7.3}{\large{Mixed}}
\note{1.5,6.9}{\large{Correlation}}

\note{1.5,3.1}{\large{Discontinuous}}

\note{14.8,15.6}{\normalsize{$I(A:E) = 0$}}
\note{15.2,15.0}{\normalsize{$I(B:E)= 2S(E)$}}
\note{14.3,14.4}{\normalsize{$I_{3}=0$}}

\note{14.8,11.6}{\normalsize{$I(A:E) = 0$}}
\note{15.55,11.0}{\normalsize{$0 \leq I(B:E) < 2S(E)$}}
\note{15.6,10.4}{\normalsize{$I_{3}= I(B:E) - 2S(E)$}}

\note{15.55,7.6}{\normalsize{$0 < I(A:E) < 2S(E)$}}
\note{15.55,7.0}{\normalsize{$0 < I(B:E) < 2S(E)$}}
\note{15,6.4}{\normalsize{$I_{3}= I(A:E)+$}}
\note{16.3,5.8}{\normalsize{$ I(B:E) - 2S(E)$}}

\note{15.55,3.6}{\normalsize{$0 < I(A:E) < 2S(E)$}}
\note{15.55,3.0}{\normalsize{$0 < I(B:E) < 2S(E)$}}
\note{15,2.4}{\normalsize{$I_{3}= I(A:E)+$}}
\note{16.3,1.8}{\normalsize{$ I(B:E) - 2S(E)$}}

\note{1.5, 17.5}{\Large{Phase}}
\note{8.25, 17.5}{\Large{Equivalent BSCs}}
\note{15.25, 17.5}{\Large{Inequalities}}


        \draw[black, thick] (0,18)--(18,18);
        \draw[black, thick] (0,17)--(18,17);
        \draw[black, thick] (0,13)--(18,13);
        \draw[black, thick] (0,9)--(18,9);
        \draw[black, thick] (0,5)--(18,5);
        \draw[black, thick] (0,1)--(18,1);
        \draw[black, thick] (0,1)--(0,18);
        \draw[black, thick] (3,1)--(3,18);
        \draw[black, thick] (13.5,1)--(13.5,18);
        \draw[black, thick] (18,1)--(18,18);

\end{tikzpicture}

\caption{Shown in the table are all the distinct BSCs of $E_Q$ and the phases in which they occur, and the values of $I(A:E)$, $I(B:E)$ and $I_3(A:B:E)$ associated with each phase. Here, the discontinuous phase includes both the $E_Q$- and $E_P$-discontinuous phases, since the BSCs for $E_Q$ are the same in both.  Note that the values of these three quantities can always distinguish between the three continuous phases, but cannot distinguish the discontinuous phase from the mixed correlation phase.  Each phase consists of two distinct BSCs which both give the numerical value of $E_Q$, an equation which fixes the locations of the optimization points for any pair of boundary regions $A$ and $B$.}
\label{fig:EQ_phases}
\end{figure}

As we did for the thermal states, we will go through each phase, describing the inequalities or equalities obeyed by various information measures, and describing the bulk surface configurations for $I(A:B)$, $E_Q$ and $E_P$. The BSCs for $E_Q$ are collected in Fig.~\ref{fig:EQ_phases}. As with thermal states, we will assume $A$ is smaller than $B$, though all the same phases exist for the exchange of the two regions. At the end of the subsection we present a plot of values for $I(A:B)$, $E_Q$ and $E_P$ for a cut through the phase diagram with $A$ and $B$ occupying a fixed total area.

 We note that for reduced thermal states all OCMs are cutoff-independent, as can be seen from their BSCs having the balance property discussed in Sec. \ref{RT}.

\bigskip
\noindent
{\bf Total correlation phase:}

In thermal states, the total correlation phase corresponded to $A$ being sufficiently small that its complement $B$ had ``jumped the horizon" and the horizon was part of $\Gamma_B$. In reduced thermal states, $A$ is no longer the boundary complement of $B$, but we can still characterize the total correlation phase (shown in red in Fig. \ref{fig:EQ Phase r0.28}) as when $B$ is large enough to include the horizon in its entanglement wedge. This implies that 
\begin{equation}
    S(BE) - S(B) = - S(E) \quad \leftrightarrow \quad I(B:E) = 2 S(E) \,,
\end{equation}
the maximum value for $I(B:E)$, and $B$ is totally correlated with $E$, hence ``total correlation" phase. Since $A$ is even smaller than the complement of $B$, we also have that $A$ has no mutual information with $E$,
\begin{equation}
   I(A:E) = 0 \,.
\end{equation}
In addition  the equation (\ref{Eq:HorizonInWedge}) applies, which combined with the above implies tripartite relations in this phase, in particular the vanishing of the tripartite information $I_3(A:B:E)$,
\begin{equation}
    I_3(A:B:E) = 0 \,,
\end{equation}
which implies all conditional mutual informations reduce to mutual informations,
\begin{equation}
    I(A:E|B) = I(A:E) = 0 \,, \quad  I(B:E|A) = I(B:E)= 2 S(E)\,, \quad I(A:B|E)= I(A:B)\,.
\end{equation}
This phase occurs when the sizes of the two regions are highly asymmetric.

\begin{figure}[H]
\centering
\begin{tikzpicture}
\node[inner sep=0pt] at (5,0)
    {\includegraphics[scale = 0.32]{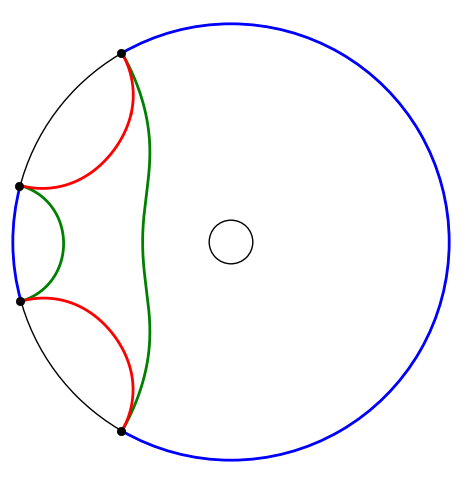}};
\note{2.9,0}{\large{$A$}}
\note{7.1,0}{\large{$B$}}
\end{tikzpicture}
\caption{The BSC for $I(A:B)$ ($=2 E_{\rm sq}(A:B))$ in the total correlation phase.  Region $B$ is large enough that its entanglement wedge contains the horizon, which is the defining property of the total correlation phase.}
\label{fig:E_sq and I(A:B)}
\end{figure}

Consider now the presentation of the OCMs in the total correlation phase. As for thermal states, this is the only phase for which $I(A:B)$ has a BSC that does not surround the horizon, as shown in Fig.~\ref{fig:E_sq and I(A:B)}. Unlike the case for thermal states, however, reduced thermal states have $\frac{1}{2} I(A:B) < S(A)$.

\begin{figure}[H]
\centering
\begin{tikzpicture}
\node[inner sep=0pt] at (5,0)
    {\includegraphics[scale = 0.32]{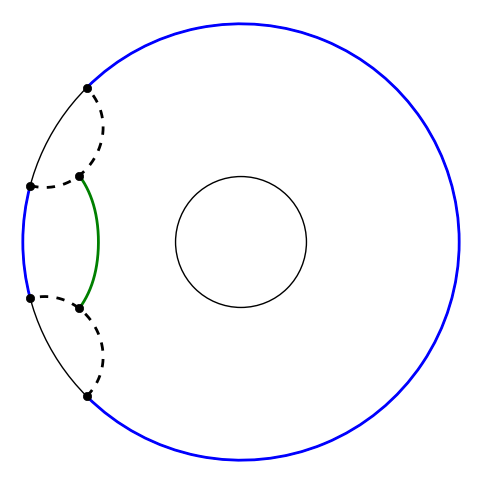}};
\note{2.9,0}{\large{$A$}}
\note{7.1,0}{\large{$B$}}
\note{3.5,0.31}{\small{$a$}}  
\note{3.5,-0.31}{\small{$a$}}
\note{4,.95}{\small{$b$}}  
\note{4,-.95}{\small{$b$}}
\note{5.7,0}{\small{$b$}}
\end{tikzpicture}
\caption{The BSC for $E_P$ in every phase except $E_P$-discontinuous.}
    \label{fig:EP and ER diff regions}
\end{figure}

$E_P$ is calculated by minimizing $S(Aa)$, with $a$ now having a part on the non-horizon components of $\Gamma_{AB}$, as well as on the horizon. The result is a combination of the situation depicted in the reduced pure state case of Fig.~\ref{fig:AdS_SSC} and the situation in the thermal state as in Fig.~\ref{Fig:ThermalResults}. The BSC is still given by the EWCS.  In all phases except the $E_P$-discontinuous phase, this is a continuous curve, having assigned the entire horizon to the ancilla $b$, as shown in Fig.~\ref{fig:EP and ER diff regions}.

We turn now to $E_Q$, which as described above has two equivalent BSCs in each phase. In the total correlation phase, shown in row 1 of Fig.~\ref{fig:EQ_phases}, neither of the two equivalent BSCs enclose the horizon, nor does the horizon contribute to either configuration.  These configurations are topologically identical to the case in pure AdS, as shown in Fig.~\ref{AdS EQ}; the two equivalent presentations of $S(Ba)$ are $a + \Gamma_B$ and $b + \Gamma_A$.

\bigskip
\noindent
{\bf Partial correlation phase:}

Since $A$ is no longer the complement of $B$ in reduced thermal states, it becomes possible to choose regions such that $B$ is not large enough to have jumped the horizon, but the complement of $A$ is. This constitutes a new phase not present in the thermal state phase diagram, the partial correlation phase (shown in blue in Fig. \ref{fig:EQ Phase r0.28}). In this phase $A$ is still small enough that
\begin{equation}
    I(A:E) = 0 \,,
\end{equation}
but $B$ has become smaller such that now
\begin{equation}
    S(BE) - S(B) > - S(E) \quad \leftrightarrow \quad I(B:E) < 2 S(E) \,.
\end{equation}
Thus although $E$ remains uncorrelated with $A$, it is no longer maximally correlated with $B$, hence the partial correlation name. The tripartite information no longer vanishes, but satisfies
\begin{equation}
    I_3(A:B:E) = I(B:E)-2S(E) \,,
\end{equation}
and is sensitive only to $B$. This phase occurs when the boundary regions $A$ and $B$ are moderately asymmetric.

\begin{figure}[H]
\centering
\begin{tikzpicture}
\node[inner sep=0pt] at (11,0)
    {\includegraphics[scale = 0.32]{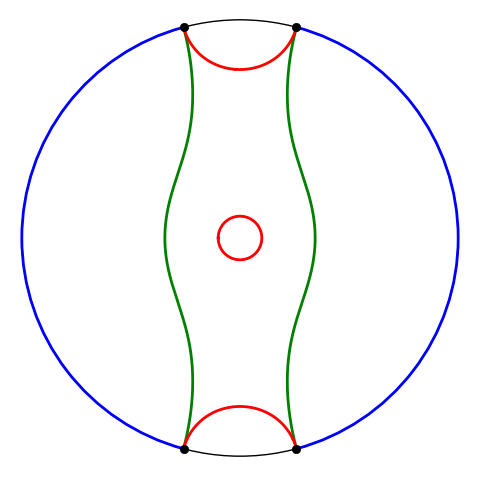}};
\note{8.9,0}{\large{$A$}}  
\note{13.1,0}{\large{$B$}}  
\end{tikzpicture}
\caption{The BSC for $I(A:B)$ ($=2 E_{\rm sq}(A:B))$ in all phases except the total correlation phase.  Both boundary regions are small enough that their entanglement wedges do not contain the horizon, which is the defining property of being outside the total correlation phase.}
    
    \label{fig:E_sq and I(A:B) 2}
\end{figure}

Outside the total correlation phase, where $\Gamma_B$ does not include the horizon, $I(A:B)$ transitions to a presentation that surrounds the horizon, as in Fig.~\ref{fig:E_sq and I(A:B) 2}; this holds in the partial correlation phase as well as all those that follow. $E_P$ remains continuous, as in Fig.~\ref{fig:EP and ER diff regions}.

The equivalent BSCs of $E_Q$ undergo a change in this phase. As shown in row 2 of Fig.~\ref{fig:EQ_phases}, the first presentation has the same topology as in the total correlation phase, but the second one now surrounds the horizon, and subtracts off the horizon length; the horizon is still assigned entirely to $b$. The two presentations of $S(Ba)$ are still $a + \Gamma_B$ and $b + \Gamma_A$, but now the horizon is no longer part of $\Gamma_B$, though it is still part of $b$.

The boundary between the total correlation and partial correlation phases depends only on the size of $B$, not that of $A$, and is therefore a straight line in the phase diagram.

\bigskip
\noindent
{\bf Mixed correlation phase:}

In the mixed correlation phase (shown in green in Fig. \ref{fig:EQ Phase r0.28}), neither $B$ nor the complement of $A$ is large enough to have jumped the horizon, and thus this phase occurs when the boundary regions are more symmetric.   In this phase, neither of the two quantities $I(A:E)$ and $I(B:E)$ are zero and both lie strictly between 0 and $2S(E)$:
\begin{equation}
    0 < I(A:E), I(B:E) < 2 S(E) \,,
\end{equation}
hence the name mixed correlation. Due to Eqn.~(\ref{Eq:HorizonInWedge}) the tripartite information satisfies
\begin{equation}
    I_3(A:B:E) = I(A:E) + I(B:E) - 2 S(E) \,,
\end{equation}
depending on both boundary regions.

In the mixed correlation phase $I(A:B)$ retains the presentation in Fig.~\ref{fig:E_sq and I(A:B) 2}, and $E_P$ remains continuous, as in Fig.~\ref{fig:EP and ER diff regions}.

The mixed correlation BSCs of $E_Q$ are shown in row 3 of Fig.~\ref{fig:EQ_phases}. The first presentation retains the same topology as in the total and partial correlation phases, and still results from $S(Ba)$ being realized by $a + \Gamma_B$. However, the second presentation is considerably more intricate, enclosing the horizon between a negative and a positive curve.  In this case $S(Ba)$ is realized not by $a + \Gamma_A$, but by a configuration like the third image in Fig.~\ref{fig:Ba}b, but with no horizon component since the horizon still belongs entirely to $b$. Neither equivalent presentation contains a horizon component.

The boundary between the partial correlation and mixed correlation phases depends only on the size of $A$, a straight line in the phase diagram, perpendicular to the boundary between the total and partial correlation phases. The two lines meet at the edge of the phase diagram where the thermal states are, at the transition between the total and mixed correlation phases, with the partial correlation phase disappearing, forming a triple point.

\bigskip
\noindent
{\bf $E_Q$-discontinuous phase:}

The remaining phases involve a transition to $S(Aa)$ becoming discontinuous for one or more measures, which may happen when the sizes of the two regions are similar. The inequalities satisfied by $I(A:E)$, $I(B:E)$ and $I_3(A:B:E)$ are the same as in the mixed correlation phase. 

In the $E_Q$-discontinuous phase (shown in yellow in Fig. \ref{fig:EQ Phase r0.28}), $I(A:B)$ retains the presentation shown in Fig.~\ref{fig:E_sq and I(A:B) 2} in this phase, and $E_P$ is still continuous, as in Fig.~\ref{fig:EP and ER diff regions}. When $E_P$ becomes discontinuous as well we enter the $E_P$-discontinuous phase. 

In the $E_Q$-discontinuous phase, the two presentations for $E_Q$ shift to row 4 of Fig.~\ref{fig:EQ_phases}, which are symmetric and similar to the case with no black hole shown in figure~\ref{AdS EQ}. Now the horizon is shared between $a$ and $b$ in the optimizing purification for $E_Q$, and either the $a$ part or the $b$ part of the horizon is subtracted in calculating the measure.

The discontinuous phases do not have straight-line boundaries on the phase diagram, but are preferred when $A$ and $B$ together fill more of the diagram, with the phase region opening towards the edge of the phase diagram where the thermal states are. This phase only exists for $r_H$ sufficiently large, and vanishes for horizon smaller than $r_H \approx 0.37$. As $r_H$ is increased past this value the phase appears first for the thermal states at the edge of the diagram and then penetrates further into the reduced thermal states with similar sized regions.

\bigskip
\noindent
{\bf $E_P$-discontinuous phase:}

\begin{figure}[H]
\centering
\begin{tikzpicture}
\node[inner sep=0pt] at (11,0)
    {\includegraphics[scale = 0.32]{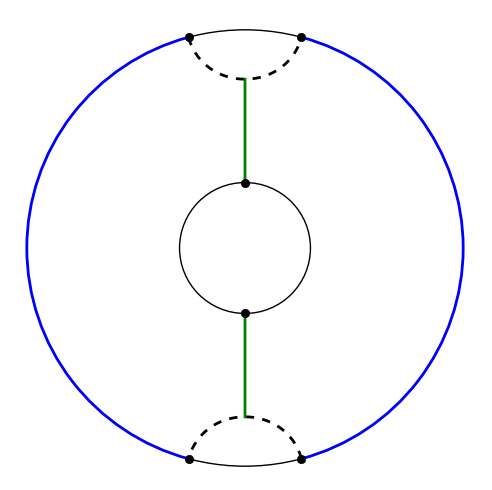}};

\note{8.9,0}{\large{$A$}}  

\note{10.6,1.3}{\small{$a$}}  
\note{10.6,-1.3}{\small{$a$}}
\note{10.3,0}{\small{$a$}} 
\note{11.4,1.3}{\small{$b$}}  
\note{11.4,-1.3}{\small{$b$}}
\note{11.7,0}{\small{$b$}} 
   
\note{13.1,0}{\large{$B$}}  
\end{tikzpicture}
\caption{The BSC for $E_P$ in the $E_P$-discontinuous phase.}
    \label{fig:EP and ER diff rad}
\end{figure}

Finally, for most symmetric boundary regions and $r_H$ sufficiently large, the presentation of $E_P$ (the EWCS) becomes discontinuous as well, as shown in Fig.~\ref{fig:EP and ER diff rad}, and we enter the $E_P$-discontinuous phase (shown in orange in Fig. \ref{fig:EQ Phase r0.28}). Now the optimal purification for $E_P$ also involves partitioning the horizon between $a$ and $b$. Like the $E_Q$-discontinuous phase, this phase only exists for $r_H$ sufficiently large (greater than $r_H \approx 0.58$) and appears preferentially near the boundary, surrounded by the $E_Q$-discontinuous phase. This phase satisfies the same inequalities for $I(A:E)$, $I(B:E)$ and $I_3(A:B:E)$ as the mixed correlation and $E_Q$-discontinuous phases, $E_Q$ remains discontinuous as well, and $I(A:B)$ retains the presentation shown in Fig.~\ref{fig:E_sq and I(A:B) 2}.

\begin{figure}[H]
\centering
\begin{tikzpicture}

\node[inner sep=0pt] at (3.3,0.25)
    {\includegraphics[trim = 115 43 68 36, clip, scale = 0.56]{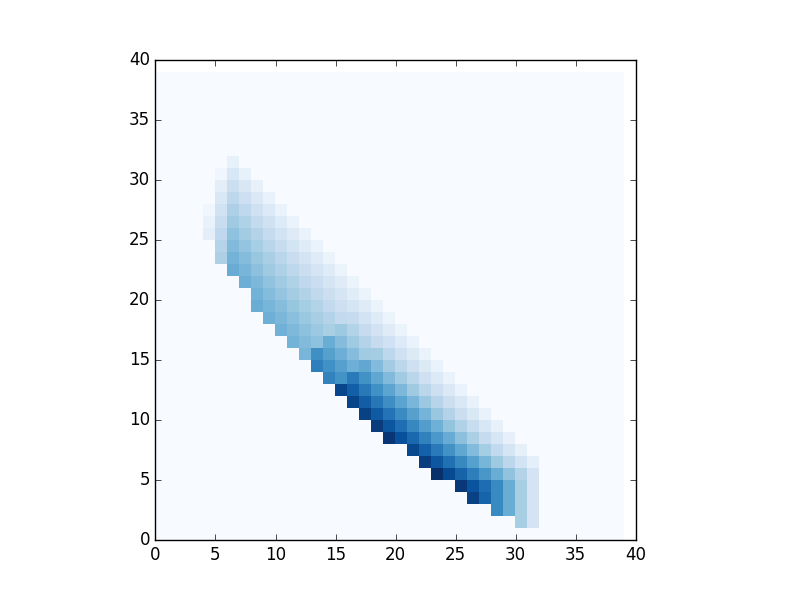}};
\node[inner sep=0pt] at (12.3,0.22)
    {\includegraphics[trim = 110 39 68 36, clip, scale = 0.585]{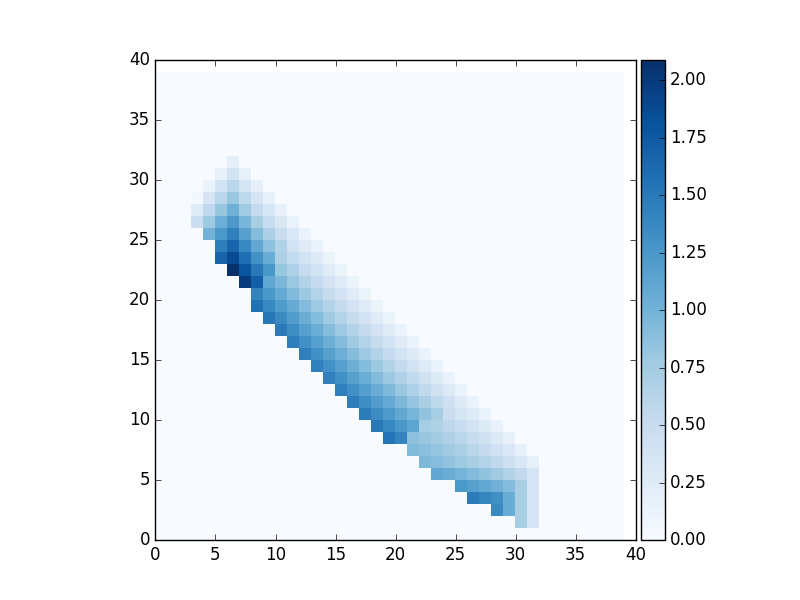}};
\note{2.5,-4.25}{\large{$\theta_{A,\rm wid}$ ~ [rad]}}
\node[rotate=90] at (-1.75,0) {\large{$\theta_{B,\rm wid}$ ~ [rad]}};

\note{-0.95,-3.55}{\large{0}}
\note{-0.95,0.2}{\large{$\pi$}}
\note{-1.05,3.75}{\large{$2\pi$}}
\note{2.8,-3.55}{\large{$\pi$}}
\note{6.3,-3.6}{\large{$2\pi$}}

\note{8.05,-3.55}{\large{0}}
\note{8.05,0.2}{\large{$\pi$}}
\note{7.95,3.75}{\large{$2\pi$}}
\note{11.8,-3.55}{\large{$\pi$}}
\note{15.3,-3.6}{\large{$2\pi$}}

\note{11.5,-4.25}{\large{$\theta_{A,\rm wid}$ ~ [rad]}}
\node[rotate=90] at (7.2,0) {\large{$\theta_{B,\rm wid}$ ~ [rad]}};
\note{2.8,4}{\Large{$I(a':E|A)$ for $E_P$}}
\note{11.8,4}{\Large{$I(a':E|A)$ for $E_Q$}}
\end{tikzpicture}
 \caption{The figure shows the discontinuity in $I(a':E|A)$ for the $E_P$-optimal purification on the left, and the $E_Q$-optimal purification on the right. These discontinuities mark the boundaries where $E_P$ and $E_Q$, respectively, become discontinuous.}
    \label{discontinuity}
\end{figure}

Unlike the total, partial and mixed correlation phases, which correspond to particular relations for $I(A:E)$ and $I(B:E)$, the transitions to the $E_Q$-discontinuous and $E_P$-discontinuous phases are not associated to any straightforward information measures involving only $A$, $B$ and $E$. However, we can see that they are accompanied by a discontinuity in the quantity $I(a':E|A)$, where $a' \equiv a\setminus E$ are the parts of $a$ not including the horizon, as can be seen in Fig.~\ref{discontinuity}. 

%
%
%

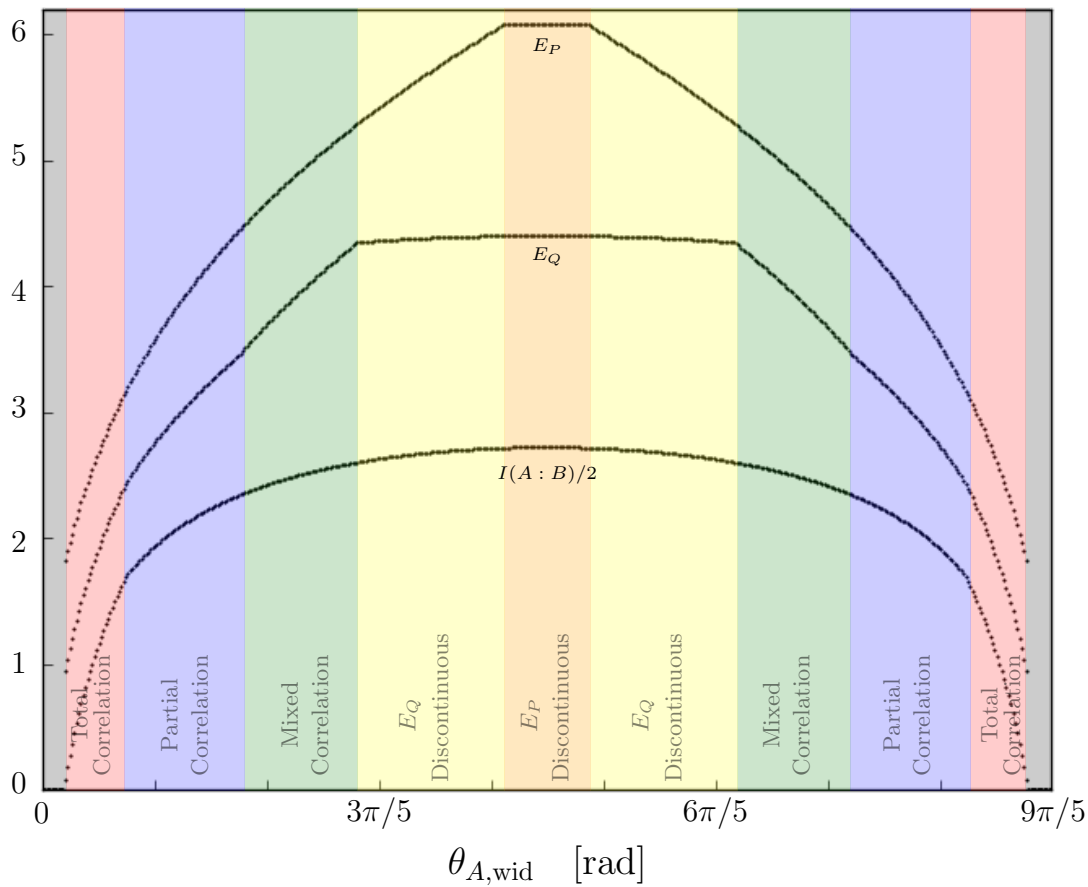
\begin{figure}[H]

\centering
\begin{tikzpicture}
\node[inner sep=0pt] at (0,0)
    {\includegraphics[scale = 0.85]{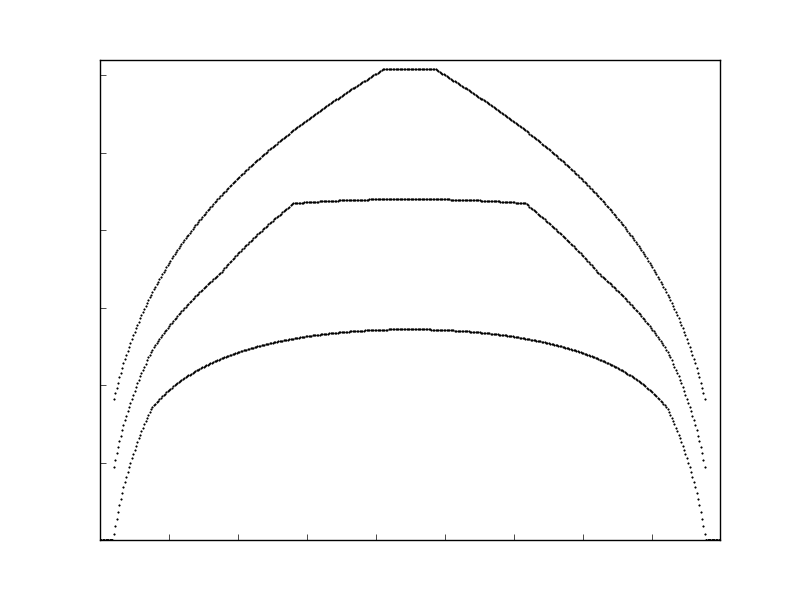}};\
    \definecolor{darkgreen}{rgb}{0,0.5,0}
\definecolor{yellow2}{rgb}{1,1,0}
\definecolor{orange2}{rgb}{1,0.643,0}
    \draw [ultra thin, draw=black, fill=black, opacity=0.2]
       (-6.45,-5.2) -- (-6.17,-5.2) -- (-6.17,5.2) -- (-6.45,5.2);
    \draw [ultra thin, draw=black, fill=black, opacity=0.2]
       (6.58,-5.2) -- (6.9,-5.2) -- (6.9,5.2) -- (6.58,5.2);
    
    \draw [ultra thin, draw=red, fill=red, opacity=0.2]
       (-6.17,-5.2) -- (-5.4,-5.2) -- (-5.4,5.2) -- (-6.17,5.2);
    \draw [ultra thin, draw=red, fill=red, opacity=0.2]
       (5.85,-5.2) -- (6.58,-5.2) -- (6.58,5.2) -- (5.85,5.2);
    
    \draw [ultra thin, draw=darkgreen, fill=darkgreen, opacity=0.2]
       (-3.8,-5.2) -- (-2.3,-5.2) -- (-2.3,5.2) -- (-3.8,5.2);
    \draw [ultra thin, draw=darkgreen, fill=darkgreen, opacity=0.2]
       (2.75,-5.2) -- (4.25,-5.2) -- (4.25,5.2) -- (2.75,5.2);
    \draw [ultra thin, draw=yellow2, fill=yellow2, opacity=0.2]
       (-2.3,-5.2) -- (-0.35,-5.2) -- (-0.35,5.2) -- (-2.3,5.2);
    \draw [ultra thin, draw=yellow2, fill=yellow2, opacity=0.2]
       (0.8,-5.2) -- (2.75,-5.2) -- (2.75,5.2) -- (0.8,5.2);
    \draw [ultra thin, draw=orange2, fill=orange2, opacity=0.25]
       (-0.35,-5.2) -- (0.8,-5.2) -- (0.8,5.2) -- (-0.35,5.2);
       
    \draw [ultra thin, draw=blue, fill=blue, opacity=0.2]
       (-5.4,-5.2) -- (-3.8,-5.2) -- (-3.8,5.2) -- (-5.4,5.2);
    \draw [ultra thin, draw=blue, fill=blue, opacity=0.2]
       (4.25,-5.2) -- (5.85,-5.2) -- (5.85,5.2) -- (4.25,5.2);

    \note{ -6.5,-5.5}{\large{$0$}}
    \note{ -2.01,-5.5}{\large{$3\pi/5$}}
    \note{ 2.46,-5.5}{\large{$6\pi/5$}}
    \note{ 6.95,-5.5}{\large{$9\pi / 5$}}
    \note{ -6.8,-5.1}{\large{$0$}}
    \note{ -6.8,-3.5}{\large{$1$}}
    \note{ -6.8,-1.9}{\large{$2$}}
    \note{ -6.8,-0.1}{\large{$3$}}
    \note{ -6.8,1.45}{\large{$4$}}
    \note{ -6.8,3.25}{\large{$5$}}
    \note{ -6.8,4.9}{\large{$6$}}

     \note{ 0.225,4.7}{$E_P$}
        \note{ 0.225,1.9}{$E_Q$}
        \note{ 0.225,-1}{$I(A:B)/2$}
        
    \node[rotate=90, text opacity = 0.5] at (-6,-4.2) {Total};
    \node[rotate=90, text opacity = 0.5] at (-5.7,-4.2) {Correlation};
    
    \node[rotate=90, text opacity = 0.5] at (-4.8,-4.2) {Partial};
    \node[rotate=90, text opacity = 0.5] at (-4.4,-4.2) {Correlation};
    
    \node[rotate=90, text opacity = 0.5] at (-3.2,-4.2) {Mixed};
    \node[rotate=90, text opacity = 0.5] at (-2.8,-4.2) {Correlation};
    
    \node[rotate=90, text opacity = 0.5] at (-1.6,-4.2) {$E_Q$};
    \node[rotate=90, text opacity = 0.5] at (-1.2,-4.1) {Discontinuous};
    
    \node[rotate=90, text opacity = 0.5] at (0,-4.2) {$E_P$};
    \node[rotate=90, text opacity = 0.5] at (0.4,-4.1) {Discontinuous};
    
    \node[rotate=90, text opacity = 0.5] at (1.5,-4.2) {$E_Q$};
    \node[rotate=90, text opacity = 0.5] at (1.9,-4.1) {Discontinuous};
    
    \node[rotate=90, text opacity = 0.5] at (3.2,-4.2) {Mixed};
    \node[rotate=90, text opacity = 0.5] at (3.6,-4.2) {Correlation};
    
    \node[rotate=90, text opacity = 0.5] at (4.8,-4.2) {Partial};
    \node[rotate=90, text opacity = 0.5] at (5.2,-4.2) {Correlation};
    
    \node[rotate=90, text opacity = 0.5] at (6.1,-4.2) {Total};
    \node[rotate=90, text opacity = 0.5] at (6.4,-4.2) {Correlation};
    
\note{0.225,-6.2}{\Large{$\theta_{A,\rm wid}$ ~ [rad]}}
\note{0.225,6}{\LARGE{Correlation measures in reduced thermal states}}

\end{tikzpicture}
\caption{Optimized correlation measures $E_P$ and $E_Q$, along with their lower bound $I(A:B)/2$, plotted for reduced thermal states as a function of $\theta_{A,\rm wid}$, with $\theta_{A,\rm wid} + \theta_{B,\rm wid} = 9\pi/5$ and $r_H\approx 0.61$.  The entanglement wedge is disconnected in the two black shaded regions, so all correlation measures vanish in these regions. The upper bound $\min(S(A), S(B))$, not shown, is an order of magnitude larger than the quantities which are plotted.}
\label{EQ_EP}
\end{figure}

The numerical values of $I(A:B)/2$, $E_Q$, and $E_P$ are plotted in Fig. \ref{EQ_EP} for a cut through the $r_H\approx 0.61$ phase diagram with $\theta_{A,\rm wid} + \theta_{B,\rm wid} = 9\pi/5$. Again the validity of Ineq. (\ref{Inequalities}) is clearly visible.  Also, similar to the thermal state case, the phase transitions are accompanied by discontinuities in the first derivative of at least one OCM, as one BSM becomes preferred over another one. Thus we see that $E_P$ has a discontinuity at the boundary of the $E_P$- and $E_Q$- discontinuous phases, $E_Q$ has discontinuities both at the boundary of the $E_Q$-discontinuous and mixed correlation phases, and at the boundary between the mixed and partial correlation phases, and $E_Q$ and $I(A:B)/2$ both have a discontinuity at the boundary between the partial and total correlation phases.  $E_P$ is constant in the $E_P$-discontinuous phase for this slice through the phase diagram, as it was for thermal states. The upper bound $\min(S(A), S(B))$ is not plotted, since it is cutoff dependent and for most reasonable values of the cutoff it is much larger than the quantities plotted.

\subsection{Additional phase diagrams}

In the previous subsection we explored in detail the phase diagram for bipartite optimized correlation measures in the case of two diametrically opposite regions with a black hole of horizon size $r_H \approx 0.61$. This showed us five nontrivial phases, the total, partial and mixed correlation phases and the $E_Q$- and $E_P$-discontinuous phases, as well as the trivial phase where the entanglement wedge of $AB$ is disconnected and all the correlation measures vanish, and displayed most of the interesting structure of the phase diagram. However, it is also interesting to consider phase diagrams as we vary the size $r_H$ of the horizon, and as we allow the regions to no longer be diametrically opposite by varying the distance between the centers, parameterized by the location $\theta_{B, {\rm cen}}$ of the center of region $B$.

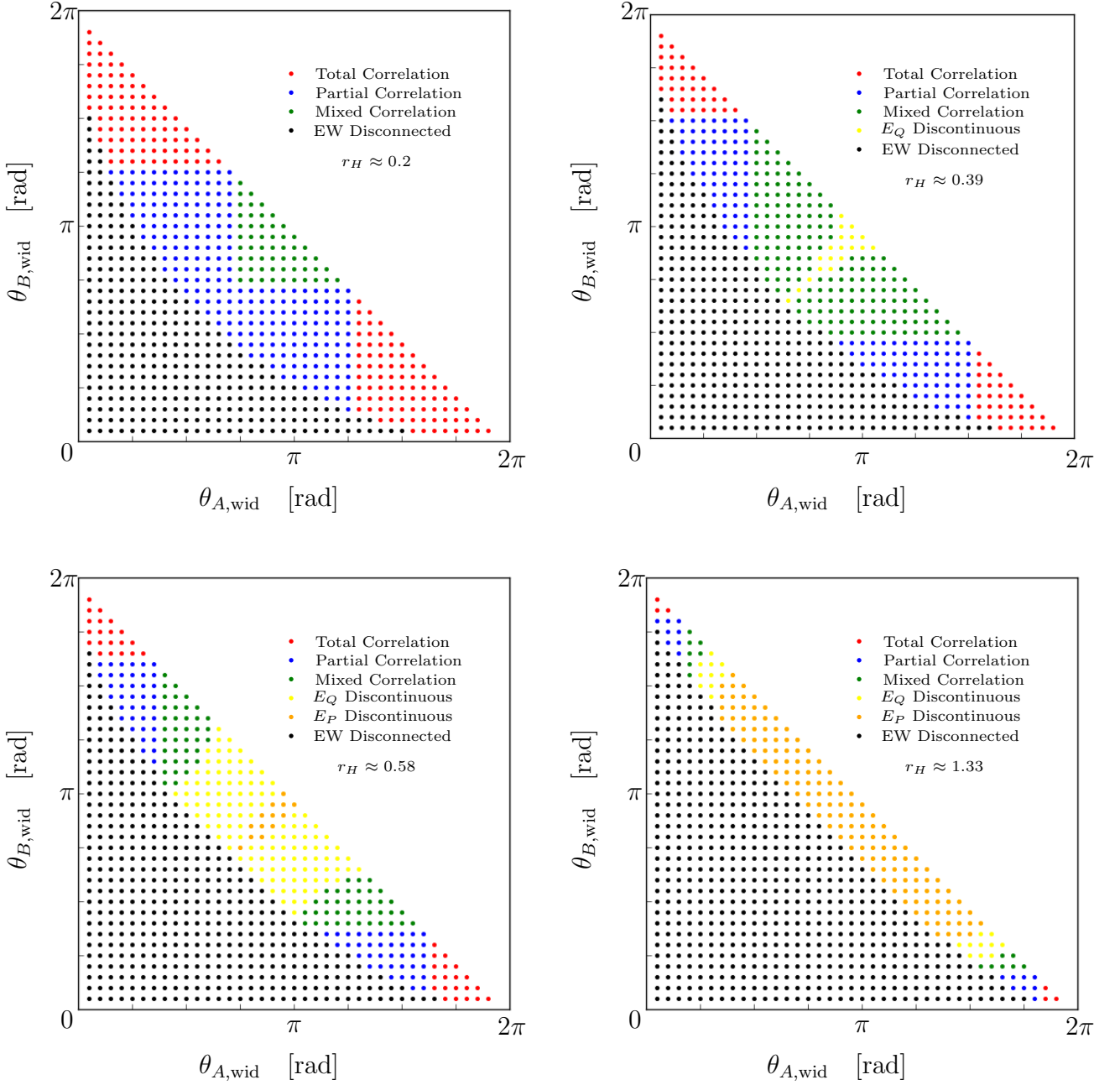
\begin{figure}[H]
\centering
\begin{tikzpicture}

\node[inner sep=0pt] at (3,0)
    {\includegraphics[trim = 93 27 68 36, clip, scale = 0.56]{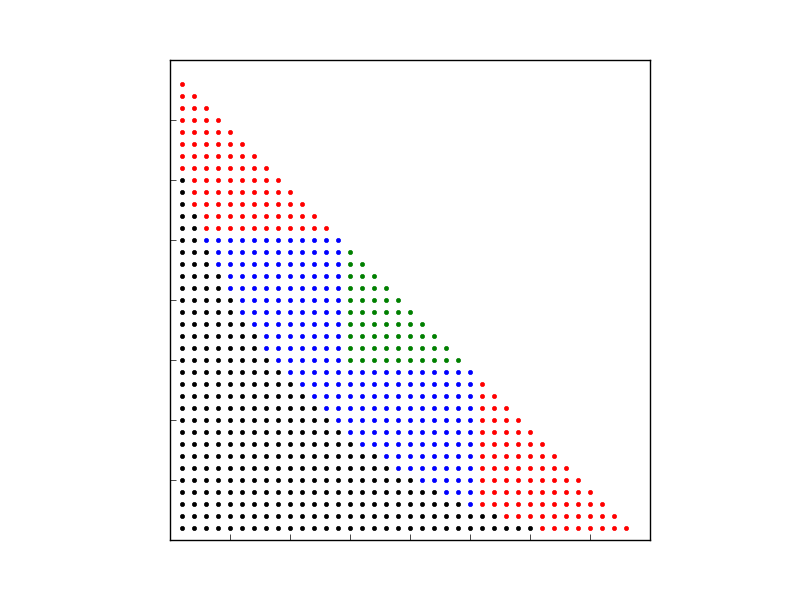}};
\node[inner sep=0pt] at (12,0)
    {\includegraphics[trim = 93 27 68 36, clip, scale = 0.55]{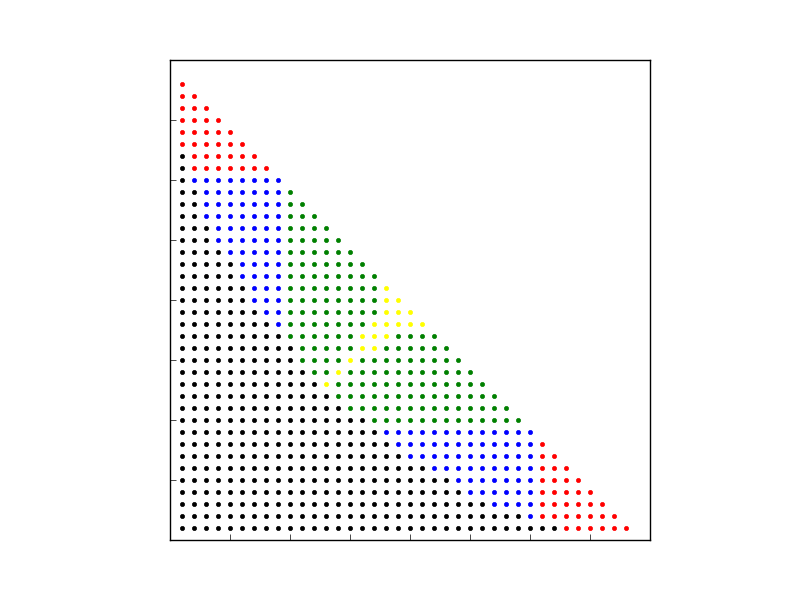}};
    
    \node[inner sep=0pt] at (3,-9)
    {\includegraphics[trim = 93 27 68 36, clip, scale = 0.56]{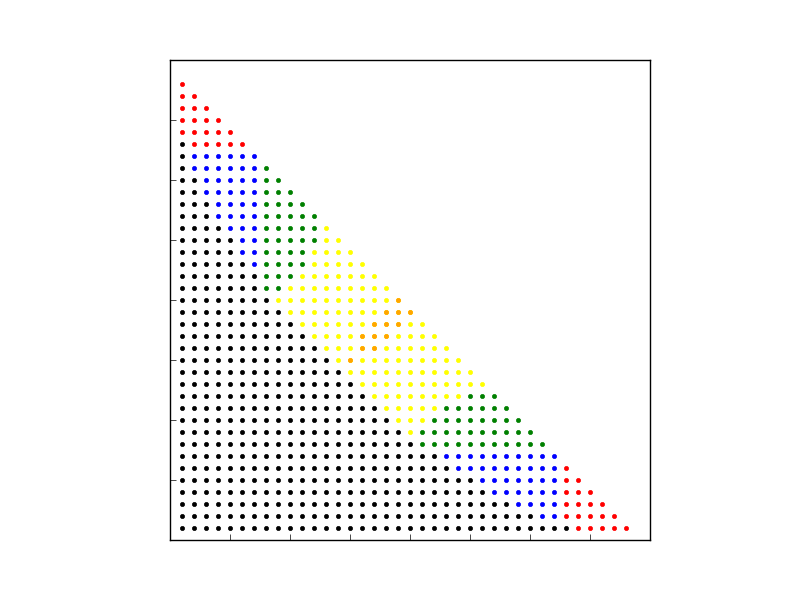}};
\node[inner sep=0pt] at (12,-9)
    {\includegraphics[trim = 93 27 68 36, clip, scale = 0.56]{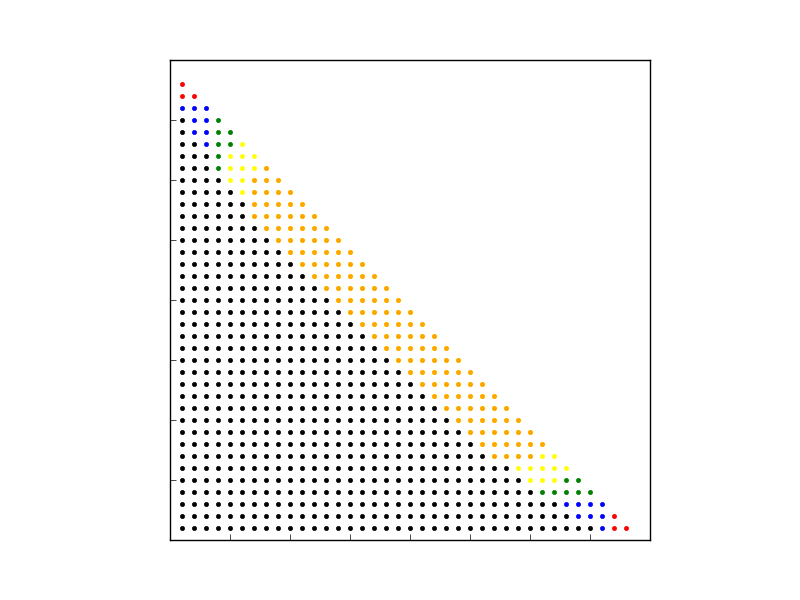}};

\definecolor{darkgreen}{rgb}{0,0.5,0}
\definecolor{yellow2}{rgb}{1,1,0}
\definecolor{orange2}{rgb}{1,0.643,0}
\fill[red] (2.85,2.5) circle (1.0pt);
\fill[blue] (2.85,2.2) circle (1.0pt);
\fill[darkgreen] (2.85,1.9) circle (1.0pt);
\fill[black] (2.85,1.6) circle (1.0pt);
\note{4.3,2.5}{{Total Correlation}}
\note{4.4,2.2}{{Partial Correlation}}
\note{4.35,1.9}{{Mixed Correlation}}
\note{4.3,1.6}{{EW Disconnected}}
\note{4.2,1.1}{{$r_H \approx 0.2$}}
\note{-0.7,-3.5}{\large{$0$}}
\note{2.9,-3.6}{\large{$\pi$}}
\note{6.35,-3.6}{\large{$2\pi$}}
\note{-0.7,0.08}{\large{$\pi$}}
\note{-0.75,3.5}{\large{$2\pi$}}

\fill[red] (2.85,-6.5) circle (1.0pt);
\fill[blue] (2.85,-6.8) circle (1.0pt);
\fill[darkgreen] (2.85,-7.1) circle (1.0pt);
\fill[yellow2] (2.85,-7.4) circle (1.0pt);
\fill[orange2] (2.85,-7.7) circle (1.0pt);
\fill[black] (2.85,-8) circle (1.0pt);
\note{4.3,-6.5}{{Total Correlation}}
\note{4.4,-6.8}{{Partial Correlation}}
\note{4.35,-7.1}{{Mixed Correlation}}
\note{4.3,-7.4}{{$E_Q$ Discontinuous}}
\note{4.3,-7.7}{{$E_P$ Discontinuous}}
\note{4.3,-8}{{EW Disconnected}}
\note{4.2,-8.5}{{$r_H \approx 0.58$}}
\note{-0.7,-12.5}{\large{$0$}}
\note{2.9,-12.6}{\large{$\pi$}}
\note{6.35,-12.6}{\large{$2\pi$}}
\note{-0.7,-8.92}{\large{$\pi$}}
\note{-0.75,-5.5}{\large{$2\pi$}}


\fill[red] (11.85,2.5) circle (1.0pt);
\fill[blue] (11.85,2.2) circle (1.0pt);
\fill[darkgreen] (11.85,1.9) circle (1.0pt);
\fill[yellow2] (11.85,1.6) circle (1.0pt);
\fill[black] (11.85,1.3) circle (1.0pt);
\note{13.3,2.5}{{Total Correlation}}
\note{13.4,2.2}{{Partial Correlation}}
\note{13.35,1.9}{{Mixed Correlation}}
\note{13.3,1.6}{{$E_Q$ Discontinuous}}
\note{13.3,1.3}{{EW Disconnected}}
\note{13.2,0.8}{{$r_H \approx 0.39$}}
\note{8.3,-3.5}{\large{$0$}}
\note{11.9,-3.6}{\large{$\pi$}}
\note{15.35,-3.6}{\large{$2\pi$}}
\note{8.3,0.08}{\large{$\pi$}}
\note{8.3,3.5}{\large{$2\pi$}}

\fill[red] (11.85,-6.5) circle (1.0pt);
\fill[blue] (11.85,-6.8) circle (1.0pt);
\fill[darkgreen] (11.85,-7.1) circle (1.0pt);
\fill[yellow2] (11.85,-7.4) circle (1.0pt);
\fill[orange2] (11.85,-7.7) circle (1.0pt);
\fill[black] (11.85,-8) circle (1.0pt);
\note{13.3,-6.5}{{Total Correlation}}
\note{13.4,-6.8}{{Partial Correlation}}
\note{13.35,-7.1}{{Mixed Correlation}}
\note{13.3,-7.4}{{$E_Q$ Discontinuous}}
\note{13.3,-7.7}{{$E_P$ Discontinuous}}
\note{13.3,-8}{{EW Disconnected}}
\note{13.2,-8.5}{{$r_H \approx 1.33$}}
\note{8.3,-12.5}{\large{$0$}}
\note{11.9,-12.6}{\large{$\pi$}}
\note{15.35,-12.6}{\large{$2\pi$}}
\note{8.3,-8.92}{\large{$\pi$}}
\note{8.25,-5.5}{\large{$2\pi$}}

\note{7.5,4.5}{\LARGE{Phases of reduced thermal states with varying $r_H$}}

\note{2.5,-4.25}{\large{$\theta_{A,\rm wid}$ ~ [rad]}}
\node[rotate=90] at (-1.45,0) {\large{$\theta_{B,\rm wid}$ ~ [rad]}};

\note{11.5,-4.25}{\large{$\theta_{A,\rm wid}$ ~ [rad]}}
\node[rotate=90] at (7.5,0) {\large{$\theta_{B,\rm wid}$ ~ [rad]}};

\definecolor{darkgreen}{rgb}{0,0.5,0}

\note{2.5,-13.25}{\large{$\theta_{A,\rm wid}$ ~ [rad]}}
\node[rotate=90] at (-1.45,-9) {\large{$\theta_{B,\rm wid}$ ~ [rad]}};

\note{11.5,-13.25}{\large{$\theta_{A,\rm wid}$ ~ [rad]}}
\node[rotate=90] at (7.5,-9) {\large{$\theta_{B,\rm wid}$ ~ [rad]}};
\end{tikzpicture}
\caption{Phase diagrams for four different values of $r_H$, with $\theta_{A,cen}=0$.  The $E_Q$-discontinuous phase exists for $r_H \gtrsim 0.37$, and the $E_P$-discontinuous phase exists for $r_H \gtrsim 0.58$.  As $r_H$ increases, more entanglement wedges become disconnected (black region grows), and the discontinuous phases expand at the expense of the total, partial and mixed correlation phases, as the larger black hole  makes it more favorable for the EWCS to break into two pieces which end at the horizon.} 
    \label{fig:varyrH}
\end{figure}

Figure \ref{fig:varyrH} shows phase diagrams for diametrically opposite regions $A$ and $B$ with four different values of $r_H$. For smaller $r_H$, the $E_Q$- and $E_P$-discontinuous regions disappear, and the size of the phase with disconnected entanglement wedge and vanishing correlation measures grows smaller as well, with much of the phase diagram taken up by the total, partial and mixed correlation phases. As $r_H$ grows larger, the region with disconnected entanglement wedge grows, and the discontinuous phases appear, eventually between them largely squeezing out the total, partial and mixed correlation phases. For large values of $r_H$, most of the diagram is either disconnected, or $E_P$-discontinuous.

The dominance of the disconnected region for large black hole can be understood that as the horizon radius increases, the Hawking temperature increases as $T_H = r_H/2\pi\ell^2$, so all mutual informations must decrease; an increasing system temperature is destroying spatial correlations in the state. Thus entanglement wedges which were connected at a smaller $r_H$ become disconnected at some larger radius when their mutual information reaches zero.

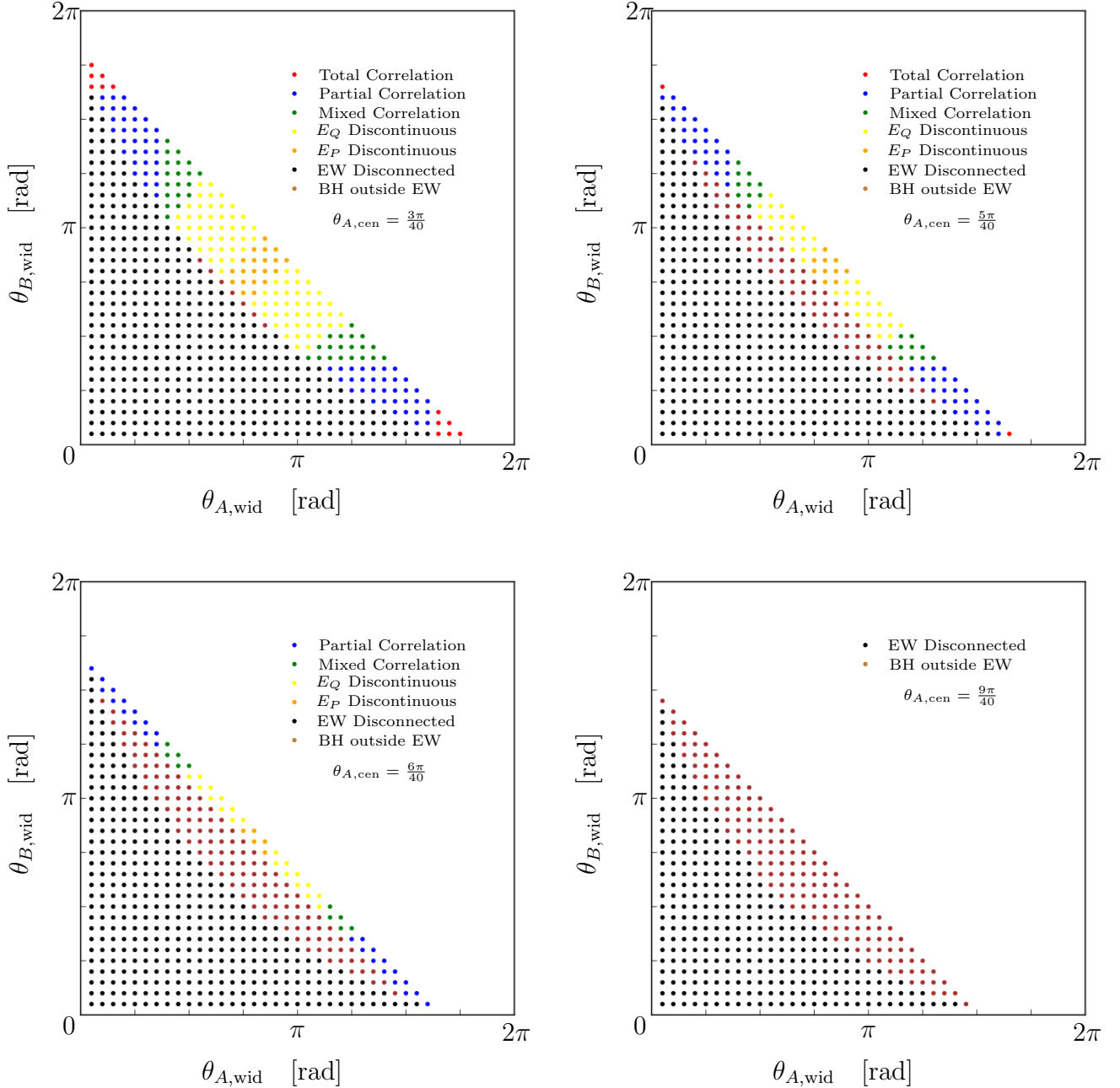
\begin{figure}[H]
\centering
\begin{tikzpicture}

\node[inner sep=0pt] at (3,0)
    {\includegraphics[trim = 93 27 68 36, clip, scale = 0.56]{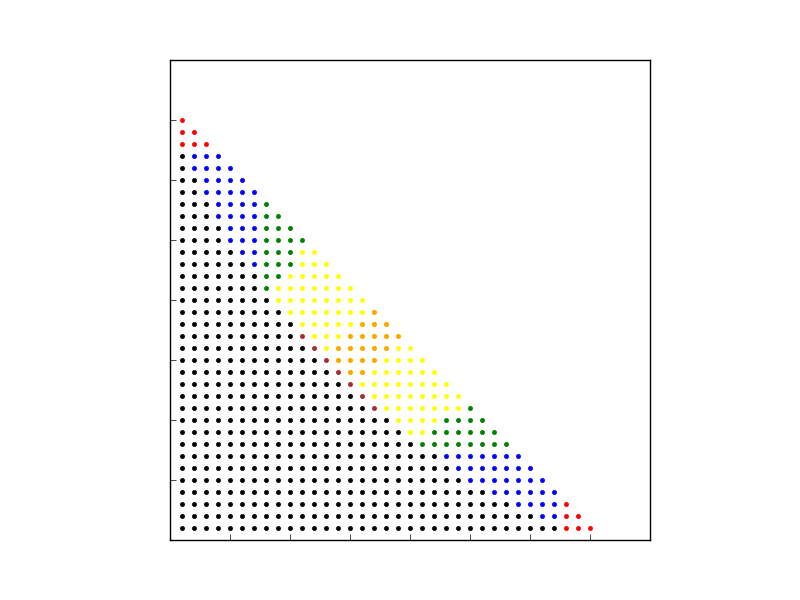}};
\node[inner sep=0pt] at (12,0)
    {\includegraphics[trim = 93 27 68 36, clip, scale = 0.56]{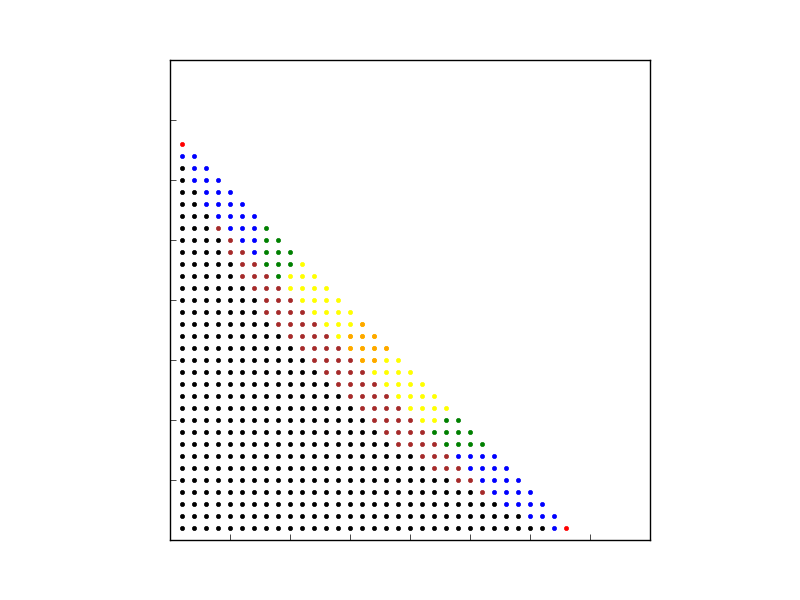}};
    
    \node[inner sep=0pt] at (3,-9)
    {\includegraphics[trim = 93 27 68 36, clip, scale = 0.56]{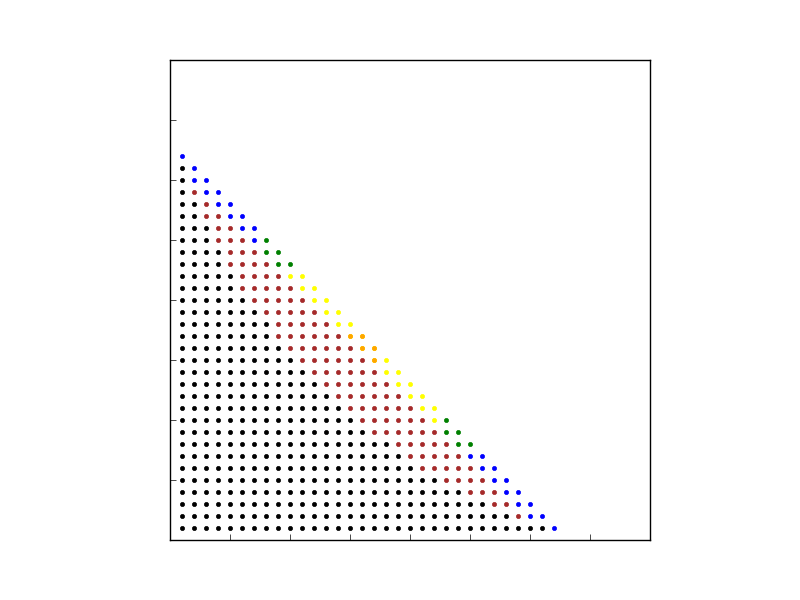}};
\node[inner sep=0pt] at (12,-9)
    {\includegraphics[trim = 93 27 68 36, clip, scale = 0.56]{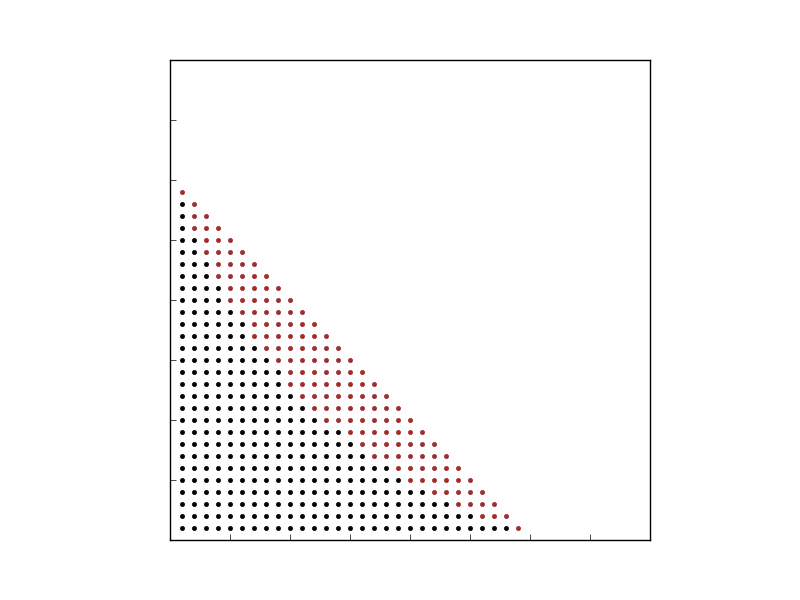}};

\definecolor{darkgreen}{rgb}{0,0.5,0}
\definecolor{yellow2}{rgb}{1,1,0}
\definecolor{orange2}{rgb}{1,0.643,0}
\fill[red] (2.85,2.5) circle (1.0pt);
\fill[blue] (2.85,2.2) circle (1.0pt);
\fill[darkgreen] (2.85,1.9) circle (1.0pt);
\fill[yellow2] (2.85,1.6) circle (1.0pt);
\fill[orange2] (2.85,1.3) circle (1.0pt);
\fill[black] (2.85,1.0) circle (1.0pt);
\fill[brown] (2.85,0.7) circle (1.0pt);

\note{4.3,2.5}{{Total Correlation}}
\note{4.4,2.2}{{Partial Correlation}}
\note{4.35,1.9}{{Mixed Correlation}}
\note{4.3,1.6}{{$E_Q$ Discontinuous}}
\note{4.3,1.3}{{$E_P$ Discontinuous}}
\note{4.3,1.0}{{EW Disconnected}}
\note{4.2,0.7}{{BH outside EW}}
\note{4.2,0.2}{{$\theta_{A,\rm cen} = \frac{3 \pi}{40}$}}
\note{-0.7,-3.5}{\large{$0$}}
\note{2.9,-3.6}{\large{$\pi$}}
\note{6.35,-3.6}{\large{$2\pi$}}
\note{-0.7,0.08}{\large{$\pi$}}
\note{-0.75,3.5}{\large{$2\pi$}}

\fill[blue] (2.85,-6.5) circle (1.0pt);
\fill[darkgreen] (2.85,-6.8) circle (1.0pt);
\fill[yellow2] (2.85,-7.1) circle (1.0pt);
\fill[orange2] (2.85,-7.4) circle (1.0pt);
\fill[black] (2.85,-7.7) circle (1.0pt);
\fill[brown] (2.85,-8) circle (1.0pt);
\note{4.4,-6.5}{{Partial Correlation}}
\note{4.35,-6.8}{{Mixed Correlation}}
\note{4.3,-7.1}{{$E_Q$ Discontinuous}}
\note{4.3,-7.4}{{$E_P$ Discontinuous}}
\note{4.3,-7.7}{{EW Disconnected}}
\note{4.2,-8}{{BH outside EW}}
\note{4.2,-8.5}{{$\theta_{A,\rm cen} = \frac{6 \pi}{40}$}}
\note{-0.7,-12.5}{\large{$0$}}
\note{2.9,-12.6}{\large{$\pi$}}
\note{6.35,-12.6}{\large{$2\pi$}}
\note{-0.7,-8.92}{\large{$\pi$}}
\note{-0.75,-5.5}{\large{$2\pi$}}


\fill[red] (11.85,2.5) circle (1.0pt);
\fill[blue] (11.85,2.2) circle (1.0pt);
\fill[darkgreen] (11.85,1.9) circle (1.0pt);
\fill[yellow2] (11.85,1.6) circle (1.0pt);
\fill[orange2] (11.85,1.3) circle (1.0pt);
\fill[black] (11.85,1) circle (1.0pt);
\fill[brown] (11.85,0.7) circle (1.0pt);
\note{13.3,2.5}{{Total Correlation}}
\note{13.4,2.2}{{Partial Correlation}}
\note{13.35,1.9}{{Mixed Correlation}}
\note{13.3,1.6}{{$E_Q$ Discontinuous}}
\note{13.3,1.3}{{$E_P$ Discontinuous}}
\note{13.3,1}{{EW Disconnected}}
\note{13.2,0.7}{{BH outside EW}}
\note{13.2,0.2}{{$\theta_{A,\rm cen} = \frac{5 \pi}{40}$}}
\note{8.3,-3.5}{\large{$0$}}
\note{11.9,-3.6}{\large{$\pi$}}
\note{15.35,-3.6}{\large{$2\pi$}}
\note{8.3,0.08}{\large{$\pi$}}
\note{8.3,3.5}{\large{$2\pi$}}

\fill[black] (11.85,-6.5) circle (1.0pt);
\fill[brown] (11.85,-6.8) circle (1.0pt);
\note{13.3,-6.5}{{EW Disconnected}}
\note{13.2,-6.8}{{BH outside EW}}
\note{13.2,-7.3}{{$\theta_{A,\rm cen} = \frac{9 \pi}{40}$}}
\note{8.3,-12.5}{\large{$0$}}
\note{11.9,-12.6}{\large{$\pi$}}
\note{15.35,-12.6}{\large{$2\pi$}}
\note{8.3,-8.92}{\large{$\pi$}}
\note{8.25,-5.5}{\large{$2\pi$}}

\note{2.5,-4.25}{\large{$\theta_{A,\rm wid}$ ~ [rad]}}
\node[rotate=90] at (-1.45,0) {\large{$\theta_{B,\rm wid}$ ~ [rad]}};

\note{11.5,-4.25}{\large{$\theta_{A,\rm wid}$ ~ [rad]}}
\node[rotate=90] at (7.5,0) {\large{$\theta_{B,\rm wid}$ ~ [rad]}};

\definecolor{darkgreen}{rgb}{0,0.5,0}

\note{2.5,-13.25}{\large{$\theta_{A,\rm wid}$ ~ [rad]}}
\node[rotate=90] at (-1.45,-9) {\large{$\theta_{B,\rm wid}$ ~ [rad]}};

\note{11.5,-13.25}{\large{$\theta_{A,\rm wid}$ ~ [rad]}}
\node[rotate=90] at (7.5,-9) {\large{$\theta_{B,\rm wid}$ ~ [rad]}};

\note{7.5,4.5}{\LARGE{Phases of reduced thermal states with varying $\theta_{A,\rm cen}$}}

\end{tikzpicture}
\caption{Phase diagrams with $r_H \approx 0.61$  and four different values of $\theta_{A,cen}$.  The new brown phase represents where the entanglement wedge is connected but does not include the black hole. As the regions get closer, eventually  all entanglement wedges are either disconnected or do not contain the black hole.}
    \label{fig:EQ Phase Ac}
\end{figure}
We can also consider fixing the horizon radius and varying $\theta_{B, {\rm cen}}$ so the centers of the regions grow closer together, as shown in Fig.~\ref{fig:EQ Phase Ac}. The closer the centers of the regions get, the less total area they can take up, so the edge of the phase diagram recedes towards the lower left. Here a novel phase appears as the centers become closer: the entanglement wedge can be connected but not include the horizon, as in Fig.~\ref{BTZ AB}b. This phase is shown in brown. The BSCs of the correlation measures in this phase have the same topology as the cases with no horizon. As the centers get closer this phase expands, until after a certain point the  phases with the horizon inside the entanglement wedge are no longer present, as all entanglement wedges are either disconnected, or don't contain the black hole.

\section{Discussion and future work}\label{Discussion}

The primary goal of the research program of \cite{LS20, LSD20, LDS20} (inspired by works such as \cite{TU18,Swing18,umemoto2018entanglement}) is to find new ways to probe the entanglement structure of holographic states using their dual geometries. While simple linear entropic formulas can capture some aspects of the correlations in a quantum state, optimized correlation measures provide access to more detailed information about these correlations.  Therefore, the ability to evaluate them in a variety of different holographic states is key to better understanding how geometry encodes entanglement.  The method presented in this paper for evaluating OCMs in spacetimes with horizons pushes that agenda one step further, and displays the richness of the phase structure that emerges. Obtaining the results involves incorporating the surface-state correspondence prescription for assigning entropies to bulk surfaces, and the reasonable geometric picture that emerges suggests that this prescription has more to tell us about the holographic picture of entanglement.

Quite aside from our goal of understanding the interplay of geometry and entanglement, holographic states provide an avenue for better understanding quantum information itself.  They offer a highly nontrivial class of states for which we can nevertheless often evaluate optimized correlation measures.  For instance, the reduced thermal states we consider are arguably the most complex set of states for which we can evaluate all extremal bipartite OCMs.  The additional geometrical structure present allows the explicit evaluation of an a priori unbounded optimization. For example, using the relationship between $E_Q$ and $I^{ss}$ this provides a rare class of non-degradable states for which we can nevertheless evaluate the symmetric side-channel assisted distillable entanglement \cite{SSW06}. We are also lead to new conjectures, such as the possibility that $E_R$ is equal to the regularized entanglement of purification \cite{LDS20}.

There is a physical interpretation of the three $E_Q$- and $E_P$-continuous phases (total, partial, and mixed correlation), using the values of $I(A:E)$ and $I(B:E)$ in these phases.  The boundary state is a thermal state, which consists of a mixture of all energy eigenstates.  For a subsystem $A$ to have zero mutual information with the purifying environment means that an observer will be unable to distinguish the energy eigenstates by measuring any observable on $A$. The condition that $I(A:E)=0$ may therefore be seen as a criterion for when the system $A$ is small enough to satisfy the eigenstate thermalization hypothesis \cite{PhysRevA.43.2046,PhysRevE.50.888,Rigol2008}.

There are several questions that remain to be answered, which we leave to future work.  First, we would like to find some quantity or set of quantities which can be evaluated on the bipartite state $\rho_{AB}$, whose values can always determine with certainty whether $\rho_{AB}$ is in the $E_P$- or $E_Q$-discontinuous phase, or one of the continuous phases.  Here, we have identified a quantity which is discontinuous (as a function of $A$ and $B$) at the boundaries of the discontinuous phases, but its value alone cannot determine which phase $\rho_{AB}$ is in.  Second, in this work we have only considered correlations between boundary regions that both live on the same side of the horizon.  Since the BTZ geometry is dual to a thermal CFT state, it can be extended beyond the horizon to form a wormhole geometry dual to the thermofield double state.  One could then ask whether correlations between regions on opposite sides of the thermofield double have interesting properties which may be captured by OCMs.  The same method used here could then be used to evaluate OCMs on such states.  Finally, since the method presented here gives not only the value of an OCM, but also the pure state that achieves it, it would be interesting to know whether the optimizing state can tell us anything about the protocol which achieves the value of the OCM, for OCMs with operational interpretations.

\section*{Acknowledgements}

OD is supported by the Department of Energy under grant DE-SC0010005. SA, OD, JL, and GS are supported by the Department of Energy under grant DE-SC0020386.

\bibliography{main}

\begin{thebibliography}{61}%
\makeatletter
\providecommand \@ifxundefined [1]{%
 \@ifx{#1\undefined}
}%
\providecommand \@ifnum [1]{%
 \ifnum #1\expandafter \@firstoftwo
 \else \expandafter \@secondoftwo
 \fi
}%
\providecommand \@ifx [1]{%
 \ifx #1\expandafter \@firstoftwo
 \else \expandafter \@secondoftwo
 \fi
}%
\providecommand \natexlab [1]{#1}%
\providecommand \enquote  [1]{``#1''}%
\providecommand \bibnamefont  [1]{#1}%
\providecommand \bibfnamefont [1]{#1}%
\providecommand \citenamefont [1]{#1}%
\providecommand \href@noop [0]{\@secondoftwo}%
\providecommand \href [0]{\begingroup \@sanitize@url \@href}%
\providecommand \@href[1]{\@@startlink{#1}\@@href}%
\providecommand \@@href[1]{\endgroup#1\@@endlink}%
\providecommand \@sanitize@url [0]{\catcode `\\12\catcode `\$12\catcode
  `\&12\catcode `\#12\catcode `\^12\catcode `\_12\catcode `\%12\relax}%
\providecommand \@@startlink[1]{}%
\providecommand \@@endlink[0]{}%
\providecommand \url  [0]{\begingroup\@sanitize@url \@url }%
\providecommand \@url [1]{\endgroup\@href {#1}{\urlprefix }}%
\providecommand \urlprefix  [0]{URL }%
\providecommand \Eprint [0]{\href }%
\providecommand \doibase [0]{http://dx.doi.org/}%
\providecommand \selectlanguage [0]{\@gobble}%
\providecommand \bibinfo  [0]{\@secondoftwo}%
\providecommand \bibfield  [0]{\@secondoftwo}%
\providecommand \translation [1]{[#1]}%
\providecommand \BibitemOpen [0]{}%
\providecommand \bibitemStop [0]{}%
\providecommand \bibitemNoStop [0]{.\EOS\space}%
\providecommand \EOS [0]{\spacefactor3000\relax}%
\providecommand \BibitemShut  [1]{\csname bibitem#1\endcsname}%
\let\auto@bib@innerbib\@empty
\bibitem [{\citenamefont {{Terhal}}\ \emph {et~al.}(2002)\citenamefont
  {{Terhal}}, \citenamefont {{Horodecki}}, \citenamefont {{Leung}},\ and\
  \citenamefont {{DiVincenzo}}}]{EP02}%
  \BibitemOpen
  \bibfield  {author} {\bibinfo {author} {\bibfnamefont {B.~M.}\ \bibnamefont
  {{Terhal}}}, \bibinfo {author} {\bibfnamefont {M.}~\bibnamefont
  {{Horodecki}}}, \bibinfo {author} {\bibfnamefont {D.~W.}\ \bibnamefont
  {{Leung}}}, \ and\ \bibinfo {author} {\bibfnamefont {D.~P.}\ \bibnamefont
  {{DiVincenzo}}},\ }\href {\doibase 10.1063/1.1498001} {\bibfield  {journal}
  {\bibinfo  {journal} {Journal of Mathematical Physics}\ }\textbf {\bibinfo
  {volume} {43}},\ \bibinfo {pages} {4286} (\bibinfo {year} {2002})},\ \Eprint
  {http://arxiv.org/abs/quant-ph/0202044} {quant-ph/0202044} \BibitemShut
  {NoStop}%
\bibitem [{\citenamefont {Tucci}(2002)}]{Tucci}%
  \BibitemOpen
  \bibfield  {author} {\bibinfo {author} {\bibfnamefont {R.~R.}\ \bibnamefont
  {Tucci}},\ }\href@noop {} {\bibfield  {journal} {\bibinfo  {journal}
  {ar{X}iv:quant-ph/0202144}\ } (\bibinfo {year} {2002})}\BibitemShut {NoStop}%
\bibitem [{\citenamefont {Christandl}\ and\ \citenamefont
  {Winter}(2004)}]{Christandl}%
  \BibitemOpen
  \bibfield  {author} {\bibinfo {author} {\bibfnamefont {M.}~\bibnamefont
  {Christandl}}\ and\ \bibinfo {author} {\bibfnamefont {A.}~\bibnamefont
  {Winter}},\ }\href@noop {} {\bibfield  {journal} {\bibinfo  {journal} {J.
  Math. Phys.}\ }\textbf {\bibinfo {volume} {45}},\ \bibinfo {pages} {829}
  (\bibinfo {year} {2004})}\BibitemShut {NoStop}%
\bibitem [{\citenamefont {Levin}\ and\ \citenamefont {Smith}(2020)}]{LS20}%
  \BibitemOpen
  \bibfield  {author} {\bibinfo {author} {\bibfnamefont {J.}~\bibnamefont
  {Levin}}\ and\ \bibinfo {author} {\bibfnamefont {G.}~\bibnamefont {Smith}},\
  }\href@noop {} {\bibfield  {journal} {\bibinfo  {journal} {IEEE Transactions
  on Information Theory}\ }\textbf {\bibinfo {volume} {66}},\ \bibinfo {pages}
  {3520} (\bibinfo {year} {2020})}\BibitemShut {NoStop}%
\bibitem [{\citenamefont {Maldacena}(1999)}]{maldacena1999}%
  \BibitemOpen
  \bibfield  {author} {\bibinfo {author} {\bibfnamefont {J.}~\bibnamefont
  {Maldacena}},\ }\href@noop {} {\bibfield  {journal} {\bibinfo  {journal}
  {International journal of theoretical physics}\ }\textbf {\bibinfo {volume}
  {38}},\ \bibinfo {pages} {1113} (\bibinfo {year} {1999})}\BibitemShut
  {NoStop}%
\bibitem [{\citenamefont {Umemoto}\ and\ \citenamefont
  {Takayanagi}(2018)}]{TU18}%
  \BibitemOpen
  \bibfield  {author} {\bibinfo {author} {\bibfnamefont {K.}~\bibnamefont
  {Umemoto}}\ and\ \bibinfo {author} {\bibfnamefont {T.}~\bibnamefont
  {Takayanagi}},\ }\href@noop {} {\bibfield  {journal} {\bibinfo  {journal}
  {Nature Physics}\ }\textbf {\bibinfo {volume} {14}},\ \bibinfo {pages} {573}
  (\bibinfo {year} {2018})}\BibitemShut {NoStop}%
\bibitem [{\citenamefont {Nguyen}\ \emph {et~al.}(2018)\citenamefont {Nguyen},
  \citenamefont {Devakul}, \citenamefont {Halbasch}, \citenamefont {Zaletel},\
  and\ \citenamefont {Swingle}}]{Swing18}%
  \BibitemOpen
  \bibfield  {author} {\bibinfo {author} {\bibfnamefont {P.}~\bibnamefont
  {Nguyen}}, \bibinfo {author} {\bibfnamefont {T.}~\bibnamefont {Devakul}},
  \bibinfo {author} {\bibfnamefont {M.~G.}\ \bibnamefont {Halbasch}}, \bibinfo
  {author} {\bibfnamefont {M.~P.}\ \bibnamefont {Zaletel}}, \ and\ \bibinfo
  {author} {\bibfnamefont {B.}~\bibnamefont {Swingle}},\ }\href@noop {}
  {\bibfield  {journal} {\bibinfo  {journal} {Journal of High Energy Physics}\
  }\textbf {\bibinfo {volume} {2018}},\ \bibinfo {pages} {98} (\bibinfo {year}
  {2018})}\BibitemShut {NoStop}%
\bibitem [{\citenamefont {Dutta}\ and\ \citenamefont
  {Faulkner}(2019)}]{dutta2019canonical}%
  \BibitemOpen
  \bibfield  {author} {\bibinfo {author} {\bibfnamefont {S.}~\bibnamefont
  {Dutta}}\ and\ \bibinfo {author} {\bibfnamefont {T.}~\bibnamefont
  {Faulkner}},\ }\href@noop {} {\bibfield  {journal} {\bibinfo  {journal}
  {arXiv preprint arXiv:1905.00577}\ } (\bibinfo {year} {2019})}\BibitemShut
  {NoStop}%
\bibitem [{\citenamefont {Engelhardt}\ and\ \citenamefont
  {Wall}(2019)}]{engelhardt2019coarse}%
  \BibitemOpen
  \bibfield  {author} {\bibinfo {author} {\bibfnamefont {N.}~\bibnamefont
  {Engelhardt}}\ and\ \bibinfo {author} {\bibfnamefont {A.~C.}\ \bibnamefont
  {Wall}},\ }\href@noop {} {\bibfield  {journal} {\bibinfo  {journal} {Journal
  of High Energy Physics}\ }\textbf {\bibinfo {volume} {2019}},\ \bibinfo
  {pages} {160} (\bibinfo {year} {2019})}\BibitemShut {NoStop}%
\bibitem [{\citenamefont {Engelhardt}\ and\ \citenamefont
  {Wall}(2018)}]{engelhardt2018decoding}%
  \BibitemOpen
  \bibfield  {author} {\bibinfo {author} {\bibfnamefont {N.}~\bibnamefont
  {Engelhardt}}\ and\ \bibinfo {author} {\bibfnamefont {A.~C.}\ \bibnamefont
  {Wall}},\ }\href@noop {} {\bibfield  {journal} {\bibinfo  {journal} {Physical
  review letters}\ }\textbf {\bibinfo {volume} {121}},\ \bibinfo {pages}
  {211301} (\bibinfo {year} {2018})}\BibitemShut {NoStop}%
\bibitem [{\citenamefont {Kudler-Flam}\ and\ \citenamefont
  {Ryu}(2019)}]{kudler2019entanglement}%
  \BibitemOpen
  \bibfield  {author} {\bibinfo {author} {\bibfnamefont {J.}~\bibnamefont
  {Kudler-Flam}}\ and\ \bibinfo {author} {\bibfnamefont {S.}~\bibnamefont
  {Ryu}},\ }\href@noop {} {\bibfield  {journal} {\bibinfo  {journal} {Physical
  Review D}\ }\textbf {\bibinfo {volume} {99}},\ \bibinfo {pages} {106014}
  (\bibinfo {year} {2019})}\BibitemShut {NoStop}%
\bibitem [{\citenamefont {Vidal}\ and\ \citenamefont
  {Werner}(2002)}]{vidal2002computable}%
  \BibitemOpen
  \bibfield  {author} {\bibinfo {author} {\bibfnamefont {G.}~\bibnamefont
  {Vidal}}\ and\ \bibinfo {author} {\bibfnamefont {R.~F.}\ \bibnamefont
  {Werner}},\ }\href@noop {} {\bibfield  {journal} {\bibinfo  {journal}
  {Physical Review A}\ }\textbf {\bibinfo {volume} {65}},\ \bibinfo {pages}
  {032314} (\bibinfo {year} {2002})}\BibitemShut {NoStop}%
\bibitem [{\citenamefont {Tamaoka}(2019)}]{tamaoka2019entanglement}%
  \BibitemOpen
  \bibfield  {author} {\bibinfo {author} {\bibfnamefont {K.}~\bibnamefont
  {Tamaoka}},\ }\href@noop {} {\bibfield  {journal} {\bibinfo  {journal}
  {Physical review letters}\ }\textbf {\bibinfo {volume} {122}},\ \bibinfo
  {pages} {141601} (\bibinfo {year} {2019})}\BibitemShut {NoStop}%
\bibitem [{\citenamefont {Bao}\ and\ \citenamefont
  {Halpern}(2018)}]{bao2018holographic}%
  \BibitemOpen
  \bibfield  {author} {\bibinfo {author} {\bibfnamefont {N.}~\bibnamefont
  {Bao}}\ and\ \bibinfo {author} {\bibfnamefont {I.~F.}\ \bibnamefont
  {Halpern}},\ }\href@noop {} {\bibfield  {journal} {\bibinfo  {journal}
  {Journal of High Energy Physics}\ }\textbf {\bibinfo {volume} {2018}},\
  \bibinfo {pages} {6} (\bibinfo {year} {2018})}\BibitemShut {NoStop}%
\bibitem [{\citenamefont {Hirai}\ \emph {et~al.}(2018)\citenamefont {Hirai},
  \citenamefont {Tamaoka},\ and\ \citenamefont {Yokoya}}]{hirai2018towards}%
  \BibitemOpen
  \bibfield  {author} {\bibinfo {author} {\bibfnamefont {H.}~\bibnamefont
  {Hirai}}, \bibinfo {author} {\bibfnamefont {K.}~\bibnamefont {Tamaoka}}, \
  and\ \bibinfo {author} {\bibfnamefont {T.}~\bibnamefont {Yokoya}},\
  }\href@noop {} {\bibfield  {journal} {\bibinfo  {journal} {Exp. Phys}\ }
  (\bibinfo {year} {2018})}\BibitemShut {NoStop}%
\bibitem [{\citenamefont {Esp{\'\i}ndola}\ \emph {et~al.}(2018)\citenamefont
  {Esp{\'\i}ndola}, \citenamefont {G{\"u}ijosa},\ and\ \citenamefont
  {Pedraza}}]{espindola2018entanglement}%
  \BibitemOpen
  \bibfield  {author} {\bibinfo {author} {\bibfnamefont {R.}~\bibnamefont
  {Esp{\'\i}ndola}}, \bibinfo {author} {\bibfnamefont {A.}~\bibnamefont
  {G{\"u}ijosa}}, \ and\ \bibinfo {author} {\bibfnamefont {J.~F.}\ \bibnamefont
  {Pedraza}},\ }\href@noop {} {\bibfield  {journal} {\bibinfo  {journal} {The
  European Physical Journal C}\ }\textbf {\bibinfo {volume} {78}},\ \bibinfo
  {pages} {646} (\bibinfo {year} {2018})}\BibitemShut {NoStop}%
\bibitem [{\citenamefont {Bao}\ and\ \citenamefont
  {Halpern}(2019)}]{bao2019conditional}%
  \BibitemOpen
  \bibfield  {author} {\bibinfo {author} {\bibfnamefont {N.}~\bibnamefont
  {Bao}}\ and\ \bibinfo {author} {\bibfnamefont {I.~F.}\ \bibnamefont
  {Halpern}},\ }\href@noop {} {\bibfield  {journal} {\bibinfo  {journal}
  {Physical Review D}\ }\textbf {\bibinfo {volume} {99}},\ \bibinfo {pages}
  {046010} (\bibinfo {year} {2019})}\BibitemShut {NoStop}%
\bibitem [{\citenamefont {Bao}\ \emph {et~al.}(2019)\citenamefont {Bao},
  \citenamefont {Chatwin-Davies},\ and\ \citenamefont
  {Remmen}}]{bao2019entanglement}%
  \BibitemOpen
  \bibfield  {author} {\bibinfo {author} {\bibfnamefont {N.}~\bibnamefont
  {Bao}}, \bibinfo {author} {\bibfnamefont {A.}~\bibnamefont {Chatwin-Davies}},
  \ and\ \bibinfo {author} {\bibfnamefont {G.~N.}\ \bibnamefont {Remmen}},\
  }\href@noop {} {\bibfield  {journal} {\bibinfo  {journal} {Journal of High
  Energy Physics}\ }\textbf {\bibinfo {volume} {2019}},\ \bibinfo {pages} {110}
  (\bibinfo {year} {2019})}\BibitemShut {NoStop}%
\bibitem [{\citenamefont {Cheng}(2020)}]{cheng2020optimized}%
  \BibitemOpen
  \bibfield  {author} {\bibinfo {author} {\bibfnamefont {N.}~\bibnamefont
  {Cheng}},\ }\href@noop {} {\bibfield  {journal} {\bibinfo  {journal}
  {Physical Review D}\ }\textbf {\bibinfo {volume} {101}},\ \bibinfo {pages}
  {066009} (\bibinfo {year} {2020})}\BibitemShut {NoStop}%
\bibitem [{\citenamefont {Ghodrati}\ \emph {et~al.}(2019)\citenamefont
  {Ghodrati}, \citenamefont {Kuang}, \citenamefont {Wang}, \citenamefont
  {Zhang},\ and\ \citenamefont {Zhou}}]{ghodrati2019connection}%
  \BibitemOpen
  \bibfield  {author} {\bibinfo {author} {\bibfnamefont {M.}~\bibnamefont
  {Ghodrati}}, \bibinfo {author} {\bibfnamefont {X.-M.}\ \bibnamefont {Kuang}},
  \bibinfo {author} {\bibfnamefont {B.}~\bibnamefont {Wang}}, \bibinfo {author}
  {\bibfnamefont {C.-Y.}\ \bibnamefont {Zhang}}, \ and\ \bibinfo {author}
  {\bibfnamefont {Y.-T.}\ \bibnamefont {Zhou}},\ }\href@noop {} {\bibfield
  {journal} {\bibinfo  {journal} {Journal of High Energy Physics}\ }\textbf
  {\bibinfo {volume} {2019}},\ \bibinfo {pages} {9} (\bibinfo {year}
  {2019})}\BibitemShut {NoStop}%
\bibitem [{\citenamefont {Chu}\ \emph {et~al.}(2020)\citenamefont {Chu},
  \citenamefont {Qi},\ and\ \citenamefont {Zhou}}]{chu2020generalizations}%
  \BibitemOpen
  \bibfield  {author} {\bibinfo {author} {\bibfnamefont {J.}~\bibnamefont
  {Chu}}, \bibinfo {author} {\bibfnamefont {R.}~\bibnamefont {Qi}}, \ and\
  \bibinfo {author} {\bibfnamefont {Y.}~\bibnamefont {Zhou}},\ }\href@noop {}
  {\bibfield  {journal} {\bibinfo  {journal} {Journal of High Energy Physics}\
  }\textbf {\bibinfo {volume} {2020}},\ \bibinfo {pages} {1} (\bibinfo {year}
  {2020})}\BibitemShut {NoStop}%
\bibitem [{\citenamefont {Velni}\ \emph {et~al.}(2019)\citenamefont {Velni},
  \citenamefont {Mozaffar},\ and\ \citenamefont {Vahidinia}}]{velni2019some}%
  \BibitemOpen
  \bibfield  {author} {\bibinfo {author} {\bibfnamefont {K.~B.}\ \bibnamefont
  {Velni}}, \bibinfo {author} {\bibfnamefont {M.~R.~M.}\ \bibnamefont
  {Mozaffar}}, \ and\ \bibinfo {author} {\bibfnamefont {M.}~\bibnamefont
  {Vahidinia}},\ }\href@noop {} {\bibfield  {journal} {\bibinfo  {journal}
  {Journal of High Energy Physics}\ }\textbf {\bibinfo {volume} {2019}},\
  \bibinfo {pages} {200} (\bibinfo {year} {2019})}\BibitemShut {NoStop}%
\bibitem [{\citenamefont {Velni}\ \emph {et~al.}(2020)\citenamefont {Velni},
  \citenamefont {Mozaffar},\ and\ \citenamefont
  {Vahidinia}}]{velni2020evolution}%
  \BibitemOpen
  \bibfield  {author} {\bibinfo {author} {\bibfnamefont {K.~B.}\ \bibnamefont
  {Velni}}, \bibinfo {author} {\bibfnamefont {M.~R.~M.}\ \bibnamefont
  {Mozaffar}}, \ and\ \bibinfo {author} {\bibfnamefont {M.}~\bibnamefont
  {Vahidinia}},\ }\href@noop {} {\bibfield  {journal} {\bibinfo  {journal}
  {Journal of High Energy Physics}\ }\textbf {\bibinfo {volume} {2020}},\
  \bibinfo {pages} {1} (\bibinfo {year} {2020})}\BibitemShut {NoStop}%
\bibitem [{\citenamefont {Ag{\'o}n}\ \emph {et~al.}(2019)\citenamefont
  {Ag{\'o}n}, \citenamefont {De~Boer},\ and\ \citenamefont
  {Pedraza}}]{agon2019geometric}%
  \BibitemOpen
  \bibfield  {author} {\bibinfo {author} {\bibfnamefont {C.~A.}\ \bibnamefont
  {Ag{\'o}n}}, \bibinfo {author} {\bibfnamefont {J.}~\bibnamefont {De~Boer}}, \
  and\ \bibinfo {author} {\bibfnamefont {J.~F.}\ \bibnamefont {Pedraza}},\
  }\href@noop {} {\bibfield  {journal} {\bibinfo  {journal} {Journal of High
  Energy Physics}\ }\textbf {\bibinfo {volume} {2019}},\ \bibinfo {pages} {75}
  (\bibinfo {year} {2019})}\BibitemShut {NoStop}%
\bibitem [{\citenamefont {Caputa}\ \emph {et~al.}(2019)\citenamefont {Caputa},
  \citenamefont {Miyaji}, \citenamefont {Takayanagi},\ and\ \citenamefont
  {Umemoto}}]{caputa2019holographic}%
  \BibitemOpen
  \bibfield  {author} {\bibinfo {author} {\bibfnamefont {P.}~\bibnamefont
  {Caputa}}, \bibinfo {author} {\bibfnamefont {M.}~\bibnamefont {Miyaji}},
  \bibinfo {author} {\bibfnamefont {T.}~\bibnamefont {Takayanagi}}, \ and\
  \bibinfo {author} {\bibfnamefont {K.}~\bibnamefont {Umemoto}},\ }\href@noop
  {} {\bibfield  {journal} {\bibinfo  {journal} {Physical Review Letters}\
  }\textbf {\bibinfo {volume} {122}},\ \bibinfo {pages} {111601} (\bibinfo
  {year} {2019})}\BibitemShut {NoStop}%
\bibitem [{\citenamefont {Kudler-Flam}\ \emph {et~al.}(2019)\citenamefont
  {Kudler-Flam}, \citenamefont {MacCormack},\ and\ \citenamefont
  {Ryu}}]{kudler2019holographic}%
  \BibitemOpen
  \bibfield  {author} {\bibinfo {author} {\bibfnamefont {J.}~\bibnamefont
  {Kudler-Flam}}, \bibinfo {author} {\bibfnamefont {I.}~\bibnamefont
  {MacCormack}}, \ and\ \bibinfo {author} {\bibfnamefont {S.}~\bibnamefont
  {Ryu}},\ }\href@noop {} {\bibfield  {journal} {\bibinfo  {journal} {Journal
  of Physics A: Mathematical and Theoretical}\ }\textbf {\bibinfo {volume}
  {52}},\ \bibinfo {pages} {325401} (\bibinfo {year} {2019})}\BibitemShut
  {NoStop}%
\bibitem [{\citenamefont {Jokela}\ and\ \citenamefont
  {P{\"o}nni}(2019)}]{jokela2019notes}%
  \BibitemOpen
  \bibfield  {author} {\bibinfo {author} {\bibfnamefont {N.}~\bibnamefont
  {Jokela}}\ and\ \bibinfo {author} {\bibfnamefont {A.}~\bibnamefont
  {P{\"o}nni}},\ }\href@noop {} {\bibfield  {journal} {\bibinfo  {journal}
  {Journal of High Energy Physics}\ }\textbf {\bibinfo {volume} {2019}},\
  \bibinfo {pages} {87} (\bibinfo {year} {2019})}\BibitemShut {NoStop}%
\bibitem [{\citenamefont {Kudler-Flam}\ \emph {et~al.}(2020)\citenamefont
  {Kudler-Flam}, \citenamefont {Nozaki}, \citenamefont {Ryu},\ and\
  \citenamefont {Tan}}]{kudler2020quantum}%
  \BibitemOpen
  \bibfield  {author} {\bibinfo {author} {\bibfnamefont {J.}~\bibnamefont
  {Kudler-Flam}}, \bibinfo {author} {\bibfnamefont {M.}~\bibnamefont {Nozaki}},
  \bibinfo {author} {\bibfnamefont {S.}~\bibnamefont {Ryu}}, \ and\ \bibinfo
  {author} {\bibfnamefont {M.~T.}\ \bibnamefont {Tan}},\ }\href@noop {}
  {\bibfield  {journal} {\bibinfo  {journal} {Journal of High Energy Physics}\
  }\textbf {\bibinfo {volume} {2020}},\ \bibinfo {pages} {31} (\bibinfo {year}
  {2020})}\BibitemShut {NoStop}%
\bibitem [{\citenamefont {Kusuki}\ and\ \citenamefont
  {Tamaoka}(2020{\natexlab{a}})}]{kusuki2019dynamics}%
  \BibitemOpen
  \bibfield  {author} {\bibinfo {author} {\bibfnamefont {Y.}~\bibnamefont
  {Kusuki}}\ and\ \bibinfo {author} {\bibfnamefont {K.}~\bibnamefont
  {Tamaoka}},\ }\href@noop {} {\bibfield  {journal} {\bibinfo  {journal}
  {Journal of High Energy Physics}\ }\textbf {\bibinfo {volume} {2020}},\
  \bibinfo {pages} {1} (\bibinfo {year} {2020}{\natexlab{a}})}\BibitemShut
  {NoStop}%
\bibitem [{\citenamefont {Kusuki}\ \emph {et~al.}(2019)\citenamefont {Kusuki},
  \citenamefont {Kudler-Flam},\ and\ \citenamefont
  {Ryu}}]{kusuki2019derivation}%
  \BibitemOpen
  \bibfield  {author} {\bibinfo {author} {\bibfnamefont {Y.}~\bibnamefont
  {Kusuki}}, \bibinfo {author} {\bibfnamefont {J.}~\bibnamefont {Kudler-Flam}},
  \ and\ \bibinfo {author} {\bibfnamefont {S.}~\bibnamefont {Ryu}},\
  }\href@noop {} {\bibfield  {journal} {\bibinfo  {journal} {Physical review
  letters}\ }\textbf {\bibinfo {volume} {123}},\ \bibinfo {pages} {131603}
  (\bibinfo {year} {2019})}\BibitemShut {NoStop}%
\bibitem [{\citenamefont {Jeong}\ \emph {et~al.}(2019)\citenamefont {Jeong},
  \citenamefont {Kim},\ and\ \citenamefont {Nishida}}]{jeong2019reflected}%
  \BibitemOpen
  \bibfield  {author} {\bibinfo {author} {\bibfnamefont {H.-S.}\ \bibnamefont
  {Jeong}}, \bibinfo {author} {\bibfnamefont {K.-Y.}\ \bibnamefont {Kim}}, \
  and\ \bibinfo {author} {\bibfnamefont {M.}~\bibnamefont {Nishida}},\
  }\href@noop {} {\bibfield  {journal} {\bibinfo  {journal} {Journal of High
  Energy Physics}\ }\textbf {\bibinfo {volume} {2019}},\ \bibinfo {pages} {1}
  (\bibinfo {year} {2019})}\BibitemShut {NoStop}%
\bibitem [{\citenamefont {Kusuki}\ and\ \citenamefont
  {Tamaoka}(2020{\natexlab{b}})}]{kusuki2020entanglement}%
  \BibitemOpen
  \bibfield  {author} {\bibinfo {author} {\bibfnamefont {Y.}~\bibnamefont
  {Kusuki}}\ and\ \bibinfo {author} {\bibfnamefont {K.}~\bibnamefont
  {Tamaoka}},\ }\href@noop {} {\bibfield  {journal} {\bibinfo  {journal}
  {Journal of High Energy Physics}\ }\textbf {\bibinfo {volume} {2020}},\
  \bibinfo {pages} {17} (\bibinfo {year} {2020}{\natexlab{b}})}\BibitemShut
  {NoStop}%
\bibitem [{\citenamefont {Umemoto}(2019)}]{umemoto2019quantum}%
  \BibitemOpen
  \bibfield  {author} {\bibinfo {author} {\bibfnamefont {K.}~\bibnamefont
  {Umemoto}},\ }\href@noop {} {\bibfield  {journal} {\bibinfo  {journal}
  {Physical Review D}\ }\textbf {\bibinfo {volume} {100}},\ \bibinfo {pages}
  {126021} (\bibinfo {year} {2019})}\BibitemShut {NoStop}%
\bibitem [{\citenamefont {Akers}\ and\ \citenamefont {Rath}(2020)}]{akers2020}%
  \BibitemOpen
  \bibfield  {author} {\bibinfo {author} {\bibfnamefont {C.}~\bibnamefont
  {Akers}}\ and\ \bibinfo {author} {\bibfnamefont {P.}~\bibnamefont {Rath}},\
  }\href@noop {} {\bibfield  {journal} {\bibinfo  {journal} {Journal of High
  Energy Physics}\ }\textbf {\bibinfo {volume} {2020}},\ \bibinfo {pages} {1}
  (\bibinfo {year} {2020})}\BibitemShut {NoStop}%
\bibitem [{\citenamefont {Bao}\ and\ \citenamefont
  {Cheng}(2019)}]{bao2019multipartite}%
  \BibitemOpen
  \bibfield  {author} {\bibinfo {author} {\bibfnamefont {N.}~\bibnamefont
  {Bao}}\ and\ \bibinfo {author} {\bibfnamefont {N.}~\bibnamefont {Cheng}},\
  }\href@noop {} {\bibfield  {journal} {\bibinfo  {journal} {Journal of High
  Energy Physics}\ }\textbf {\bibinfo {volume} {2019}},\ \bibinfo {pages} {1}
  (\bibinfo {year} {2019})}\BibitemShut {NoStop}%
\bibitem [{\citenamefont {Marolf}(2019)}]{Marolf2019CFTSA}%
  \BibitemOpen
  \bibfield  {author} {\bibinfo {author} {\bibfnamefont {D.}~\bibnamefont
  {Marolf}},\ }\href@noop {} {\bibfield  {journal} {\bibinfo  {journal}
  {Journal of High Energy Physics}\ }\textbf {\bibinfo {volume} {2020}},\
  \bibinfo {pages} {1} (\bibinfo {year} {2019})}\BibitemShut {NoStop}%
\bibitem [{\citenamefont {Mollabashi}\ and\ \citenamefont
  {Tamaoka}(2020)}]{mollabashi2020field}%
  \BibitemOpen
  \bibfield  {author} {\bibinfo {author} {\bibfnamefont {A.}~\bibnamefont
  {Mollabashi}}\ and\ \bibinfo {author} {\bibfnamefont {K.}~\bibnamefont
  {Tamaoka}},\ }\href@noop {} {\bibfield  {journal} {\bibinfo  {journal}
  {Journal of High Energy Physics}\ }\textbf {\bibinfo {volume} {2020}},\
  \bibinfo {pages} {1} (\bibinfo {year} {2020})}\BibitemShut {NoStop}%
\bibitem [{\citenamefont {Du}\ \emph {et~al.}(2020)\citenamefont {Du},
  \citenamefont {Shu},\ and\ \citenamefont {Zhu}}]{du2020inequalities}%
  \BibitemOpen
  \bibfield  {author} {\bibinfo {author} {\bibfnamefont {D.-H.}\ \bibnamefont
  {Du}}, \bibinfo {author} {\bibfnamefont {F.-W.}\ \bibnamefont {Shu}}, \ and\
  \bibinfo {author} {\bibfnamefont {K.-X.}\ \bibnamefont {Zhu}},\ }\href@noop
  {} {\bibfield  {journal} {\bibinfo  {journal} {The European Physical Journal
  C}\ }\textbf {\bibinfo {volume} {80}},\ \bibinfo {pages} {1} (\bibinfo {year}
  {2020})}\BibitemShut {NoStop}%
\bibitem [{\citenamefont {Ghodrati}(2021)}]{ghodrati2021correlations}%
  \BibitemOpen
  \bibfield  {author} {\bibinfo {author} {\bibfnamefont {M.}~\bibnamefont
  {Ghodrati}},\ }\href@noop {} {\bibfield  {journal} {\bibinfo  {journal}
  {arXiv preprint arXiv:2110.12970}\ } (\bibinfo {year} {2021})}\BibitemShut
  {NoStop}%
\bibitem [{\citenamefont {Umemoto}\ and\ \citenamefont
  {Zhou}(2018)}]{umemoto2018entanglement}%
  \BibitemOpen
  \bibfield  {author} {\bibinfo {author} {\bibfnamefont {K.}~\bibnamefont
  {Umemoto}}\ and\ \bibinfo {author} {\bibfnamefont {Y.}~\bibnamefont {Zhou}},\
  }\href@noop {} {\bibfield  {journal} {\bibinfo  {journal} {Journal of High
  Energy Physics}\ }\textbf {\bibinfo {volume} {2018}},\ \bibinfo {pages} {152}
  (\bibinfo {year} {2018})}\BibitemShut {NoStop}%
\bibitem [{\citenamefont {Levin}\ \emph {et~al.}(2020)\citenamefont {Levin},
  \citenamefont {DeWolfe},\ and\ \citenamefont {Smith}}]{LSD20}%
  \BibitemOpen
  \bibfield  {author} {\bibinfo {author} {\bibfnamefont {J.}~\bibnamefont
  {Levin}}, \bibinfo {author} {\bibfnamefont {O.}~\bibnamefont {DeWolfe}}, \
  and\ \bibinfo {author} {\bibfnamefont {G.}~\bibnamefont {Smith}},\ }\href
  {\doibase 10.1103/PhysRevD.101.046015} {\bibfield  {journal} {\bibinfo
  {journal} {Phys. Rev. D}\ }\textbf {\bibinfo {volume} {101}},\ \bibinfo
  {pages} {046015} (\bibinfo {year} {2020})}\BibitemShut {NoStop}%
\bibitem [{\citenamefont {Miyaji}\ and\ \citenamefont
  {Takayanagi}(2015)}]{SurfaceState}%
  \BibitemOpen
  \bibfield  {author} {\bibinfo {author} {\bibfnamefont {M.}~\bibnamefont
  {Miyaji}}\ and\ \bibinfo {author} {\bibfnamefont {T.}~\bibnamefont
  {Takayanagi}},\ }\href@noop {} {\bibfield  {journal} {\bibinfo  {journal}
  {Progress of Theoretical and Experimental Physics}\ }\textbf {\bibinfo
  {volume} {2015}} (\bibinfo {year} {2015})}\BibitemShut {NoStop}%
\bibitem [{\citenamefont {DeWolfe}\ \emph {et~al.}(2020)\citenamefont
  {DeWolfe}, \citenamefont {Levin},\ and\ \citenamefont {Smith}}]{LDS20}%
  \BibitemOpen
  \bibfield  {author} {\bibinfo {author} {\bibfnamefont {O.}~\bibnamefont
  {DeWolfe}}, \bibinfo {author} {\bibfnamefont {J.}~\bibnamefont {Levin}}, \
  and\ \bibinfo {author} {\bibfnamefont {G.}~\bibnamefont {Smith}},\ }\href
  {\doibase 10.1103/PhysRevD.102.066001} {\bibfield  {journal} {\bibinfo
  {journal} {Phys. Rev. D}\ }\textbf {\bibinfo {volume} {102}},\ \bibinfo
  {pages} {066001} (\bibinfo {year} {2020})}\BibitemShut {NoStop}%
\bibitem [{\citenamefont {Ba\~nados}\ \emph {et~al.}(1992)\citenamefont
  {Ba\~nados}, \citenamefont {Teitelboim},\ and\ \citenamefont
  {Zanelli}}]{BTZ}%
  \BibitemOpen
  \bibfield  {author} {\bibinfo {author} {\bibfnamefont {M.}~\bibnamefont
  {Ba\~nados}}, \bibinfo {author} {\bibfnamefont {C.}~\bibnamefont
  {Teitelboim}}, \ and\ \bibinfo {author} {\bibfnamefont {J.}~\bibnamefont
  {Zanelli}},\ }\href {\doibase 10.1103/PhysRevLett.69.1849} {\bibfield
  {journal} {\bibinfo  {journal} {Phys. Rev. Lett.}\ }\textbf {\bibinfo
  {volume} {69}},\ \bibinfo {pages} {1849} (\bibinfo {year}
  {1992})}\BibitemShut {NoStop}%
\bibitem [{\citenamefont {{Chen}}\ and\ \citenamefont
  {{Winter}}(2012)}]{NonAdd}%
  \BibitemOpen
  \bibfield  {author} {\bibinfo {author} {\bibfnamefont {J.}~\bibnamefont
  {{Chen}}}\ and\ \bibinfo {author} {\bibfnamefont {A.}~\bibnamefont
  {{Winter}}},\ }\href@noop {} {\bibfield  {journal} {\bibinfo  {journal}
  {ArXiv e-prints}\ } (\bibinfo {year} {2012})},\ \Eprint
  {http://arxiv.org/abs/1206.1307} {arXiv:1206.1307 [quant-ph]} \BibitemShut
  {NoStop}%
\bibitem [{\citenamefont {Smith}\ \emph {et~al.}(2008)\citenamefont {Smith},
  \citenamefont {Smolin},\ and\ \citenamefont {Winter}}]{SSW06}%
  \BibitemOpen
  \bibfield  {author} {\bibinfo {author} {\bibfnamefont {G.}~\bibnamefont
  {Smith}}, \bibinfo {author} {\bibfnamefont {J.}~\bibnamefont {Smolin}}, \
  and\ \bibinfo {author} {\bibfnamefont {A.}~\bibnamefont {Winter}},\
  }\href@noop {} {\bibfield  {journal} {\bibinfo  {journal} {IEEE Trans. Info.
  Theory}\ }\textbf {\bibinfo {volume} {54}},\ \bibinfo {pages} {4208}
  (\bibinfo {year} {2008})}\BibitemShut {NoStop}%
\bibitem [{\citenamefont {Ryu}\ and\ \citenamefont {Takayanagi}(2006)}]{RT}%
  \BibitemOpen
  \bibfield  {author} {\bibinfo {author} {\bibfnamefont {S.}~\bibnamefont
  {Ryu}}\ and\ \bibinfo {author} {\bibfnamefont {T.}~\bibnamefont
  {Takayanagi}},\ }\href@noop {} {\bibfield  {journal} {\bibinfo  {journal}
  {Phys. Rev. Lett.}\ }\textbf {\bibinfo {volume} {96}},\ \bibinfo {pages}
  {181602} (\bibinfo {year} {2006})}\BibitemShut {NoStop}%
\bibitem [{\citenamefont {Hubeny}\ \emph {et~al.}(2007)\citenamefont {Hubeny},
  \citenamefont {Rangamani},\ and\ \citenamefont {Takayanagi}}]{Hubeny:2007xt}%
  \BibitemOpen
  \bibfield  {author} {\bibinfo {author} {\bibfnamefont {V.~E.}\ \bibnamefont
  {Hubeny}}, \bibinfo {author} {\bibfnamefont {M.}~\bibnamefont {Rangamani}}, \
  and\ \bibinfo {author} {\bibfnamefont {T.}~\bibnamefont {Takayanagi}},\
  }\href {\doibase 10.1088/1126-6708/2007/07/062} {\bibfield  {journal}
  {\bibinfo  {journal} {JHEP}\ }\textbf {\bibinfo {volume} {07}},\ \bibinfo
  {pages} {062} (\bibinfo {year} {2007})},\ \Eprint
  {http://arxiv.org/abs/0705.0016} {arXiv:0705.0016 [hep-th]} \BibitemShut
  {NoStop}%
\bibitem [{\citenamefont {Faulkner}\ \emph {et~al.}(2013)\citenamefont
  {Faulkner}, \citenamefont {Lewkowycz},\ and\ \citenamefont
  {Maldacena}}]{Faulkner:2013ana}%
  \BibitemOpen
  \bibfield  {author} {\bibinfo {author} {\bibfnamefont {T.}~\bibnamefont
  {Faulkner}}, \bibinfo {author} {\bibfnamefont {A.}~\bibnamefont {Lewkowycz}},
  \ and\ \bibinfo {author} {\bibfnamefont {J.}~\bibnamefont {Maldacena}},\
  }\href {\doibase 10.1007/JHEP11(2013)074} {\bibfield  {journal} {\bibinfo
  {journal} {JHEP}\ }\textbf {\bibinfo {volume} {11}},\ \bibinfo {pages} {074}
  (\bibinfo {year} {2013})},\ \Eprint {http://arxiv.org/abs/1307.2892}
  {arXiv:1307.2892 [hep-th]} \BibitemShut {NoStop}%
\bibitem [{\citenamefont {Engelhardt}\ and\ \citenamefont
  {Wall}(2015)}]{engelhardt2015quantum}%
  \BibitemOpen
  \bibfield  {author} {\bibinfo {author} {\bibfnamefont {N.}~\bibnamefont
  {Engelhardt}}\ and\ \bibinfo {author} {\bibfnamefont {A.~C.}\ \bibnamefont
  {Wall}},\ }\href@noop {} {\bibfield  {journal} {\bibinfo  {journal} {Journal
  of High Energy Physics}\ }\textbf {\bibinfo {volume} {2015}},\ \bibinfo
  {pages} {1} (\bibinfo {year} {2015})}\BibitemShut {NoStop}%
\bibitem [{\citenamefont {Nakata}\ \emph {et~al.}(2021)\citenamefont {Nakata},
  \citenamefont {Takayanagi}, \citenamefont {Taki}, \citenamefont {Tamaoka},\
  and\ \citenamefont {Wei}}]{nakata2021new}%
  \BibitemOpen
  \bibfield  {author} {\bibinfo {author} {\bibfnamefont {Y.}~\bibnamefont
  {Nakata}}, \bibinfo {author} {\bibfnamefont {T.}~\bibnamefont {Takayanagi}},
  \bibinfo {author} {\bibfnamefont {Y.}~\bibnamefont {Taki}}, \bibinfo {author}
  {\bibfnamefont {K.}~\bibnamefont {Tamaoka}}, \ and\ \bibinfo {author}
  {\bibfnamefont {Z.}~\bibnamefont {Wei}},\ }\href@noop {} {\bibfield
  {journal} {\bibinfo  {journal} {Physical Review D}\ }\textbf {\bibinfo
  {volume} {103}},\ \bibinfo {pages} {026005} (\bibinfo {year}
  {2021})}\BibitemShut {NoStop}%
\bibitem [{\citenamefont {Hayden}\ \emph {et~al.}(2013)\citenamefont {Hayden},
  \citenamefont {Headrick},\ and\ \citenamefont {Maloney}}]{Hayden:2011ag}%
  \BibitemOpen
  \bibfield  {author} {\bibinfo {author} {\bibfnamefont {P.}~\bibnamefont
  {Hayden}}, \bibinfo {author} {\bibfnamefont {M.}~\bibnamefont {Headrick}}, \
  and\ \bibinfo {author} {\bibfnamefont {A.}~\bibnamefont {Maloney}},\ }\href
  {\doibase 10.1103/PhysRevD.87.046003} {\bibfield  {journal} {\bibinfo
  {journal} {Phys. Rev.}\ }\textbf {\bibinfo {volume} {D87}},\ \bibinfo {pages}
  {046003} (\bibinfo {year} {2013})},\ \Eprint {http://arxiv.org/abs/1107.2940}
  {arXiv:1107.2940 [hep-th]} \BibitemShut {NoStop}%
\bibitem [{\citenamefont {{Bao}}\ and\ \citenamefont {{\em
  et~al.}}(2015)}]{HEC15}%
  \BibitemOpen
  \bibfield  {author} {\bibinfo {author} {\bibfnamefont {N.}~\bibnamefont
  {{Bao}}}\ and\ \bibinfo {author} {\bibnamefont {{\em et~al.}}},\ }\href
  {\doibase 10.1007/JHEP09(2015)130} {\bibfield  {journal} {\bibinfo  {journal}
  {Journal of High Energy Physics}\ }\textbf {\bibinfo {volume} {9}},\ \bibinfo
  {eid} {130} (\bibinfo {year} {2015})},\ \Eprint
  {http://arxiv.org/abs/1505.07839} {arXiv:1505.07839 [hep-th]} \BibitemShut
  {NoStop}%
\bibitem [{\citenamefont {Swingle}(2012)}]{Swing12}%
  \BibitemOpen
  \bibfield  {author} {\bibinfo {author} {\bibfnamefont {B.}~\bibnamefont
  {Swingle}},\ }\href@noop {} {\bibfield  {journal} {\bibinfo  {journal} {Phys.
  Rev. D}\ }\textbf {\bibinfo {volume} {86}},\ \bibinfo {pages} {065007}
  (\bibinfo {year} {2012})}\BibitemShut {NoStop}%
\bibitem [{\citenamefont {{Pastawski}}\ \emph {et~al.}(2015)\citenamefont
  {{Pastawski}}, \citenamefont {{Yoshida}}, \citenamefont {{Harlow}},\ and\
  \citenamefont {{Preskill}}}]{Happy15}%
  \BibitemOpen
  \bibfield  {author} {\bibinfo {author} {\bibfnamefont {F.}~\bibnamefont
  {{Pastawski}}}, \bibinfo {author} {\bibfnamefont {B.}~\bibnamefont
  {{Yoshida}}}, \bibinfo {author} {\bibfnamefont {D.}~\bibnamefont {{Harlow}}},
  \ and\ \bibinfo {author} {\bibfnamefont {J.}~\bibnamefont {{Preskill}}},\
  }\href@noop {} {\bibfield  {journal} {\bibinfo  {journal} {Journal of High
  Energy Physics}\ }\textbf {\bibinfo {volume} {6}},\ \bibinfo {eid} {149}
  (\bibinfo {year} {2015})}\BibitemShut {NoStop}%
\bibitem [{\citenamefont {Hayden}\ \emph {et~al.}(2016)\citenamefont {Hayden},
  \citenamefont {Nezami}, \citenamefont {Qi}, \citenamefont {Thomas},
  \citenamefont {Walter},\ and\ \citenamefont {Yang}}]{hayden2016holographic}%
  \BibitemOpen
  \bibfield  {author} {\bibinfo {author} {\bibfnamefont {P.}~\bibnamefont
  {Hayden}}, \bibinfo {author} {\bibfnamefont {S.}~\bibnamefont {Nezami}},
  \bibinfo {author} {\bibfnamefont {X.-L.}\ \bibnamefont {Qi}}, \bibinfo
  {author} {\bibfnamefont {N.}~\bibnamefont {Thomas}}, \bibinfo {author}
  {\bibfnamefont {M.}~\bibnamefont {Walter}}, \ and\ \bibinfo {author}
  {\bibfnamefont {Z.}~\bibnamefont {Yang}},\ }\href@noop {} {\bibfield
  {journal} {\bibinfo  {journal} {Journal of High Energy Physics}\ }\textbf
  {\bibinfo {volume} {2016}},\ \bibinfo {pages} {9} (\bibinfo {year}
  {2016})}\BibitemShut {NoStop}%
\bibitem [{\citenamefont {Nezami}\ and\ \citenamefont
  {Walter}(2020)}]{nezami2020multipartite}%
  \BibitemOpen
  \bibfield  {author} {\bibinfo {author} {\bibfnamefont {S.}~\bibnamefont
  {Nezami}}\ and\ \bibinfo {author} {\bibfnamefont {M.}~\bibnamefont
  {Walter}},\ }\href@noop {} {\bibfield  {journal} {\bibinfo  {journal}
  {Physical Review Letters}\ }\textbf {\bibinfo {volume} {125}},\ \bibinfo
  {pages} {241602} (\bibinfo {year} {2020})}\BibitemShut {NoStop}%
\bibitem [{\citenamefont {Deutsch}(1991)}]{PhysRevA.43.2046}%
  \BibitemOpen
  \bibfield  {author} {\bibinfo {author} {\bibfnamefont {J.~M.}\ \bibnamefont
  {Deutsch}},\ }\href {\doibase 10.1103/PhysRevA.43.2046} {\bibfield  {journal}
  {\bibinfo  {journal} {Phys. Rev. A}\ }\textbf {\bibinfo {volume} {43}},\
  \bibinfo {pages} {2046} (\bibinfo {year} {1991})}\BibitemShut {NoStop}%
\bibitem [{\citenamefont {Srednicki}(1994)}]{PhysRevE.50.888}%
  \BibitemOpen
  \bibfield  {author} {\bibinfo {author} {\bibfnamefont {M.}~\bibnamefont
  {Srednicki}},\ }\href {\doibase 10.1103/PhysRevE.50.888} {\bibfield
  {journal} {\bibinfo  {journal} {Phys. Rev. E}\ }\textbf {\bibinfo {volume}
  {50}},\ \bibinfo {pages} {888} (\bibinfo {year} {1994})}\BibitemShut
  {NoStop}%
\bibitem [{\citenamefont {Rigol}\ \emph {et~al.}(2008)\citenamefont {Rigol},
  \citenamefont {Dunjko},\ and\ \citenamefont {Olshanii}}]{Rigol2008}%
  \BibitemOpen
  \bibfield  {author} {\bibinfo {author} {\bibfnamefont {M.}~\bibnamefont
  {Rigol}}, \bibinfo {author} {\bibfnamefont {V.}~\bibnamefont {Dunjko}}, \
  and\ \bibinfo {author} {\bibfnamefont {M.}~\bibnamefont {Olshanii}},\ }\href
  {\doibase 10.1038/nature06838} {\bibfield  {journal} {\bibinfo  {journal}
  {Nature}\ }\textbf {\bibinfo {volume} {452}},\ \bibinfo {pages} {854–858}
  (\bibinfo {year} {2008})}\BibitemShut {NoStop}%
\bibitem [{\citenamefont {Brill}(2000)}]{Brill:1999xm}%
  \BibitemOpen
  \bibfield  {author} {\bibinfo {author} {\bibfnamefont {D.}~\bibnamefont
  {Brill}},\ }\href {\doibase
  10.1002/(SICI)1521-3889(200005)9:3/5<217::AID-ANDP217>3.0.CO;2-H} {\bibfield
  {journal} {\bibinfo  {journal} {Annalen Phys.}\ }\textbf {\bibinfo {volume}
  {9}},\ \bibinfo {pages} {217} (\bibinfo {year} {2000})},\ \Eprint
  {http://arxiv.org/abs/gr-qc/9912079} {arXiv:gr-qc/9912079} \BibitemShut
  {NoStop}%
\end{thebibliography}%

\section*{Appendix A: Inequalities for bipartite optimized correlation measures}

We can write the bipartite optimized correlation measures as
\begin{eqnarray}
	E_\alpha(A:B) = \inf_{|\psi\rangle} f^\alpha(A, B, a, b) \,,
\end{eqnarray}
with objective functions
\begin{eqnarray}
	f_P(A,B,a,b) &=& S(Aa)\nonumber\\
	f_R(A,B,a,b) &=& S(Aa)  + {1 \over 2}(S(AB) - S(a) - S(b))\\
	f_Q(A,B,a,b) &=&  {1 \over 2}(S(A) + S(B) + S(Aa) - S(Ba))\nonumber\\
	f_{\rm sq} (A,B,a,b) &=&  {1 \over 2}(S(Aa) + S(Ba) - S(a) - S(b) )\nonumber \,.
\end{eqnarray}
We may consider $I(A:B)/2  \equiv \frac{1}{2}(S(A) + S(B) - S(AB))$ as the fifth bipartite correlation measure (obeying monotonicity and symmetry); since  it it not optimized we can think of it as being equal to its own objective function. Here we summarize the inequalities obeyed by these measures.

For any given purification $ab$ of our state, we can always consider setting $a=0$, so that $ABb$ is pure. We then find $f_P(A,B,0,b) = f_R(A,B,0,b) = f_Q(A,B,0,b) = S(A)$. Similarly for $b=0$, these three objective functions equal $S(B)$. Since these are possible, though not necessarily optimal, purifications we  have
\begin{eqnarray}
	E_P, E_R, E_Q \leq \min (S(A), S(B)) \,.
\end{eqnarray}
Likewise we notice $f_{\rm sq}(A,B,0,b) = I(A:B)/2$, and thus
\begin{eqnarray}
	E_{\rm sq} \leq {1 \over 2} I(A:B) \,.
\end{eqnarray}
Next, we can use the fact that an inequality on objective functions leads to an inequality on the corresponding correlation measures,
\begin{eqnarray}
	f^\alpha(A,B,a,b) - f^\beta(A,B,a,b)  \geq 0 \quad \to \quad E_\alpha(A:B) \geq E_\beta(A;B)\,.
\end{eqnarray}
To see this, let $a^*, b^*$ be the purification optimizing $E_\alpha$, so $E_\alpha(A,B) = f^\alpha(A,B,a^*, b^*)$. Then $a^*, b^*$ is a possible but generally non-optimal purification for $E_\beta$, so $E_\beta(A,B) \leq f^\beta(A,B, a^*, b^*)$. Using  $f^\beta(A,b,a^*,b^*) \leq f^\alpha(A,b,a^*,b^*)$ then obtains the result.

Now from the definitions of the objective functions we can show that 
\begin{eqnarray}
f^P(A,B,a,b) - f^R(A,B,a,b) &=& {1 \over 2} I(a:b)\nonumber\\
	f^P(A,B,a,b) - f^Q(A,B,a,b) &=& {1 \over 2} I(a:b|A)\\
	f^R(A,B,a,b) - {1 \over 2} I(A:B) &=& {1 \over 2}I(a:B|A) + {1 \over 2}I(b:A|B)\nonumber\\
	f^Q(A,B,a,b) - {1 \over 2}I(A:B) &=&{1 \over 2}I(a:B|A) + {1 \over 2}I(b:A) \nonumber\,.
\end{eqnarray}
All of the right-hand sides are conic combinations of mutual informations and conditional mutual informations, and are hence non-negative. This implies
\begin{eqnarray}
	{1 \over 2} I(A:B) \leq E_R, E_Q \leq E_P\,,
\end{eqnarray}
and combining this with earlier results we have
\begin{eqnarray}
\label{GeneralInequality}
	E_{\rm sq} \leq {1 \over 2} I(A:B) \leq E_R, E_Q \leq E_P \leq \min(S(A), S(B)) \,,
\end{eqnarray}
establishing a hierarchy for almost every measure.

Regarding $E_Q$ and $E_R$, one can show their difference is related to the tripartite information $I_3$,  
\begin{eqnarray}
	f^R(A,B,a,b) - f^Q(A,B,a,b) &=& - {1 \over 2}I_3(A:B:a) \,,
\end{eqnarray}
which for a general quantum system has no definite sign; thus $E_Q$ and $E_R$ have no general relative relationship and (\ref{GeneralInequality}) is as far as we can go for generic states. However, for holographic states the tripartite information is known to be non-positive, known as the monogamy of mutual information. Thus for holographic states we can further refine our chain of inequalities to
\begin{eqnarray}
	E_{\rm sq} \leq {1 \over 2} I(A:B) \leq E_Q \leq  E_R \leq E_P \leq \min(S(A), S(B)) \,,
\end{eqnarray}
giving a full ordered sequence for all five measures.

\section*{Appendix B: BTZ spacetime and relationship to AdS}

In this paper we are interested in three-dimensional AdS black holes, the so-called Ba\~nados-Teitelboim-Zanelli (BTZ) geometries \cite{BTZ}. Here we review how they manifest as quotients of empty anti-de Sitter space.

The metric of three-dimensional anti-de Sitter space in global coordinates can be written as
\begin{eqnarray}
\label{PureAdS}
	ds^2 = -\left(1 + {\rho^2 \over \ell^2}\right)dt^2 + {d\rho^2 \over 1 + \rho^2/\ell^2} + \rho^2 d\phi^2 \,, 
\end{eqnarray}
where $\ell$ is the AdS length scale, and the boundary of AdS is at $\rho \to \infty$. It can be useful to map the boundary to a finite coordinate value, and one transformation that accomplishes this is
\begin{eqnarray}
	\rho \equiv {2r \over 1-r^2/\ell^2}\,, \quad \quad \phi = \theta \,,
\end{eqnarray}
which leads to the alternate AdS metric,
\begin{eqnarray}
\label{AdSDisk}
	ds^2 = - {(1 + r^2/\ell^2)^2\over (1 - r^2/\ell^2)^2}dt^2  +{4 \over (1 - r^2/\ell^2)^2} (dr^2 + r^2 d\theta^2) \,,
\end{eqnarray}
with the boundary occurring at $r = \ell$. Constant time slices of the metric (\ref{AdSDisk}) have the form of the Poincar\'e disk, a common presentation of two-dimensional hyperbolic space. Geodesics on the Poincar\'e disk take a simple form: they are segments of circles that intersect the edge of the disk at right angles.

The BTZ spacetime is a black hole geometry living inside AdS$_3$. We can write the metric as
\begin{eqnarray}
\label{BTZ}
	ds^2 = - f(\rho) dt^2 + f(\rho)^{-1} d\rho^2 + \rho^2 d\phi^2 \,,
\end{eqnarray}		
where
\begin{eqnarray}
	f(\rho) \equiv {\rho^2 \over \ell^2} - m \,.
\end{eqnarray}
The geometry is asymptotically AdS with radius $\ell$ as $\rho \to \infty$. $m$ is the mass parameter of the black hole, with the horizon located at the radius $r_H$,
\begin{eqnarray}
	f(r_H) = 0 \,, \quad\to\quad r_H \equiv \ell\sqrt{m}  \,.
\end{eqnarray}
Unlike the case for higher dimensions, the metric of AdS$_3$ (\ref{PureAdS}) is not recovered in the massless limit $m\to0$, but for $m=-1$. This is a reflection of the ``gapped" nature of AdS$_3$, with the vacuum separated from the set of black hole states by a finite mass gap.

\begin{figure}[H]
    \centering
    \includegraphics[scale = .35]{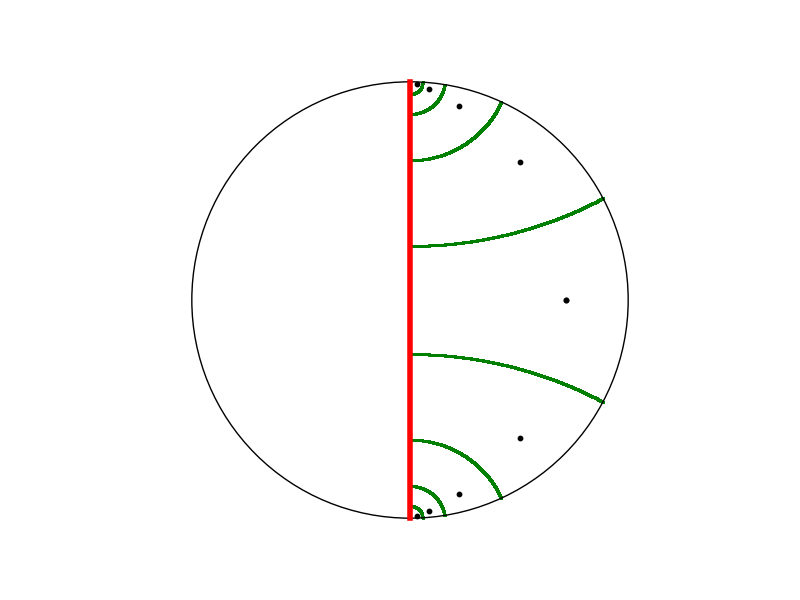}
    \caption{The Poincar\'e disk showing the embedding of multiple copies of BTZ under the coordinate map (\ref{BTZtoAdS}).  The green curves are the images of the BTZ lines $\phi = (2n+1)\pi$ for integer $n$, and each ``cell" between pairs of green lines is a complete copy of BTZ. The BTZ horizon is mapped to a segment of the red line, and the BTZ boundary to a segment of the disk boundary. The dots in each cell are an example of points that are identified under the quotient. There are an infinite number of cells, though most look very small in these coordinates.}
    \label{fig:PoincareDisk}
\end{figure}

The BTZ spacetime is locally equivalent to a piece of AdS$_3$. Under the transformation \cite{Brill:1999xm},
\begin{eqnarray}
\label{BTZtoAdS}
	{r^2 \over \ell^2} \equiv  { \rho \cosh( \sqrt{m}\phi) - \sqrt{m} \ell \over \rho \cosh( \sqrt{m}\phi) + \sqrt{m} \ell  } \,, \quad \quad
	\cos^2 \theta \equiv {\rho^2 - m \ell^2\over\rho^2 \cosh^2 (\sqrt{m} \phi)- m\ell^2 } \,,
\end{eqnarray}
with the sign of $\theta$ chosen to match the sign of $\phi$, the BTZ metric (\ref{BTZ}) becomes the metric for AdS$_3$ in Poincar\'e disk form (\ref{AdSDisk}). However, the image of BTZ under this map is only a portion of AdS$_3$. Consider a constant time slice. The boundary of the BTZ geometry is mapped to an arc of the edge of the Poincar\'e disk, while the BTZ horizon $\rho = r_H$ is mapped to a segment of the vertical line bisecting the middle of the disk. Then the lines $\phi = \pm \pi$, which coincide in BTZ, have distinct images on the disk. Thus all of BTZ, parameterized by $\phi \in (-\pi, \pi)$, is mapped to a single ``cell" or ``fundamental domain" bounded by these lines, as in  figure~\ref{fig:PoincareDisk}. This region on the Poincar\'e disk becomes arbitrarily narrow as the black hole mass vanishes $m \to 0$, while it expands to fill the right-hand-side of the disk as $m \to \infty$.

The BTZ geometry for other values of $\phi \in (-\infty, \infty)$ fills out an infinite number of other cells that span the right-hand-side of the Poincar\'e disk, as in figure~\ref{fig:PoincareDisk}. Thus a single copy of BTZ corresponds to a quotient of AdS$_3$ under the image of $\phi \sim \phi + 2\pi$, identifying all these cells. Meanwhile, the left-hand-side of the Poincar\'e disk can be thought of as a duplicate asymptotic region beyond the horizon, corresponding to the maximally extended black hole spacetime, and its images.

\end{document}